\documentclass[11pt]{article}
\pdfoutput=1
\usepackage{jcapmod}

\usepackage{shorthand}
\usepackage{mathtools}
\usepackage{booktabs}
\usepackage[english]{babel}
\usepackage{amsmath,amssymb,amsbsy,amstext, amsthm, simplewick, amsfonts}
\usepackage{hyperref}
\usepackage{graphicx}
\usepackage[small]{caption}
\usepackage{siunitx}
\usepackage{upgreek}
\usepackage{framed}
\usepackage{wrapfig}
\usepackage{multirow}
\usepackage{bbm}
\usepackage[svgnames,dvipsnames,x11names]{xcolor}

\usepackage{selinput}

\usepackage{bm}
\usepackage{float}
\usepackage{geometry}
\usepackage{yfonts}
\usepackage{caption}
\usepackage{subcaption}
\usepackage{sidecap}
\usepackage{longtable}
\usepackage{anyfontsize}
\usepackage{dsfont}
\usepackage{tikz}
\usepackage{relsize}
\usepackage{tcolorbox}

\usepackage{xcolor}
\usepackage{xparse}

\usepackage{slashed}
\usepackage{simpler-wick}

\NewDocumentCommand{\colornucleus}{omme{_^}}{%
  \begingroup\colorlet{currcolor}{.}%
  \IfValueTF{#1}
   {\textcolor[#1]{#2}}
   {\textcolor{#2}}
    {%
     #3% the nucleus
     \IfValueT{#4}{_{\textcolor{currcolor}{#4}}}% subscript
     \IfValueT{#5}{^{\textcolor{currcolor}{#5}}}% superscript
    }%
  \endgroup
}

\usepackage{array}
\newcolumntype{L}[1]{>{\raggedright\let\newline\\\arraybackslash\hspace{0pt}}m{#1}}
\newcolumntype{C}[1]{>{\centering\let\newline\\\arraybackslash\hspace{0pt}}m{#1}}
\newcolumntype{R}[1]{>{\raggedleft\let\newline\\\arraybackslash\hspace{0pt}}m{#1}}

\usepackage[framemethod=default]{mdframed}
\newmdenv[skipabove=7pt,
skipbelow=7pt,
rightline=false,
leftline=false,
topline=false,
bottomline=false,
backgroundcolor=gray!10,
linecolor=gray,
innerleftmargin=5pt,
innerrightmargin=5pt,
innertopmargin=5pt,
innerbottommargin=5pt,
leftmargin=0cm,
rightmargin=0cm,
linewidth=4pt]{eBox}
\newmdenv[skipabove=7pt,
skipbelow=7pt,
rightline=false,
leftline=false,
topline=false,
bottomline=false,
backgroundcolor=gray!10,
linecolor=gray,
innerleftmargin=5pt,
innerrightmargin=5pt,
innertopmargin=-5pt,
innerbottommargin=5pt,
leftmargin=0cm,
rightmargin=0cm,
linewidth=4pt]{eBox2}
\usepackage[framemethod=default]{mdframed}
\newmdenv[skipabove=7pt,
skipbelow=7pt,
rightline=true,
leftline=true,
topline=true,
bottomline=true,
backgroundcolor=gray!15,
linecolor=gray,
innerleftmargin=5pt,
innerrightmargin=5pt,
innertopmargin=5pt,
innerbottommargin=5pt,
leftmargin=0cm,
rightmargin=0cm,
linewidth=0.75pt]{eBox3}

\definecolor{Red}{RGB}{214, 39, 40}
\definecolor{Blue}{RGB} {31, 119, 180}
\definecolor{Orange}{RGB}{255, 153, 51}
\definecolor{Purple}{RGB}{178, 102, 255}
\definecolor{Green}{RGB}{44, 160, 44}

\definecolor{vio}{RGB}{19, 130, 164}
\definecolor{vioo}{RGB}{89, 2, 155}
\newcommand{\Comment}[1]{{}}
\definecolor{darkblue}{rgb}{0.15,0.35,0.55}
\definecolor{reddish}{rgb}{0.65, 0.2, 0.2}
\definecolor{darkgreen}{RGB}{50,150,0}
\definecolor{greyish}{rgb}{.90,.90,.90}
\definecolor{greyish2}{rgb}{.96,.96,.96}
\definecolor{greyish3}{rgb}{.37,.37,.37}
\definecolor{darkblue2}{rgb}{0.3,0.4,0.9}
\definecolor{Blue3}{RGB}{31, 119, 180}
\usepackage[linktocpage=true]{hyperref}
\hypersetup{
colorlinks=true,
citecolor=darkblue,
linkcolor=reddish,
urlcolor=darkblue,
pdfauthor={},
pdftitle={},
pdfsubject={}
}

\newcommand{\db}[1]{\textcolor{Red}{{\bf DB}: #1}} 
\usepackage{colortbl}
\definecolor{lightgreen}{cmyk}{0.2, 0, 0.2, 0.2}
\definecolor{lightgray2}{cmyk}{0.1,0.1,0,0.1}
\definecolor{Red2}{RGB}{214, 39, 40}
\definecolor{Blue2}{RGB} {31, 119, 180}
\definecolor{Orange2}{RGB}{255, 127, 14}
\definecolor{Green2}{RGB}{44, 160, 44}

\makeatletter
\newlength{\apb@width}
\newcommand{\autoparbox}[2][c]{\settowidth{\apb@width}{#2}\parbox[#1]{\apb@width}{#2}}

\makeatother


\def\hs{\hskip 1pt}

\def\beq{\begin{equation}}
\def\eeq{\end{equation}}
\def\be{\begin{equation}}
\def\ee{\end{equation}}

\def\a{{\hat \alpha}}

\DeclarePairedDelimiter\floor{\lfloor}{\rfloor}

\allowdisplaybreaks[1]
\begin{document}

\newgeometry{top=2cm, bottom=2cm, left=2cm, right=2cm}

\begin{titlepage}
\setcounter{page}{1} \baselineskip=15.5pt 
\thispagestyle{empty}

\begin{center}
{\fontsize{21}{18} \bf A New Twist on Spinning (A)dS Correlators }
\end{center}

\vskip 20pt
\begin{center}
\noindent
{\fontsize{14}{18}\selectfont 
Daniel Baumann\hs$^{1,2,3,4}$, Gr\'egoire Mathys\hs$^{5}$,\\[8pt]
Guilherme L.~Pimentel\hs$^{6}$  and Facundo Rost\hs$^{1,3}$}
\end{center}

\begin{center}
  \vskip8pt
\textit{$^1$  Leung Center for Cosmology and Particle Astrophysics,
Taipei 10617, Taiwan}

  \vskip8pt
\textit{$^2$  Center for Theoretical Physics,
National Taiwan University, Taipei 10617, Taiwan}

  \vskip8pt
\textit{$^3$ Institute of Physics, University of Amsterdam, Amsterdam, 1098 XH, The Netherlands}

  \vskip8pt
\textit{$^4$ Max-Planck-Institut f\"ur Physik, Werner-Heisenberg-Institut, 85748 Garching bei M\"unchen, Germany}

\vskip 8pt
\textit{$^5$ Institute of Physics,
Ecole Polytechnique F\'ed\'eral de Lausanne,
CH-1015 Lausanne, Switzerland }

\vskip 8pt
\textit{$^6$ Scuola Normale Superiore and INFN, Piazza dei Cavalieri 7, 56126, Pisa, Italy}
\end{center}

%=========================================
\vspace{0.4cm}
\begin{center}{\bf Abstract}
\end{center}
\noindent
Massless spinning correlators in cosmology are extremely complicated. In contrast,
the scattering amplitudes of massless particles with spin are very simple. We propose that the reason for the unreasonable complexity of these correlators lies in the use of inconvenient kinematic variables. For example, in de Sitter space, consistency with unitarity and the background isometries imply that the correlators must be conformally covariant and also conserved. However, the commonly used kinematic variables for correlators do not make all of these properties manifest.
In this paper, we introduce twistor space as a powerful way to satisfy all kinematic constraints.
We show that conformal correlators of conserved currents can be written as twistor integrals, where the conservation condition 
translates into holomorphicity of the integrand. The functional form of the twistor-space correlators is very simple and easily bootstrapped. For the case of three-point functions, we verify explicitly that this reproduces known results in embedding space. We also perform a half-Fourier transform of the twistor-space correlators to obtain their counterparts in momentum space. We conclude that twistors provide a promising new avenue to study conformal correlation functions that exposes their hidden simplicity. 
\end{titlepage}
\restoregeometry

\newpage
\setcounter{tocdepth}{3}
\setcounter{page}{2}

\linespread{1.2}
\tableofcontents
\linespread{1.1}

\newpage
\section{Introduction}

Many exciting advances in theoretical physics had their beginnings in a significant improvement in the foundations of the subject. For example, the recent renaissance of the conformal bootstrap~\cite{Rattazzi:2008pe} 
can be traced  to a better understanding of its kinematic building blocks, such as three-point functions \cite{osborn1994implications,Erdmenger:1996yc,osborncft,Weinberg:2010fx,Costa:2011mg} and conformal blocks \cite{Dolan:2000ut,Dolan:2003hv,Costa:2011dw}, which 
has been essential for harnessing the full power of conformal symmetry. With kinematics under much better control, many interesting dynamical questions became answerable through a powerful mix of numerical and analytical techniques~\cite{Poland:2018epd}.

\vskip 4pt
Similarly, the S-matrix bootstrap was powered in its revival by an improved understanding of the kinematics of scattering amplitudes \cite{Dixon:1996wi}. 
In particular, the three-point amplitudes of massless particles reveal their essential features only in spinor helicity variables~\cite{Cheung:2017pzi, Elvang:2015rqa}. 
From these three-point building blocks, all consistent theories of long-range interactions can then be derived by demanding proper factorization of the amplitudes at four points~\cite{Benincasa:2007xk, McGady:2013sga}. Moreover, as the kinematic constraints were made more obvious, the kinematic data started inhabiting mathematical spaces called positive geometries~\cite{Arkani-Hamed:2013jha,Arkani-Hamed:2017tmz}. 
This has led to entirely different ways of conceptualizing what scattering amplitudes are by thinking of them as volumes of these geometries.

\vskip 4pt
These two subjects---conformal field theory and scattering amplitudes---come together in 
primordial cosmology. First, all cosmological correlators contain singularities when the total energy of a graph (or subgraph) vanishes and the residues of these singularities are the corresponding flat-space amplitudes~\cite{Raju:2012zr, Maldacena:2011nz}. Second, in the case of slow-roll inflation with weak interactions, these correlators inherit an approximate conformal symmetry  from the isometries of the de Sitter spacetime. It is therefore natural to expect that the tools and perspectives developed in conformal field theory and for scattering amplitudes are relevant in the cosmological context~\cite{Baumann:2022jpr}. 

\vskip 4pt
In contrast to conformal field theory and scattering amplitudes, however, the foundations of primordial cosmology aren't yet fully developed.
A particularly embarrassing fact is that the correlators of spinning fields in de Sitter space are still poorly understood. Even for three-point correlators, there is currently no presentation that makes all of their symmetries evident and the kinematically allowed structures ``inevitable."  The problem gets worse at higher points, where one finds an explosion of complexity for gluon and graviton correlators  (see e.g.~\cite{Baumann:2020dch, Baumann:2021fxj, Albayrak:2020fyp, Bonifacio:2022vwa}). As we will argue in this paper, the reason for the unreasonable complexity of spinning correlators lies in an inconvenient choice of kinematic variables. Moreover, we will show that formulating these correlators using the language of twistors exposes their hidden simplicity. 

\vskip 4pt
The natural habitat of cosmological correlators is {\it Fourier space}. 
The reasons that cosmologists like to work in Fourier space are that first, the distinct Fourier modes of perturbations, $\vec k_i$, evolve independently at linear order, and second, the homogeneity of the background spacetime implies three-dimensional momentum conservation, $\sum_i \vec{k}_i = 0$.  Another advantage of Fourier space is that current conservation is easy to implement as a transversality condition on polarization vectors. However, an important drawback  of working in Fourier space is that conformal symmetry is hard to enforce, taking the form of a partial differential equation---the conformal Ward identity~\cite{Bzowski:2013sza}.

\begin{figure}[t!]
   \centering
      \includegraphics[width=0.9\textwidth]{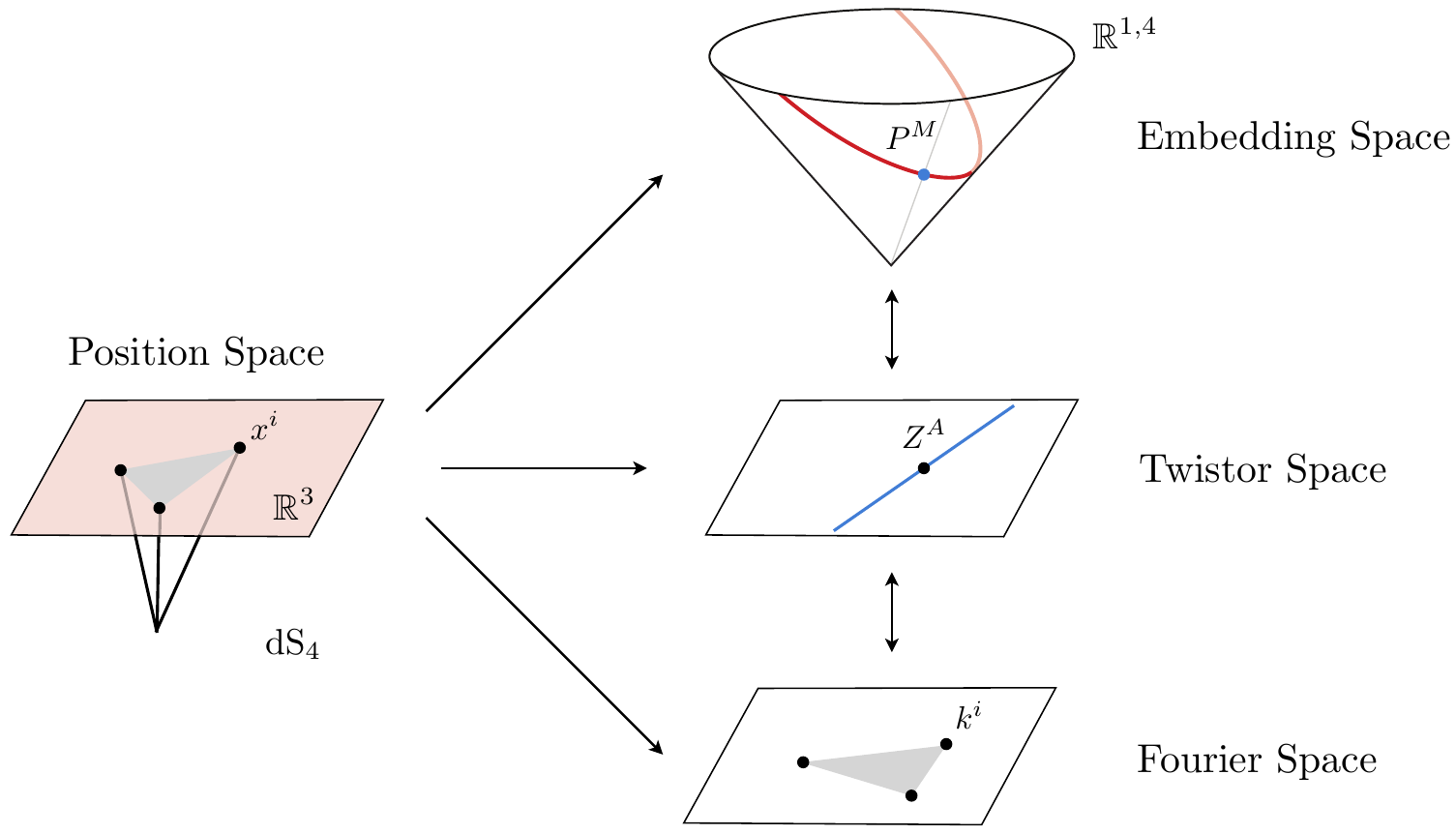} 
      \caption{Outline of the different kinematic spaces discussed in this paper. The natural arena in cosmology is Fourier space, but conformal symmetry is hard to implement there. In conformal field theory, we therefore prefer to work in embedding space, but current conservation is then a nontrivial constraint. In this paper, we study conformal correlators of conserved currents in twistor space, in which all kinematic constraints can be made manifest.}
      \label{fig:RoadMap}
\end{figure}

\vskip 4pt
Conformal field theories are therefore often expressed in the so-called {\it embedding space} (introduced by Dirac~\cite{Dirac:1936fq} in 1936 and reinvigorated in \cite{Costa:2011mg}). 
This starts from the observation that the conformal algebra on $\mathbb{R}^d$ is isomorphic to the algebra of Lorentz transformations on~$\mathbb{R}^{1,d+1}$. 
By defining a suitable embedding of $\mathbb{R}^d$ into~$\mathbb{R}^{1,d+1}$ (see Fig.\,\ref{fig:RoadMap}), the $d$-dimensional conformal transformations (which are nonlinear and complicated) can be uplifted to $(d+2)$-dimensional Lorentz transformations (which are linear and simple). However, while conformal symmetry becomes much simpler to enforce in embedding space, current conservation is now harder to implement~\cite{Costa:2011mg}. In practice, one first has to define the space of conformally-invariant structures and then impose a differential constraint to select the sub-space of conserved correlators. This two-step procedure is conceptually unsatisfying. It would be much nicer if we could write down an expression for the correlator that from the start makes both conformal symmetry and current conservation manifest.
In this paper, we will show that this can be achieved in {\it twistor space}\footnote{We would like to thank Mariana Carrillo Gonz\'alez for first suggesting to us the connection between conformal correlators and (mini) twistor space \cite{Ward:1989vja,Adamo:2017xaf,CarrilloGonzalez:2022ggn}.} (see~\cite{Neiman:2013hca,Neiman:2017mel} for important prior work). 

\vskip 4pt
First, we will introduce embedding-space spinors $\Lambda_a^A$~\cite{Binder:2020raz,Chiodaroli:2022ssi} 
and then define twistor coordinates as a specific linear combination of these spinors~$Z^A \equiv \pi^a \Lambda_a^A$. 
This allows us to write an integral representation for correlators that automatically ensures both conformal symmetry and current conservation. Conserved correlators in twistor space are holomorphic functions of the coordinates $Z^A_i$ (for each field $i$) whose precise form is easily bootstrapped. For the case of three-point functions, we verify explicitly that this reproduces known results in embedding space. Unlike in embedding space, however, current conservation is part of the ansatz and doesn't have to be imposed by hand as an additional constraint. 

\vskip 4pt
As a simple illustration of the power of these ideas, consider the three-point function of gravitons in twistor space:
\begin{equation}
\langle T(Z_1)T(Z_2)T(Z_3)\rangle^{\pm} =\int {\rm d} c_{ij} ~{\rm exp}\left(i\sum_{i,j} c_{ij}Z_i \cdot Z_j\right) (c_{12}c_{23}c_{31})^{\pm 2}\, ,
\end{equation}
where $Z_i$ are the twistor coordinates and $c_{ij}$ are Schwinger parameters.
The choice of $-$ or $+$ in the exponent corresponds to Einstein gravity and higher-derivative  $R^3$ interactions, respectively. Note that the Schwinger-parameterized correlator, $A^\pm(c_{ij}) = (c_{12}c_{23}c_{31})^{\pm 2}$, is remarkably simple.

\vskip 4pt
Following~\cite{Nair:1988bq,Witten:2003nn,Arkani-Hamed:2009hub,Mason:2009sa}, we also perform half-Fourier transforms of our twistor-space results to connect them to the corresponding correlators in momentum space. Interestingly, this half-Fourier transform is somewhat subtle, and requires careful analytic continuation to Euclidean momenta. These computations will demonstrate that our correlators are indeed correct, and hopefully convince the reader that the first presentation, in terms of twistors, is much simpler.

\vskip 4pt
Twistors are extremely useful variables for scattering amplitudes in flat space \cite{Witten:2003nn,Arkani-Hamed:2009hub,Mason:2009sa}, making the four-dimensional conformal symmetry of gauge-theory amplitudes manifest. Our proposed usage of twistors here is more primitive --- we are simply trying to make all the constraints from (Anti-)de Sitter isometries manifest. In that sense, twistors in (A)dS are more akin to flat-space spinor helicity variables. Their usage will therefore be more widely applicable than in flat space, which relies to some extent in additional symmetries beyond Poincar\'e.

\vskip 4pt
Although, in this paper, we only showcase the power of twistors for three-point functions, the simplicity and inevitability of our results suggests that we have much to look forward to. For example, a natural next goal is to obtain four-point functions from consistent factorization, thus simplifying the known expressions in momentum space and bootstrapping them from basic principles. We furthermore believe that we will then be geared up to find cosmological recursion relations that are as simple as those of flat space.  Finally, we are optimistic that, within twistor space, we are coming closer to an all-multiplicity formula for gluon correlators in de Sitter space like the Parke--Taylor formula for gluon amplitudes in flat space.  We will pursue these avenues in future work.

\vspace{0.25cm}
\paragraph{Outline} The outline of the paper is as follows: In  Section~\ref{sec:EmbeddingSpace}, we review the embedding-space formalism for conformal field theories, focusing especially on the challenge of finding conformal correlators of conserved currents. As a bridge to our new formulation in twistor space, we also introduce a representation of the embedding-space positions in terms of spinor variables~\cite{Binder:2020raz,Chiodaroli:2022ssi}. In Section~\ref{sec:TwistorSpace}, we introduce twistor coordinates as a linear combination of the embedding-space spinors and present an integral representation for spinning correlators in (A)dS where current conservation is made manifest as holomorphicity of the integrand. For the case of three-point functions, we confirm explicitly that our twistor-space results reproduce the known answers in embedding space. In Section~\ref{sec:FourierSpace}, we perform a half-Fourier transform of the twistor-space correlators to obtain their counterparts in momentum space.  Finally, we summarize our conclusions in Section~\ref{sec:Conclusions}.

\vskip 4pt
A number of appendices contain technical details: In Appendix~\ref{sec-discrete-symm}, we show how the twistor-space correlators transform under parity and time reversal. In Appendix~\ref{app:computations}, we explicitly compute the  twistor integrals introduced in Section~\ref{ssec:boot}, proving that this leads to the known results in embedding space.
In Appendix~\ref{app:computations2}, we discuss the half-Fourier transforms and the dispersive integrals appearing in Section~\ref{ssec:tranforms}. Lastly, in Appendix~\ref{sec:TwoPoints}, we bootstrap the two-point functions of spin-$S$ currents in twistor space.

\paragraph{Notation}  Throughout the paper, we use natural units, $\hbar = c \equiv 1$, and the mostly plus convention for the metric. Positions on the spatial boundary are denoted by $x^\mu$, with $\mu=1,\cdots, d$, and their conjugate momenta by $k^\mu$. The corresponding positions in embedding space are $P^M$, with $M=0,1,\cdots, d+1$, and we use $W^M$ for polarization vectors. For  $d=3$, we will work with CFTs in Lorentzian signature, whose embedding space is $\mathbb{R}^{2,3}$, with the metric given by $\eta_{MN}=\text{diag}(-1,1,1,1,-1)$. We define spinor variables for the embedding-space position $P^M$ via
\beq
\slashed{P}_A^{~B}=P^M (\Gamma_M)_A^{~B}=\epsilon^{ab}\Lambda_{a,A}\Lambda^{B}_b\,,
\eeq
where $(\Gamma_M)_A^{~B}$ are the Gamma matrices defined in \eqref{equ:Gamma-matrices}, $a,b=1,2$ are little group indices and $A,B=1,\cdots,4$ are spinor indices. The latter are raised and lowered 
by the anti-symmetric matrix $\Omega_{AB}$ defined in \eqref{equ:Omega},  
while the former are raised and lowered by the Levi-Civita symbol~$\epsilon^{ab}$ (with $\epsilon^{12}=\epsilon_{21}=1$). All indices are raised and lowered with the left-multiplication convention, $V^A=\Omega^{AB} V_B$ and $V_A=\Omega_{AB}V^B$,
and similarly for the little group indices $a,b$. Furthermore, we will denote contractions of the spinor indices $A,B$ as
\be\label{equ:notation-contraction}
U\cdot V\equiv U^A V_A \quad\text{and}\quad U\cdot M\cdot V \equiv U^A M_{A}^{~B} V_B\,,
\ee 
with the understanding that they are always contracted from upper left to lower right. The polarization vector $W^M$ can be written as
\be 
\slashed{W}_{A}^{~B}=(\Gamma_M)_A^{~B} W^M=\Upsilon_A \Upsilon^{*,B}-\Upsilon^*_A\Upsilon^B\,,
\label{equ:polvector-W-notation}
\ee 
where $\Upsilon_A$ obeys $\slashed{P}_A^{~B} \Upsilon_B=0$ and $\Upsilon^{*,A}$ is defined as a spinor that satisfies $\Upsilon^{*,A}\slashed{P}_A^{~B}=\Upsilon^B$.

\vskip4pt
Coordinates in twistor ($Z^A$) and dual twistor space ($W_A$) can be written as 
\be\label{equ:twistors-notation}
Z^A \equiv \pi^a \Lambda_a^A\quad\text{and}\quad W_A\equiv \pi^a\Lambda_{a,A}\,,    
\ee
where $Z^A \slashed{P}_A^{~B}=0$ and  $ \slashed{P}_A^{~B}W_B=0$. Twistor integrals 
have the projective measure $DZ=D\pi\equiv d\pi^a \hs\pi^b \hs\epsilon_{ab}$, and similarly for $DW=D\pi$.

\newpage
\section{Spinning Correlators in Embedding Space}
\label{sec:EmbeddingSpace}

In inflationary cosmology, we are interested in the correlations of fluctuations living on the future boundary of an approximate de Sitter spacetime.  If the bulk interactions aren't too strong, these correlations inherit a conformal symmetry from the isometries of the spacetime. This makes the study of conformal field theories (CFTs) of interest to cosmologists. 

\vskip 4pt
In this section, we introduce embedding space as a powerful way to make conformal symmetry manifest. We also describe the challenge of imposing current conservation for massless spinning fields within this formalism.

\subsection{Review of Embedding Space}

We will begin with a lightning review of the embedding-space approach to conformal field theories. Experts should skip this section, while novices can find more details in~\cite{Rychkov:2016iqz}.

 \paragraph{Projective null cone} It is a well-known fact that the conformal algebra on $d$-dimensional Euclidean space, $\mathbb{R}^d$, is isomorphic to the algebra of Lorentz transformations on $(d+2)$-dimensional Minkowski space, $\mathbb{R}^{1,d+1}$. This suggests that it should be possible to find a suitable embedding of $\mathbb{R}^d$
into $\mathbb{R}^{1,d+1}$, so that the Lorentz transformations in the higher-dimensional space
become conformal transformations on the lower-dimensional slice.

\begin{figure}[t!]
   \centering
      \includegraphics[scale=1.0]{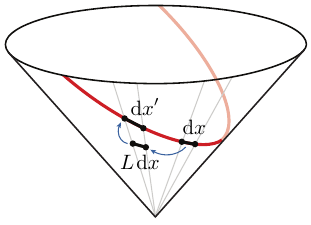} 
      \caption{Graphical illustration of the fact that Lorentz transformations on the projective null cone become conformal transformations on the Euclidean section.}
      \label{fig:Embedding}
\end{figure}

\vskip 4pt
 We define the coordinates on the higher-dimensional space as $ P^M$, with $M=0,1,\cdots,d+1$ and a metric given by $\eta^{MN}=\text{diag}(-1,1,\cdots,1)$.  Under a Lorentz transformation, these spacetime coordinates transform as
 $P^M \mapsto L^M{}_N P^N$. 
 We also define a {\it projective null cone} in the embedding space as
\begin{align}
P^2 &=0\,,\\
P^M &\sim \rho P^M\,, \label{equ:rescaling}
\end{align}
where $\rho$ is a rescaling parameter.
The {\it Euclidean section} is then defined by the constraint
\be
(P^+, P^-, P^\mu) = (1,x^2,x^\mu)\,,
\label{equ:section}
\ee
where $P^\pm \equiv P^0 \pm P^{d+1}$ 
are lightcone coordinates and $x^{\mu}$, with $\mu=1,\ldots,d$, are the coordinates on~$\mathbb{R}^d$. In Figure~\ref{fig:Embedding}, we illustrate  the action of a Lorentz transformation on an infinitesimal interval~${\rm d} x$. At first, it looks like this transformation moves the interval off the Euclidean section, but, under the identification (\ref{equ:rescaling}), we get the new interval ${\rm d} x^\prime$ on the section. It can also be shown that the constraint (\ref{equ:section}) is designed in such a way that the combined Lorentz transformation and rescaling becomes a conformal transformation on the Euclidean slice~\cite{Rychkov:2016iqz} .

 \paragraph{Tensors in embedding space}
 Consider symmetric, traceless and transverse tensors $O_{M_1 \ldots M_S}$ defined on the projective null cone. Contracting the tensor components with auxiliary null polarization vectors $W^M$, we can write these tensors
in index-free notation, 
\be
O^{(S)}(P,W) = O_{M_1 \ldots M_S}(P) \,W^{M_1} \cdots W^{M_S}\,,
\ee
where $W^M$ satisfies $W^2=P\cdot W=0$. Under
rescalings of the embedding-space coordinates and the polarization vectors, we have
\beq
O^{(S)}(\rho P, \alpha W) = \rho^{-\Delta} \alpha^S \,O^{(S)}(P,W)\, ,
\label{equ:scaling}
\eeq
where $\Delta$ and $S$ are the dimension and spin of the field, respectively.

 \paragraph{Conformal correlators} Conformal correlators in embedding space are simply the most general Lorentz-invariant expressions with the correct scaling behavior according to (\ref{equ:scaling}).
 It is convenient to define the following (parity-even) conformally-invariant structures 
 \beq
\begin{aligned}
P_{ij} &\equiv -P_i \cdot P_j\,,\\[6pt]
H_{ij} &\equiv - 2 \big[(W_i \cdot W_j)(P_i \cdot P_j) - (W_i \cdot P_j)(W_j \cdot P_i)\big]\,,\\
V_{i,jk} &\equiv \frac{(W_i \cdot P_j)(P_k \cdot P_i) - (W_i \cdot P_k)(P_j \cdot P_i)}{P_j \cdot P_k}\,,
\end{aligned}
\label{equ:BuildingBlocks}
\eeq
which serve as the basic building blocks for conformal correlators. For instance, the two-point function of a spin-$S$ field is
\be
 \braket{O_1^{(S)} O_2^{(S)}} = \frac{H_{12}^S}{(P_{12})^{\Delta+S}} \, ,
 \label{equ:2pt}
 \ee
 where we have dropped an overall normalization constant. The subscript on $O_i^{(S)}$ denotes both the type of the field and its position, i.e.~$O_i^{(S)} = O_i^{(S)}(P_i^M)$. Equation~\eqref{equ:2pt}
is the only object that can be constructed out of $P_{12}$ and $H_{12}$ that has the appropriate rescaling covariance \eqref{equ:scaling} for each field. Similarly, the three-point function of two scalars and a
spin-$S$ field is
\be
\braket{O_1 O_2 O_3^{(S)}} = \frac{V_3^S}{P_{12}^{(\Delta_1+\Delta_2-\Delta_3-S)/2} P_{23}^{(\Delta_2+\Delta_3-\Delta_1+S)/2} P_{31}^{(\Delta_3+\Delta_1-\Delta_2+S)/2} } \,,
\ee
where we defined $V_i \equiv V_{i,jk}$ for $\{i,j,k\}$ a cyclic permutation of $\{1,2,3\}$, and dropped an overall constant factor.
 
\vskip 4pt
There are also examples where more than one structure is consistent with conformal invariance.
For instance, the three-point function of identical spin-$2$ tensors is  
\beq
\braket{O^{(2)}_1 O^{(2)}_2 O^{(2)}_3} = \frac{1}{(P_{12} P_{23} P_{31})^{1+\Delta/2}} \, \sum_{n=1}^5 c_n F_n\,, \quad {\rm with} \quad F_n \equiv \begin{pmatrix} V_1^2V_2^2V_3^2\\[4pt]
H_{12} V_1V_2V_3^2+\text{cyclic}\\[4pt]
H_{12}^2 V_3^2+\text{cyclic}\\[4pt]
H_{12}H_{23} V_1 V_3+\text{cyclic}\\[4pt]
H_{12} H_{23} H_{31}
\end{pmatrix} ,
\label{equ:structures-TTT}
\eeq
which has five allowed structures that can be combined with arbitrary weights $c_n$.
As we increase the spin of the fields, the number of allowed structures also increases. For example, the correlator  of two identical spin-$2$ fields and a spin-$4$ field, $\braket{O^{(2)}_1 O^{(2)}_2 O^{(4)}_3}$, has $10$ allowed structures~\cite{Costa:2011mg}.

\vskip4pt
Let us also mention that  in dimension $d=3$ (which is the case of interest in the present work), the  building blocks in (\ref{equ:BuildingBlocks}) satisfy the relation
\be 
-2H_{12}H_{23}H_{31}=\left(V_1H_{23}+V_2H_{31}+V_3H_{12}+2 V_1V_2V_3\right)^2\,,
\ee 
which implies that the last structure in \eqref{equ:structures-TTT} is not independent from the others. In this case, there are therefore only $4$ independent structures for $\braket{O^{(2)}_1 O^{(2)}_2 O^{(2)}_3}$  and  $9$ independent structures for $\braket{O^{(2)}_1 O^{(2)}_2 O^{(4)}_3}$~\cite{Kravchuk:2016qvl}.
 
\paragraph{Current conservation} Correlators of conserved tensors must satisfy an additional kinematic constraint, which in position space reads
\beq
\partial_{\mu_n}J^{\mu_1 \ldots \mu_n \ldots \mu_S}=0\,.
\eeq
As long as the scaling dimension of the conserved current takes the value 
$\Delta=S+d-2$,
this conservation condition can be uplifted to embedding space as~\cite{Costa:2011mg} 
\be 
\frac{\partial}{\partial P_{M_n}}D_{M_n} J=0\,,\quad\text{where}\quad D_M \equiv \left(\frac{1}{2}+W\cdot \frac{\partial}{\partial W}\right)\frac{\partial}{\partial W^M}-\frac{1}{2}W_M\frac{\partial^2}{\partial W\cdot \partial W}\,.
\label{equ:conservation-operator-embedding}
\ee  
To impose conservation for a certain correlator, we then deduce
 the linear combinations of the conformally-invariant structures  that are compatible with the conservation condition. 

\vskip 4pt
Let us specialize to the case $d=3$, where a conserved spin-$S$ current has dimension $\Delta=S+1$. As an example, consider the case of a conserved spin-$1$ current.   The parity-preserving three-point function is~\cite{Giombi:2011rz, Zhiboedov:2012bm,osborn1994implications} 
 \be\label{equ:JJJcorrelator-embedding}
 \braket{J^{\tilde a}_1 J^{\tilde b}_2 J^{\tilde c}_3} = f^{\tilde a\tilde b\tilde c} \hs \frac{c_1 V_1 V_2 V_3 + c_2 (V_1 H_{23} + V_2 H_{13} + V_3 H_{12})}{(P_{12} P_{13} P_{23})^{3/2}}\,,
 \ee
 where $\tilde a$, $\tilde b$, $\tilde c$ are color indices and $f^{\tilde a\tilde b\tilde c}$ are the structure constants. We see that there are two independent structures for this correlator that are compatible with both conformal symmetry and current conservation.\footnote{In addition, there is one parity-violating structure that we have not displayed.}
 In the bulk, these two structures are associated to the Yang--Mills three-point vertex ($c_1= c_2$) and a higher-derivative cubic term $F^3$ ($c_1= 5c_2$)~\cite{Caron-Huot:2021kjy}.

  \vskip 4pt
  
The (parity-even) three-point function of a conserved spin-2 tensor (like the stress tensor) can be any linear combination of the four structures in \eqref{equ:structures-TTT} that satisfies the conservation condition~\eqref{equ:conservation-operator-embedding}. This reduces the space of allowed structures from four to two:
\begin{align}
\braket{T_1 T_2 T_3}_{\rm GR} &\propto \frac{\left(6\hs V_1^2 H_{2,3}^2+16\hs V_2V_3 H_{31}H_{12}+4\hs H_{23}V_1^2V_2V_3+\text{cyclic}\right)-9\hs V_1^2V_2^2V_3^2}{(P_{12}P_{23}P_{31})^{5/2}}\,,\label{equ:TTT-E-embedding}\\
\braket{T_1 T_2 T_3}_{W^3}&\propto \frac{\left(2\hs V_1^2 H_{2,3}^2-16\hs V_2V_3 H_{31}H_{12}-52\hs H_{23}V_1^2V_2V_3+\text{cyclic}\right)-147\hs V_1^2V_2^2V_3^2}{(P_{12}P_{23}P_{31})^{5/2}}\,.\label{equ:TTT-W-embedding}
\end{align}
In the bulk, these two structures are associated to the three-point vertex of Einstein gravity and a higher-derivative Weyl-cubed term $W^3$, respectively. 

 \vskip 4pt
 Conceptually, it is somewhat unsatisfying to first write down the larger space of conformally-invariant structures and then find the specific linear combination(s) corresponding to conserved correlators. 
 It would be nicer if we could write down directly the structures where both conformal invariance and conservation are made manifest from the start. In this paper, we will show that this is possible in twistor space.
 
\subsection{Spinor Variables in Embedding Space}

Spinor helicity variables are a powerful way to describe the on-shell kinematics of  scattering amplitudes of massless particles in four dimensions; see \cite{Elvang:2015rqa, Cheung:2017pzi} for a review. This starts with the observation that
on-shell massless particles satisfy $p^2=0$, so that their four-momenta can be written as
\beq
p_{\alpha \dot \alpha} = p_\mu (\sigma^\mu)_{\alpha \dot \alpha}  = \lambda_\alpha \tilde \lambda_{\dot{\alpha}}\,,
\label{equ:4d-spinors}
\eeq
where we contract $p_\mu$ with the Pauli matrices $(\sigma^\mu)_\alpha^{~\dot\beta}=(1,\sigma^i)_\alpha^{~\dot\beta}$ with one index lowered by $\epsilon_{\dot\alpha\dot\beta}$, and $\lambda_\alpha$ and $\tilde \lambda_{\dot \alpha}$ are two-component spinors. Notice that (\ref{equ:4d-spinors}) is invariant under the little group transformation $\lambda_\alpha\mapsto t\lambda_\alpha$ and $\tilde \lambda_{\dot{\alpha}}\mapsto t^{-1}\tilde \lambda_{\dot{\alpha}}$.
Given two particles $i$ and $j$, we define their ``angle" and ``square" brackets
\be\label{equ:4d-spinor-brackets}
\braket{ij} \equiv\epsilon^{ \beta\alpha  } \lambda_{i,\alpha} \lambda_{j, \beta}\hs  \,, \quad [ij] \equiv \epsilon^{  \dot \beta \dot \alpha}\lambda_{i,\dot \alpha} \tilde \lambda_{j, \dot\beta}\,,
\ee
where  $\epsilon^{\alpha \beta}$ is the Levi-Civita symbol (with $\epsilon^{12}=+1$). These are the Lorentz-invariant and little group-covariant building blocks of the spinor helicity formalism, meaning that any function of the 4d kinematics can be written in terms of these brackets.

\vskip 4pt 
Since the embedding-space coordinates  $P^M\in \mathbb{R}^{1,d+1}$ satisfy $P^2=0$, it is natural to wonder if a spinor helicity formalism can be applied here as well. There are two important differences: First, unlike the momentum-space four-vectors, the embedding-space coordinates do {\it not} satisfy the analog of momentum conservation, i.e.~$P_1+P_2 + \cdots + P_N  \ne 0$. Second,  the little group that leaves the position vector $P^M$ invariant is slightly larger than the little group in 4d momentum space. This is because a rescaling of the spinors that amounts to the rescaling $P^M\mapsto \rho P^M$ does not change the corresponding physical position $x^\mu$ on the Euclidean slice, and hence these rescalings belong to the little group. 

\vskip4pt
Of course, these spinor variables depend on the specific dimension $d$ of the CFT. We will first present these variables for $d=2$ and then consider the case of primary interest $d=3$.

  \paragraph{\boldsymbol{$d=2$}} 
For two-dimensional CFTs, the embedding space is four-dimensional Minkowski space. 
 We can therefore write the embedding-space position $P^M$ in the same way as in~(\ref{equ:4d-spinors}):
\be 
P_{\alpha\dot\alpha}=P_M(\sigma^M)_{\alpha\dot\alpha}=\Lambda_{\alpha} \tilde \Lambda_{\dot \alpha}\,,
\label{equ:4d-spinors-embedding}
\ee 
where $\Lambda_\alpha$ and $\tilde \Lambda_{\dot \alpha}$ are two-component spinors. 
As in \eqref{equ:4d-spinor-brackets}, we can define spinor brackets as the basic building blocks for conformal correlators.
Notice that the transformations
\be 
\Lambda_\alpha\mapsto r \Lambda_\alpha\,,\,\,\,\,\tilde \Lambda_{\dot \alpha}\mapsto \bar r \tilde \Lambda_{\dot \alpha}\,,
\ee 
change the embedding-space position as $P^M\mapsto r \bar r P^M$, where $r$ is a complex parameter. Since $P^M \sim \rho P^M$, this leaves the position
$x^\mu$ invariant and thus is also a little group transformation. Unlike in the 4d flat-space case~\eqref{equ:4d-spinors}, the little group now has two independent parameters. 
We can fix the little group by taking
\be 
\Lambda_\alpha=(-\bar  w,1)\,,\,\,\,\,\tilde\Lambda_{\dot\alpha}=(1,w)\,,
\label{equ:2dcft-Section}
\ee 
where $w \equiv x^1+ix^2$ and $\bar w \equiv x^1-ix^2$ are determined by the physical position $(x^1,x^2)$ after projecting \eqref{equ:4d-spinors-embedding} to the Euclidean section~\eqref{equ:section}.

\vskip4pt
It is straightforward to check that a field $O^{(h,\bar h)}$ with {\it scaling dimension} $\Delta=h+\bar h$ and {\it planar spin} $S=h-\bar h$ (which may be positive or negative in 2d CFTs \cite{DiFrancesco:1997nk}) transforms under little group transformations as 
\be 
O^{(h,\bar h)}\mapsto r^{-2h} \bar r^{-2\bar h}O^{(h,\bar h)}\,,
\label{equ:4d-covariance}
\ee 
in terms of the holomorphic and anti-holomorphic conformal dimensions $(h,\bar h)$ \cite{DiFrancesco:1997nk}.

  \paragraph{\boldsymbol{$d=3$}}  In inflationary cosmology, we are interested in three-dimensional Euclidean CFTs in~$\mathbb{R}^3$, with a five-dimensional embedding space $\mathbb{R}^{1,4}$. To perform the calculations that are the core of this work, we will instead often consider the closely related Lorentzian CFTs in $\mathbb{R}^{1,2}$, whose embedding space is $\mathbb{R}^{2,3}$. 
  
  \vskip4pt
  
Let us start by introducing some conventions. The metric of the embedding space $\mathbb{R}^{2,3}$ is $\eta^{MN}=\text{diag}(-1,1,1,1,-1)$, and the {\it Poincaré section} is  
\be 
(P^\mu,P^+,P^-)=(x^\mu,1,x^2)\,,
\label{equ:Poincare-section}
\ee 
where $\mu=0,1,2$ is the index of $x^\mu\in\mathbb{R}^{1,2}$, $x^2=\eta_{\mu\nu}x^\mu x^\nu$, and $P^\pm\equiv P^4\pm P^3$. Moreover, we take the $\Gamma$-matrices of $\mathbb{R}^{2,3}$ to be explicitly real~\cite{Binder:2020raz}: 
\begin{align}\label{equ:Gamma-matrices}
	(\Gamma_0)_A^{~B} &=\left(
	\begin{array}{cccc}
		0 & 1 & 0 & 0 \\
		-1 & 0 & 0 & 0 \\
		0 & 0 & 0 & -1 \\
		0 & 0 & 1 & 0 \\
	\end{array}
	\right) ,~~~~(\Gamma_1)_A^{~B}=\left(
	\begin{array}{cccc}
		0 & -1 & 0 & 0 \\
		-1 & 0 & 0 & 0 \\
		0 & 0 & 0 & -1 \\
		0 & 0 & -1 & 0 \\
	\end{array}
	\right) ,~~~~(\Gamma_2)_A^{~B}=\left(
	\begin{array}{cccc}
		-1 & 0 & 0 & 0 \\
		0 & 1 & 0 & 0 \\
		0 & 0 & -1 & 0 \\
		0 & 0 & 0 & 1 \\
	\end{array}
	\right) , \nonumber \\[4pt]
	(\Gamma_3)_A^{~B} &=\left(
	\begin{array}{cccc}
		0 & 0 & 0 & -1 \\
		0 & 0 & 1 & 0 \\
		0 & 1 & 0 & 0 \\
		-1 & 0 & 0 & 0 \\
	\end{array}
	\right) ,~~~~(\Gamma_4)_A^{~B}=\left(
	\begin{array}{cccc}
		0 & 0 & 0 & 1 \\
		0 & 0 & -1 & 0 \\
		0 & 1 & 0 & 0 \\
		-1 & 0 & 0 & 0 \\
	\end{array} 
	\right) .
\end{align}
The spinor indices $A,B=1,2,3,4$ are raised and lowered 
by the following anti-symmetric matrix
\be
\Omega_{AB}=-\Omega^{AB}=\left(
\begin{array}{cccc}
	0 & 0 & 1 & 0 \\
	0 & 0 & 0 & 1 \\
	-1 & 0 & 0 & 0 \\
	0 & -1 & 0 & 0 \\
\end{array}
\right),
\label{equ:Omega}
\ee 
with the left-multiplication convention 
\be
U^A=\Omega^{AB}U_B\quad \text{and} \quad U_A=\Omega_{AB}U^B\,.
\ee 
Notice that the $\Gamma$-matrices with lowered indices, $(\Gamma^M)_{AB}=\Omega_{BC}(\Gamma^M)_A^{~C}$,  are then anti-symmetric and $\Omega$-traceless. 
We will denote contractions of the spinor indices $A,B$ by
\be
U\cdot V\equiv U^A V_A \quad\text{and}\quad U\cdot M\cdot V \equiv U^A M_{A}^{~B} V_B\,,
\ee 
with the understanding that they are always contracted from upper left to lower right.

\vskip4pt

Using these conventions, we can define spinor variables for 3d CFTs as~\cite{Binder:2020raz}\footnote{These spinor variables are similar to the spinor helicity variables for five-dimensional flat space, which were introduced recently in \cite{Chiodaroli:2022ssi}.}
\be
\slashed{P}_A^{~B}=P^M (\Gamma_M)_A^{~B}=\epsilon^{ab}\Lambda_{a,A}\Lambda^{B}_b\,,
\label{equ:defLambda}
\ee 
where $a,b=1,2$ are little group indices which are raised and lowered by $\epsilon^{ab}$. The reality condition for these spinors  is that their components are all real. Since the $\Gamma$-matrices are traceless, the spinors must satisfy the constraint $\Lambda^{a,A}\Lambda_{a,A}=0$. This implies that\footnote{Note that the $2\times 2$ matrix $\Lambda_a \cdot \Lambda_b$ is antisymmetric, and thus can be written as $\Lambda_a \cdot \Lambda_b=c\hs \epsilon_{ab}$, with $c=\Lambda^a \cdot \Lambda_a/2$. Hence, if $\Lambda^a \cdot \Lambda_a=0$, then $\Lambda_a \cdot \Lambda_b=0$.} $\Lambda_a\cdot \Lambda_b=0$, and that the two spinors $\Lambda_{a,A}$  belong to the kernel of $\slashed{P}$:
\be\label{equ:Lambda-kerP}
\slashed{P}_A^{~B}\Lambda_{a,B}=0\,. 
\ee
In fact, $\Lambda_{1,A}$ and $\Lambda_{2,A}$  form a two-dimensional basis of ${\rm ker}(\slashed{P})$. Notice that the transformation
\be 
\Lambda_a^A\mapsto r\hs V_a^{~b}\hs\Lambda_b^A\,,
\ee 
with $r>0$
and $V_a^{~b}$ a real matrix satisfying $\det(V)=1$, does not change the position $x^\mu=(x^0,x^1,x^2)\in\mathbb{R}^{1,2}$ obtained after projecting $P^M$ to the Poincaré section \eqref{equ:Poincare-section}. Hence, they are little group transformations\footnote{Note that the little group and the spacetime symmetry group commute because they act on different indices. 
In general, they are related through the so-called Howe duality, which maps their respective irreps~\cite{Basile:2020gqi,Basile:2024ydc}.} that amount simply to a change of basis of the space ${\rm ker}(\slashed{P})$. We can fix the little group by writing 
\be\label{equ:LambdaSection} 
\Lambda_a^A =\left(
\begin{array}{cccc}
	0 & 1 & -x^{2} & -x^{0}-x^{1} \\
	1 & 0 & -x^{0}+x^{1} & -x^{2} \\
\end{array}
\right) ,
\ee 
which corresponds via \eqref{equ:defLambda} to the Poincar\'e section of the embedding space.

\vskip 4pt
Furthermore, the polarization vectors $W^M$ satisfy $W^2=W\cdot P=0$ and the equivalence relation $W_M\sim W_M+c P_M$.
These properties are manifestly satisfied if we write 
\be 
\slashed{W}_{A}^{~B}=(\Gamma_M)_A^{~B} W^M=\Upsilon_A \Upsilon^{*,B}-\Upsilon^*_A\Upsilon^B\,,
\label{equ:polvector-W}
\ee 
where $\Upsilon_A$ is an element of $\rm{ker}(\slashed{P})$ and thus can be written as a linear combination of the $\Lambda_a^A$ spinors:
\be
\Upsilon_A= \zeta^a\Lambda_{a,A}\,.
\label{equ:def-zeta}
\ee 
The {\it dual spinor} $\Upsilon^{*,A}$ is defined as the mapping 
\be
\Upsilon^{*,A}\slashed{P}_A^{~B}=\Upsilon^B\,.
\label{equ:dual}
\ee 
It is easy to see that $\Upsilon^{*,A}$ is defined up to the equivalence relation $\Upsilon^{*,A}\sim \Upsilon^{*,A}+U^A$, where $U_A\in\rm{ker}(\slashed{P})$, so that $U^A\slashed{P}_A^{~B}=0$. This equivalence relation for $\Upsilon^{*,A}$ implies $W_M\sim W_M+c P_M$ for the polarization vector. 

\vskip4pt
Finally, we notice that a spin-$S$ field $J(P,W)$ in embedding space can be written in terms of the spinors $\Lambda_a^A$ and $\Upsilon^A$ as
\be \label{equ:Jwithabcindices}
J(P,W)=J(\Lambda_a^A,\Upsilon^B)=\zeta^{a_1} \cdots \zeta^{a_{2S}}\hs J_{a_1\cdots a_{2S}}(\Lambda_a^A)\,,
\ee 
where we used that the field is an homogeneous function of $\zeta^a$ of degree $2S$. We also defined the coefficients $J_{a_1\cdots a_{2S}}(\Lambda_a^A)$ in this expansion, which constitute the components of a symmetric tensor with $2S$ little group indices that only depends on $\Lambda_a^A$. We further observe that the conservation condition~\eqref{equ:conservation-operator-embedding} for this tensor can be written as~\cite{Binder:2020raz}
\be 
\partial_{a_1a_2}^M J^{a_1\cdots a_{2S}}=0\,,\quad\text{with}\quad\partial_{ab}^M\equiv \Lambda_{(a}^A (\Gamma^M)_A^{~B} \frac{\partial}{\partial \Lambda^{b),B}}\,.
\label{equ:conservation-operator-spinors}
\ee 
Although there is now a free index $M$, this equation is equivalent to imposing $\Delta=S+1$ together with \eqref{equ:conservation-operator-embedding}.

\subsection{Conservation and Holomorphicity}
\label{ssec:holomorphicity}

We would like to make manifest that a correlator of conserved tensors satisfies the differential conservation condition~\eqref{equ:conservation-operator-embedding}. We will first study the simpler case of $d=2$, where conservation arises from holomorphicity, and then discuss the challenge of generalizing this to $d=3$.

 \paragraph{\boldsymbol{$d=2$}}  Like for three-point scattering amplitudes \cite{Elvang:2015rqa,Cheung:2017pzi}, we can bootstrap three-point correlators in 2d CFTs by proposing a power-law ansatz in terms of spinor brackets
\be 
\braket{O_1 O_2 O_3}=\frac{\lambda_{123}}{\braket{12}^{n_3}[12]^{\bar{n}_3}\braket{23}^{n_1}[23]^{\bar{n}_1}\braket{31}^{n_2}[31]^{\bar{n}_2}}\,.
\label{equ:ansatz}
\ee 
Since there is no momentum conservation, we must consider both angle and square brackets.
Moreover, we have a bigger little group to exploit. Indeed, imposing the little group covariance~\eqref{equ:4d-covariance} for each of the three fields, we find
\begin{align}
n_i&=h_j+h_k-h_i\,,\,\,\,\,\qquad \bar n_i=\bar h_j+\bar h_k-\bar h_i\,.
\end{align}
As an example, consider the case of a conserved tensor $T$. In 2d CFTs, a conserved tensor of planar spin $S$ must satisfy 
$\Delta=|S|$, and the conformal weights can either be $(h,\bar{h})=(|S|,0)$ or $(h,\bar{h})=(0,|S|)$ depending on the sign of $S=h-\bar{h}$. Thus, the conserved tensor separates into a {\it holomorphic} component $T(w)$ (with $\bar h=0$) and an {\it anti-holomorphic} component $\bar T(\bar w)$ (with $h=0$), which are separately conserved, i.e.~they satisfy $\bar\partial T =0$ and $\partial \bar T=0$. Evaluating (\ref{equ:ansatz}) for either $(h,\bar{h})=(|S|,0)$ or $(h,\bar{h})=(0,|S|)$, we  find  
\begin{align}
	\braket{T_1 T_2 T_3 } &=\frac{\text{const.}}{\braket{12}^{n_3}\braket{23}^{n_1}\braket{31}^{n_2}}\,,\\
	\braket{\bar{T}_1\bar{T}_2\bar{T}_3}& =\frac{\text{const.}}{[12]^{\bar n_3}[23]^{\bar n_1}[31]^{\bar n_2}}\,,
\end{align}
where the exponents are $n_i=\bar n_i=|S_j|+|S_k|-|S_i|$. These correlators are either holomorphic or anti-holomorphic in the sense that they depend only on angle or square brackets, respectively. After projecting to the Poincaré slice using  \eqref{equ:2dcft-Section}, this translates into holomorphicity or anti-holomorphicity in physical position space, because the correlators will depend only on $w_i$ or  $\bar w_i$, respectively. We thus conclude that for 2d CFTs conservation is intimately connected to holomorphicity, and that we can make conservation manifest by taking our correlators to be holomorphic.

\paragraph{\boldsymbol{$d=3$}} How does this generalize to 3d CFTs?
Making conservation manifest is not as trivial as before, because the correlator cannot depend on only one of the two spinors $\Lambda_a$ with $a=1,2$. Instead, it will always depend on both. A quick way to see this is to note that the product
\be 
4 P_{ij}=-\Tr(\slashed{P}_i\slashed{P}_j)=-\Lambda_{i}^a\cdot \Lambda_{j,b} ~\Lambda_{j}^b\cdot \Lambda_{i,a}
\label{equ:Pij-Lambda-spinors}
\ee 
does not factorize into the product of two spinor brackets as in the 2d case, where $2P_{ij}=\braket{ij}[ij]$. We can also see this explicitly by writing the structures in (\ref{equ:BuildingBlocks}) in terms of the spinors
\begin{align} \label{equ:structure-H}
H_{ij}&=-\frac{1}{2}\hs (\Upsilon_i\cdot \Upsilon_j)^2\,,\\
V_{i,jk}&=\frac{1}{4 P_{ij}}\hs \Upsilon_i\cdot \slashed{P}_j\cdot \slashed{P}_k\cdot \Upsilon_j \,,\label{equ:structure-V}
\end{align} 
together with $\eqref{equ:Pij-Lambda-spinors}$ for $P_{ij}$. 
Consider, for example, the three-point function $\langle J_1 J_2 J_3 \rangle$ of spin-1 currents in \eqref{equ:JJJcorrelator-embedding} written completely in terms of these variables. It can easily be seen that the result depends on both spinors $\Lambda_1^A$ and $\Lambda_2^A$, and hence cannot be holomorphic in the same sense as for $d=2$. However, something special must happen for this correlator to be conserved: There should be an analog of holomorphicity for 3d CFTs that is not as trivial as in the $d=2$ case. We will learn in the next section that this hidden holomorphicity can be exposed by writing our correlators in twistor space. 

\newpage
\section{Conserved Currents in Twistor Space}
\label{sec:TwistorSpace}

As we have seen in the previous section, working in embedding space makes the conformal symmetries of the boundary correlators easy to implement. 
However, for massless particles, the challenge is to find a kinematic description that also propagates the right number of degrees of freedom, corresponding conserved to currents in the boundary theory. 
In embedding space, this conservation must be imposed by hand as an additional differential constraint. 

\vskip 4pt
We will show, in this section, that current conservation instead trivializes when we interpret linear combinations of the embedding-space spinors as points in {\it twistor space}.\footnote{See~\cite{Neiman:2013hca,Neiman:2017mel} for important prior work.}

\subsection{Discovering Twistor Space}

We will trace out a natural route that unavoidably leads us to twistors as the right variables for describing conserved currents in conformal field theory. For readers familiar with twistor space, most of this section will be introductory and can safely be skipped. However, for non-experts, we will try to give a fresh take on twistors, which connects them more easily with other kinematic spaces that are more familiar to cosmologists, holographers and bootstrappers. 

\paragraph{Holomorphicity} In Section~\ref{ssec:holomorphicity}, we described how conservation in 2d CFTs is related to holomorphicity. Our challenge is to find a similar holomorphicity property for 3d CFTs. As we will see, this will automatically lead us to twistors!

\vskip 4pt
To get some intuition, we first consider the simpler case of massless fields in flat space. As is well known, the generic solution to the free scalar equation in two dimensions is given by holomorphic functions $F(z)$. Depending on the meaning of the complex variable $z$, the solution can be interpreted as solving the Laplace or wave equation.
Remarkably, there exists a 3d analogue of this elegant construction~\cite{Ward:1989vja}: Given the three coordinates $(x_0,x_1,x_2)$, a generic solution of the Klein--Gordon equation, $(-\partial_{x_0}^2 + \partial_{x_1}^2+\partial_{x_2}^2) F =0$, is a function $F(Z)$ that only depends on 
the variable
\be\label{equ:Z-holomorphicity-3d}
Z \equiv x_0+x_1 \sin \varphi+x_2 \cos\varphi\, ,
\ee
where $\varphi$ is an arbitrary angle. The function $F(Z)$ is holomorphic in the sense that it depends on the coordinates $x_\mu$ only through the linear combination in \eqref{equ:Z-holomorphicity-3d}. 
There is one price to pay though: the introduction of the parameter $\varphi$. To get a solution that depends only on the coordinates, we have to integrate over $\varphi$ as 
\be \label{equ:integral-varphi}
f(x_\mu) \equiv \frac{1}{2}\int_{-\pi}^{\pi} d\varphi\, F(Z)\,,
\ee  
where $F(Z)$ may also have an explicit dependence on $\varphi$ that we ignore. This formula can be made more projective-looking by writing 
\be\label{equ:integral-pi-discovering}
f(x_\mu)=\int D\pi\, F(\pi^a x_{ab}\pi^b)\, , \quad {\rm where} \quad x_{ab} \equiv \begin{pmatrix}x_0+x_1&-x_2\\-x_2&x_0-x_1\end{pmatrix}\ \ {\rm and}\quad D\pi\equiv d\pi^b \pi^c\epsilon_{bc} \,.
\ee
The two components of $\pi^a$ are real, and we are therefore integrating over the projective space $\mathbb{RP}^1$ of lines in $\mathbb R^2$ that pass through the origin. In order for the integral to be invariant under the rescaling 
$\pi^a \to r\hs \pi^a$, the integrand $F(Z)$ must be a homogeneous function of degree $-1$, i.e.~$F(rZ)=r^{-1}\hs F(Z)$.  Parameterizing
$\pi^a \equiv r(1, \tan(\varphi/2))$ and integrating $\varphi$ over the interval $(-\pi,\pi)$, after modding out the scale~$r$, we recover the result~\eqref{equ:integral-varphi}.\footnote{A related integral representation for the Green function of the Laplace equation was found by Whittaker more than 120 years ago \cite{Whittaker}.} 
As explained in \cite{Ward:1989vja}, the integral in \eqref{equ:integral-pi-discovering} is a so-called {\it Penrose transform} that defines a map from mini-twistor space (i.e.~the space of null planes in $\mathbb{R}^{1,2}$ defined by $Z= {\rm const}$) to coordinate space. 

\paragraph{Integral transform}  We now return to our problem of interest. As we explained in Section~\ref{ssec:holomorphicity}, the correlators of conserved currents in a 3d CFT always depend on {\it both} embedding-space spinors $\Lambda_1^A$ and $\Lambda_2^A$, and hence it cannot be holomorphic in the same sense as for $d=2$. 

\vskip 4pt
Taking into account the lessons we have just learned about holomorphicity in 3d, we consider a function $F(Z^A)$ that depends on the spinors $\Lambda_a^A$ only through a specific linear combination
\be \label{eq-Z-pi}
Z^A \equiv \pi^a \Lambda_a^A\, .
\ee
In order for the spinors $Z^A$ to satisfy the same reality condition as the embedding-space spinors $\Lambda_a^A$ (i.e.~for their components to be purely real), the coefficients $\pi^a$ must be real as in~\eqref{equ:integral-pi-discovering}. We erase the arbitrariness in the choice of $\pi^a$ by integrating over it, using the projective measure $D\pi\equiv d\pi^b\pi^c\epsilon_{bc}$ as in~\eqref{equ:integral-pi-discovering}. For a conserved spin-$S$ current, the resulting integral transform is
\be\label{equ:Jabc-pi-integral}
J^{a_1 a_2\cdots a_{2S}}(\Lambda_a^A) = \int D\pi \,\pi^{a_1} \pi^{a_2}\cdots \pi^{a_{2S}} F( \pi^a \Lambda_a^A) \,, 
\ee
%\fr{Comment that a possible explicit dependence of $F$ on $\pi^a$ won't make sense due to little group covariance?}
where we considered the tensor with explicit little group indices as in \eqref{equ:Jwithabcindices}. 
To match the  
indices on both sides, we have included $2S$ factors of $\pi^{a}$ in the integrand. Note that there is no explicit dependence of $F$ on $\pi^a$ because it would not be consistent with the little group covariance of the current. 
As we will prove in the following section, the integral representation in \eqref{equ:Jabc-pi-integral} makes conservation manifest simply because the integrand $F(Z^A)$ is holomorphic, in the sense that it depends only on the linear combination in \eqref{eq-Z-pi}. This is precisely what we wanted to achieve in Section~\ref{ssec:holomorphicity}: Any tensor written as \eqref{equ:Jabc-pi-integral} will automatically be a conserved current!

\vskip 4pt
In order for the integral over $\pi_a$ to be well-defined projectively, 
the function $F(Z^A)$ must  be a homogeneous function of degree $-2(S+1)$, i.e.~$F(rZ^A)=r^{-2(S+1)}\hs F(Z^A)$. 
 Since $Z^2\sim \slashed{P}$, we see that the dimension of the field is therefore fixed to be $\Delta=S+1$. This is really nice. Requiring the integral transform \eqref{equ:Jabc-pi-integral} to be well-defined has fixed the conformal scaling weight of the field to be precisely that of a conserved current.\footnote{For spinning correlators, there is one more subtlety associated to the choice of helicity---the formula in \eqref{equ:Jabc-pi-integral} is for self-dual currents. This is best dealt with in $Z$ language, before committing to a parametrization as in \eqref{eq-Z-pi}. We postpone the discussion of helicities to the next subsection.}
 
\begin{figure}[t!]
	\centering
	\includegraphics[width=0.7\textwidth]{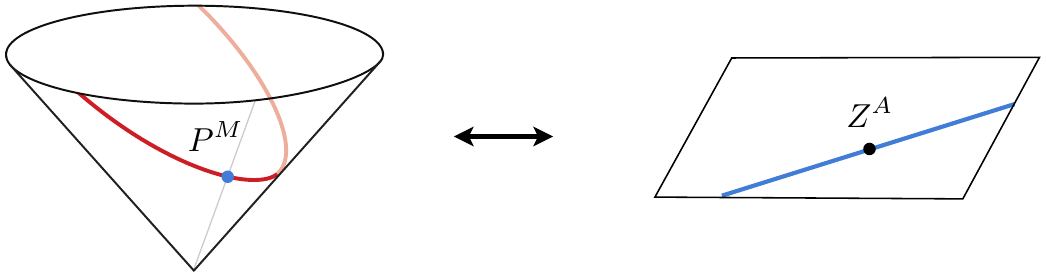}
	\caption{Illustration of the map between  embedding space and twistor space defined by the relation \eqref{equ:incidence-PZ}.}
	\label{fig:Penrose-map}
\end{figure}

\paragraph{Twistors}
Let us digress briefly to explain why the space parameterized by the coordinates $Z^A$ in \eqref{eq-Z-pi} is twistor space.

\vskip 4pt
We first recall that the spinors  $\Lambda_{a}^A$ satisfy the constraint \eqref{equ:Lambda-kerP}, i.e.~$\slashed{P}_A^{~B}\Lambda_{a,B}=0$, which defines a natural two-dimensional subspace associated to a given embedding-space position $P^M$, namely that spanned by the two spinors $\Lambda_{a}^{A}$.  
This constraint   implies that the
spinors $Z^A$
also satisfy
\be\label{equ:incidence-PZ}
\slashed{P}_A^{~B} Z^A=0\,,
\ee
i.e.~they live in the kernel of (the transpose of) $\slashed{P}_A^{~B}$. Equation~\eqref{equ:incidence-PZ} defines a map between the null positions $P^M$ in embedding space 
and the spinors $Z^A$. It is non-local because it maps one position $P^M$ to a whole two-dimensional subspace of spinors. After modding out the scale of these spinors, this subspace becomes a projective one-dimensional subspace, which can be pictorially represented as a line (see Figure~\ref{fig:Penrose-map}). 
The map in \eqref{equ:incidence-PZ} can also be interpreted as the so-called ``incidence relation"  between position space and twistor space~\cite{Adamo:2017qyl}, which in turn allows us to identify the spinor $Z^A$ as a twistor, and the integral~\eqref{equ:Jabc-pi-integral} as a ``Penrose transform." Indeed, splitting $Z^A$ into two-component spinors, $Z^A \equiv (\lambda^\alpha,\mu_{\dot\alpha})$, and projecting $P^M$ to the Poincaré slice~\eqref{equ:Poincare-section}, the relation~\eqref{equ:incidence-PZ} becomes
\be 
\mu_{\dot\alpha}=x_{\alpha\dot\alpha}\lambda^\alpha\,,\quad\text{where}\quad x_{\alpha\dot\alpha}=\begin{pmatrix}x_0+x_1&-x_2\\-x_2&x_0-x_1\end{pmatrix},
\ee 
which is the usual incidence relation described in the twistor literature \cite{Adamo:2017qyl}.\footnote{Note that every twistor $Z^A$ is mapped through \eqref{equ:incidence-PZ} to all positions $P^M$ on the null cone of embedding space that satisfy such a relation. After projecting to the Poincaré slice, it is easy to show that this forms a null line in the Lorentzian position space $\mathbb{R}^{1,2}$. Hence, a field written in twistor space will be localized along null lines in~$\mathbb{R}^{1,2}$, which seems like a potentially useful language for light-ray operators. We will comment on this connection in Section~\ref{sec:Conclusions}. \label{footnoteLightray} }

\vskip 4pt
The inner product between two twistors is defined as
\beq
 Z_i\cdot Z_j =Z_i^A \Omega_{AB} Z_j^B\,,
\eeq
where $\Omega_{AB}$ was defined in \eqref{equ:Omega}. 
Importantly, this inner product is invariant under a transformation by the real symplectic group Sp$(4,\mathbb R)$---the double cover of SO$(2,3)$---since
\beq
\begin{aligned}
Z_i^A &\mapsto Z_i^B\hs M_B^{~A} \,,\\
\Omega_{AB} &\mapsto  M_A^{~C} \Omega_{CD} M_B^{~D}=\Omega_{AB}\,.
\end{aligned}
\eeq
Moreover, the reality condition of the twistors $Z_i^A$ is preserved under this map as well. Note that SO$(2,3)$ is precisely the conformal group of the 3d Lorentzian boundary, and also the isometry group of the 4d bulk spacetime. The latter can be either Lorentzian AdS$_4$ or dS$_{2,2}$, with signature $(2,2)$, as both can be embedded in $\mathbb{R}^{2,3}$.  
Hence, the tensor $\Omega_{AB}$ can be interpreted as the ``infinity twistor," which is the only element in twistor theory that breaks the bulk conformal symmetry into the isometry group  
of the four-dimensional spacetime~\cite{Adamo:2017qyl,Arkani-Hamed:2009hub}.

\paragraph{Correlators as twistor integrals}
So far, we have explained how to trivialize conservation for a single current by writing it in terms of an integral of the form \eqref{equ:Jabc-pi-integral}. However, our observables of interest are correlators of conserved currents 
defined at multiple points. We can make conservation manifest for each current in a $n$-point correlator by considering a function $F(Z_i^A)=F(Z_1^A,\cdots,Z_n^A)$ that depends on each of the embedding-space spinors $\Lambda_{i,a}^A$ only through the combinations 
\be 
Z_i^A \equiv \pi_i^a\Lambda_{i,a}^A\, .
\ee 
As before, we must integrate over the $n$ arbitrary parameters $\pi_i^a$ to obtain a Penrose-like integral representation of the correlator. For example, we can write the three-point function of a conserved spin-1 current as
\be
\langle J^{a_1 a_2}(\Lambda_{1})  J^{b_1 b_2}(\Lambda_2) J^{c_1 c_2}(\Lambda_3) \rangle =\int D\pi_1  \,\pi_1^{a_1} \pi_1^{a_2}\int D \pi_2  \,\pi_2^{b_1} \pi_2^{b_2} \int D\pi_3  \,\pi_3^{c_1} \pi_3^{c_2}\, F(Z_i^A)\, .
\label{equ:Penrose-JJJ-discovering}
\ee
Finally, conformal symmetry in the 3d position space 
is automatically satisfied if the integrand depends on the twistors only through their inner products, $F(Z_i\cdot Z_j)$, and we can bootstrap this function by imposing how it scales with respect to each twistor $Z_i^A$.

\vskip 4pt
The upshot of this discussion is that the natural language to describe correlators, with all of their kinematic requirements made manifest, is that of holomorphic functions in twistor space! Every point in embedding space defines a pair of spinors, from which one can build an associated twistor. The resulting holomorphic functions satisfy stringent constraints, essentially from scaling, and can be bootstrapped very easily (at least for two- and three-point functions).  
The twistor formulas are extremely simple, essentially distributional. In the rest of this section, we will provide more details about the Penrose transform and build the conformal correlators of conserved currents. We will also compute the twistor integrals for various examples, to demonstrate that they reproduce the well-known formulas in the literature.

\subsection{Currents as Twistor Integrals}

In the previous subsection, we presented the twistor integral~(\ref{equ:Jabc-pi-integral}) describing a conserved spin-$S$ current. It will also be convenient to write this in index-free form~\cite{Neiman:2013hca,Neiman:2017mel}:
\be 
J(P,\Upsilon^*)=i^{-S}\int DZ \hs (\Upsilon^*\cdot Z)^{2 S}\hs F(Z^B)\,,
\label{equ:twistorintegral-product}
\ee 
where $\Upsilon^*\cdot Z=-\Upsilon^*_A Z^A$, and the factor $i^{-S}$ was introduced for later convenience. Let us explain the elements in this formula in a bit more detail: 
\begin{itemize}
\item First, we are integrating~$Z^A$ over the line in twistor space defined by $\slashed{P}_A^{~B} Z^A=0$. Due to the projective equivalence on twistors~$Z^A\sim r Z^A$, we take the measure $DZ$ to be projective. Given the parameterization in \eqref{eq-Z-pi}, we have\footnote{Equivalently, the measure can also be defined without choosing a specific parameterization such that $\slashed{P}^{AB} DZ=Z^A dZ^B-Z^B dZ^A$ \cite{Neiman:2017mel}.\label{footnote-defmeasure}} 
\be 
DZ=D\pi=  d\pi^a \pi^b\epsilon_{ab} \,.
\ee 
By expanding $\pi^a = r (u^a+\omega\hs v^a)$ in a basis normalized as $v^au_a=1$, this can be written as 
\be \label{equ:measure-omega}
DZ =r\hs d\omega \,v^a \pi_a=r^2d\omega\,,
\ee 
where we notice explicitly that there is no integral over the scale $r$. In order for the whole integral \eqref{equ:twistorintegral-product} also to be independent of this scale, we require the integrand to be a homogeneous function scaling as
\be 
F(r Z^A)=r^{-2S-2}F(Z^A)\, .
\label{equ:scaling-F}
\ee 
This puts an important constraint on the functional form of the integrand. Regarding the precise contour of integration of the $d\omega$ integral, we observe that it depends on the signature of the three-dimensional position space.  
This is clearly understood for the case of a Lorentzian position space $\mathbb{R}^{1,2}$, where the twistors are purely real and thus we have to integrate $\omega$ over the real line, or equivalently $Z$ over the real projective space of twistors, such that $\slashed{P}_A^{~B} Z^A =0$. This is the reason why we choose to work  
in a three-dimensional position space with Lorentzian signature.

\item Second, the factor $(\Upsilon^*\cdot Z)^{2 S}$ contains the $2S$ spinors $\Upsilon^*_A$ that correspond to the $S$ {\it physical} polarization vectors, related to the {\it gauge-redundant} $W^M$ through~\eqref{equ:polvector-W}. It is easy to check that \eqref{equ:dual}, together with \eqref{equ:defLambda} and \eqref{equ:def-zeta}, implies  
$\Upsilon^*\cdot Z=\zeta^a\pi_a$. The integral \eqref{equ:twistorintegral-product} can therefore be written as
\be
J(\Lambda_a,\zeta^a) 
=i^{-S}\int D\pi\, (\zeta^c \pi_c)^{2S}\hs F(\pi^d \Lambda_d^B)\,.
\label{equ:integral-pi-J}
\ee  
Notice that the scaling of $F(Z^A)$ in \eqref{equ:scaling-F} implies that the current scales as $r^{-2(S+1)}$ when we rescale $\Lambda_a^A\mapsto r \Lambda_a^A$. Hence, the current $J$ must have scaling dimension
\be  
\Delta=S+1\,,
\ee  
to admit an integral representation like \eqref{equ:integral-pi-J}. This is precisely the scaling dimension of a conserved current of spin $S$~\cite{Costa:2011mg}. Moreover, we can also check that this current is automatically conserved by showing  that it vanishes after acting on it with the conservation operator \eqref{equ:conservation-operator-spinors}:
\begin{align}
\partial_{a_1 a_2}^M J^{a_1\cdots a_{2S}}(\Lambda_a)&=i^{-S}\hs\Lambda_{(a_1}^A (\Gamma^M)_A^{~B} \int D\pi \, \pi^{a_1}\pi^{a_2}\cdots \pi^{a_{2S}}\hs \frac{\partial Z^C}{\partial \Lambda^{a_2),B}}\hs \frac{\partial}{\partial Z^C}F(Z)\nonumber\\
&=i^{-S}\hs\Lambda_{a_1}^A (\Gamma^M)_A^{~B} \int D\pi \, \pi^{a_1}\pi^{a_2}\cdots \pi^{a_{2S}}\hs (-\pi_{a_2})\hs \frac{\partial}{\partial Z^B}F(Z)\nonumber\\
&=0\,,
\end{align}
where we used the chain rule in the first line and $\pi^{a_2}\pi_{a_2}=\epsilon^{a_2b_2}\pi_{a_2}\pi_{b_2}=0$ in the last step.
\end{itemize}

\vskip4pt
Finally, let us mention that there is an alternative integral representation where we integrate instead over the \textit{dual} twistor space:
\be 
\tilde J(P,\Upsilon^*)=i^S\hs\int DW \left(\Upsilon^*\cdot\frac{\partial }{\partial W}\right)^{2S}\hs \tilde F(W_B)\,.
\label{equ:alt}
\ee 
Here, the integral is defined over the line given by $\slashed{P}_A^{~B}W_B=0$. Note that (\ref{equ:alt}) is of the same form as~\eqref{equ:twistorintegral-product} with $\Upsilon^*\cdot Z$ replaced by $\Upsilon^*\cdot\partial_W$. In order for the integral not to depend on the scale of the dual twistor coordinate, the integrand must scale as
\be \label{equ:scaling-Ftilde}
\tilde F(rW_A)=r^{2S-2} \tilde F(W_A)\,.
\ee 
We can translate this dual integral representation to the form \eqref{equ:twistorintegral-product} by expressing the integrand in terms of its Fourier transform
\begin{align}
\tilde{F}(W_A)&=\int \frac{d^4 Z}{(2\pi)^2} \hs  e^{i Z\cdot W}\hs F(Z^A)\,.
\label{equ:fouriertransform-F}
\end{align}
Plugging this into \eqref{equ:alt} gives precisely the twistor representation~\eqref{equ:twistorintegral-product} of $\tilde{J}$, where the integrand $F(Z^A)$ is the Fourier transform of $\tilde F(W_A)$; cf.~\eqref{equ:penroseoffourier} in Appendix~\ref{app:computations}. This also implies that \eqref{equ:alt} is indeed manifestly conserved.

\subsection{Spinning Correlators in Twistor Space}

We have learned that conservation can be made manifest for a current in 3d CFTs by expressing it as an integral over twistors like \eqref{equ:twistorintegral-product} or \eqref{equ:alt}. Hence, we can trivialize the conservation condition for $n$-point correlators of conserved currents by writing them in terms of $n$ integrals over twistors, one for each external point:
\begin{align}
\langle J_1J_2 \cdots J_n\rangle &= \left[ \prod_{j=1}^n i^{-S_j}\int DZ_j \hs (\Upsilon_j^*\cdot Z_j)^{2S_j} \right]
F(Z_i^A) \, ,
\label{equ:npt-ansatz} \\
\langle \tilde J_1 \tilde J_2\cdots \tilde J_n\rangle &= \left[ \prod_{j=1}^n i^{S_j} \int DW_i \left(\Upsilon_j^*\cdot \frac{\partial}{\partial W_j}\right)^{2S_j} \right]
\tilde F(W_{i,A})\, . 
\label{equ:npt-ansatz2}
\end{align}
As explained below \eqref{equ:Penrose-JJJ-discovering}, the
conformal symmetry of the correlators can be made manifest if we take the integrand to only depend on products between twistors contracted with the infinity twistor $\Omega_{AB}$:\footnote{Four-dimensional Dirac deltas like~$\delta^{(4)}(\sum_j c_j Z_j^A)$ are also invariant under conformal transformations, but do not depend only on products between twistors. \label{footnoteDiracdeltas} }
\be 
F(Z_i^A)=F(Z_i\cdot Z_j)\,, \quad {\rm where} \quad Z_i\cdot Z_j =Z_i^A \Omega_{AB} Z_j^B\,.
\ee 
Similarly, we take $\tilde F(W_{i,A})$  to depend only on the inner products $W_i\cdot W_j=-W_{i,A} \Omega^{AB} W_{j,B}$. 

\vskip4pt
Of course, we can also consider a mixed ansatz where some integrals are over twistors~$Z^A$ 
and others are over dual twistors $W_A$.  
For instance, for a three-point function, we may choose to represent the first field with a dual twistor and the remaining ones with twistors, which gives a twistor correlator of the form $F(W_1,Z_2,Z_3)$. As in \cite{Arkani-Hamed:2009hub}, we will refer to the different choices as choices of a basis. Notice that we can always translate integrals over dual twistors to integrals over twistors and vice versa by Fourier transforming the integrand as in~\eqref{equ:fouriertransform-F}, and hence we can change the basis by applying the corresponding Fourier transformations.  
In particular, we will see that all parity-even conserved structures at three points can be written either in the form \eqref{equ:npt-ansatz} or in the form \eqref{equ:npt-ansatz2} with an integrand that only depends on the inner products $Z_j\cdot Z_k$ or $W_j\cdot W_k$, respectively (i.e.~with no four-dimensional Dirac deltas like those in Footnote~\ref{footnoteDiracdeltas}).  

\subsection{Bootstrapping Three-Point Functions}
\label{ssec:boot}

In this section, we will bootstrap the three-point functions of conserved currents in twistor space.\footnote{See Appendix~\ref{sec:TwoPoints}, for a discussion of two-point functions.} More specifically, we will bootstrap the integrands $F(Z_i\cdot Z_j) $ and $\tilde F(W_i\cdot W_j)$ in~\eqref{equ:npt-ansatz} and \eqref{equ:npt-ansatz2}.

\vskip 4pt
First, we try a power-law ansatz for $F$ of the form
\be
F(Z_i\cdot Z_j) \stackrel{?}{=}  i^{S_T}\hs\frac{1}{(Z_1\cdot  Z_2)^{n_3+1}}\frac{1}{(Z_2\cdot Z_3)^{n_1+1}}\frac{1}{(Z_3\cdot Z_1)^{n_2+1}}\,,
\label{equ:powerlaw-F}
\ee
where the $ i^{S_T}$ factor, with $S_T\equiv S_1+S_2+S_3$, was introduced for later convenience, and the exponents are fixed by demanding that the projective integrals are invariant under rescalings:
\be 
\begin{aligned}
n_1 &\equiv S_2+S_3-S_1\, , \\ 
n_2 &\equiv S_1+S_3-S_2\, , \\ 
n_3 &\equiv S_1+S_2-S_3\, .
\end{aligned}
\label{equ:powers-n}
\ee 
However, as we prove in Appendix~\ref{sec-discrete-symm}, the correlator \eqref{equ:npt-ansatz} obtained from this ansatz for $F$ is odd under PT (i.e.~parity composed with time reversal). 
This is inconsistent with the CPT theorem, since the charge conjugation operator C doesn't affect the conserved currents.
 Hence, the ansatz~\eqref{equ:powerlaw-F} will not give us the expected correlators. 
 
\vskip 4pt
Instead, we will consider the following ansatz for the integrand 
\be 
F(Z_i \cdot  Z_j)= i^{S_T}\hs\delta^{[n_3]}(Z_1\cdot  Z_2)\hs\delta^{[n_1]}(Z_2\cdot Z_3)\hs\delta^{[n_2]}(Z_3\cdot Z_1)\,,
\label{equ:integrandF}
\ee 
where we define\footnote{We use the principal value prescription to write the factor $c^n$ in the integrand when $n\leq -1$, i.e.~$c^{-|n|}=((c+i\epsilon)^{-|n|}+(c-i\epsilon)^{-|n|})/2$.\label{footnote-pv-prescription}} 
\be 
\delta^{[n]}(x)\equiv i^{-n} \int_{-\infty}^{\infty} \frac{dc}{2\pi} e^{-icx} c^n=\begin{cases}\displaystyle \frac{d^n}{dx^n}\delta(x) &\text{ if }n\geq0\,,\\[8pt]
	\displaystyle\frac{1}{2(|n|-1)!}\hs\text{sign}(x) \hs x^{|n|-1}&\text{ if }n\leq-1\,.
	\end{cases}
\label{equ:def-delta-n}
\ee 
The powers $n_i$ in \eqref{equ:integrandF} are constrained to be the same as in~\eqref{equ:powers-n}. 
It is easy to check that, for all $n\in\mathbb{Z}$, the functions $\delta^{[n]}(x)$ satisfy\footnote{To prove equation \eqref{eq-derivative-diracdelta-n}, for $n=-m\leq -2$, we assume that the distributions are being integrated against a test function $g(x)$ that satisfies $x^{m-1}g(x)\to 0$ when $x\to0$ (we can achieve this by assuming that $g(x)$ doesn't diverge when $x\to 0$).} \begin{align} 
	\frac{d}{dx} \delta^{[n]}(x)&=\delta^{[n+1]}(x)\,,\label{eq-derivative-diracdelta-n}\\
	\delta^{[n]}(\rho x)&=\sign(\rho)\frac{1}{\rho^{n+1}}\hs \delta^{[n]}(x)\,.\label{eq-scaling-diracdelta-n}
\end{align}
We prove in Appendix~\ref{sec-discrete-symm} that
the integral \eqref{equ:npt-ansatz} obtained from the integrand \eqref{equ:integrandF} is even under PT, as expected. In this way, we have been able to bootstrap the integrand for the ansatz \eqref{equ:npt-ansatz} to be \eqref{equ:integrandF} after imposing conformal invariance, scaling covariance, and that it should be even under PT.  

\vskip 4pt
Following a similar logic, we find that the integrand of the ansatz~\eqref{equ:npt-ansatz2} with dual twistors is 
\be 
\tilde F(W_i  \cdot W_j)= i^{-S_T}\hs\delta^{[-n_3]}(W_1 \cdot W_2)\hs\delta^{[-n_1]}(W_2 \cdot W_3)\hs\delta^{[-n_2]}(W_3 \cdot W_1)\,,
\label{equ:integrandFt}
\ee 
where the powers $n_i$ are again the same as in~\eqref{equ:powers-n}.  Note that, for equal spins $S=1$, the result is a product of three sign factors. Amusingly, the three-point amplitude in Yang--Mills theory can also be expressed in twistor space as a product of three sign factors~\cite{Arkani-Hamed:2009hub}, albeit with a different infinity twistor. A similar correspondence holds between the correlator \eqref{equ:integrandFt} for equal spins $S=2$ and the three-point twistor amplitude in Einstein gravity, which are both given by a product of absolute values. The only difference is again in the infinity twistor because it can discern between flat space and (A)dS.

\vskip 4pt
Using the  integral representation of the delta function \eqref{equ:def-delta-n},
 the integrands \eqref{equ:integrandF} and \eqref{equ:integrandFt} can also be expressed as
\begin{align}  \label{equ:F-twistor}
F(Z_i \cdot Z_j)&=\int \frac{d^3c_{ij}}{(2\pi)^3}\hs \exp\left(-i c_{12} Z_1 \cdot Z_2+\text{cyclic}\right)\hs A(c_{ij})\,, \\
\tilde F(W_i \cdot W_j)&=\int \frac{d^3c_{ij}}{(2\pi)^3}\hs \exp\left(-i c_{12} W_1 \cdot W_2+\text{cyclic}\right)\hs \tilde A(c_{ij})\,,\label{equ:F-dualtwistor}
\end{align} 
where 
\begin{align}
A(c_{ij}) &=  c_{12}^{n_{3}} \hs c_{23}^{n_{1}} \hs c_{31}^{n_{2}} = \frac{1}{\tilde A(c_{ij})}\, .\label{equ:A-cij}
\end{align}
We see that the results take a particularly simple form in terms of the Schwinger parameters $c_{ij}$.

\vskip 4pt 
For identical currents, with spin $S$, we have
\begin{align}
A(c_{ij}) &= ( c_{12} \hs c_{23}\hs c_{31})^S = \frac{1}{\tilde A(c_{ij})}\,.
\label{equ:A-cij2}
\end{align}
We will soon show that $\tilde A(c_{ij})$ arises from the {\it leading interactions} (i.e.~YM and GR for $S=1$ and $S=2$), while $A(c_{ij})$ corresponds to {\it higher-derivative interactions} (i.e.~the higher-curvature interactions $F^3$ and $W^3$ for $S=1$ and $S=2$).\footnote{In~\cite{Baumann:2021fxj}, these were called ``minimal coupling" and ``non-minimal coupling", respectively.} 

\vskip 4pt
We see that, in terms of the Schwinger parameters, the three-point gravity ($S=2$) correlators are the square of the 
 gauge-theory ($S=1$) correlators. Although this double-copy structure looks rather trivial in twistor space, it does not work like this for correlators in embedding space or in momentum space, meaning that we do {\it not} get the gravity correlator by squaring the corresponding gauge-theory correlator~\cite{Lee:2022fgr}. The fact that this actually works for the twistor correlators written in Schwinger parameters is just an example of their simplicity compared to their  counterparts in embedding and momentum space.

\vskip 4pt
Notice that despite the similarity of the Schwinger-parametrized formulas $A(c_{ij})$ and $\tilde A(c_{ij})$, the presence of Schwinger parameters in the {\it denominator} for $\tilde A(c_{ij})$ is associated to correlators that come from the leading interactions in the bulk (YM or GR). This is precisely analogous to what happens for three-point scattering amplitudes in flat space expressed in spinor helicity variables: Despite being purely holomorphic/anti-holomorphic, they only have denominators for the cases of Yang--Mills or GR, thus signalizing a (very mild) breaking of holomorphicity. These denominators are the avatar of the pending challenge of building higher-point functions, showcasing the ``non-local" nature of the few consistent four-particle amplitudes of massless particles.  

\begin{center}
***
\end{center}

In Appendix~\ref{app:computations}, we explicitly compute the three-point twistor integrals \eqref{equ:npt-ansatz} and \eqref{equ:npt-ansatz2} with the integrands given by \eqref{equ:integrandF} and \eqref{equ:integrandFt}, respectively, and verify that this leads to known results in embedding space.\footnote{The computations will be performed for totally spacelike configurations, i.e.~all pairs of points are spacelike separated. The restriction to such configurations is because Wightman functions are holomorphic in this region (see Footnote~8 in \cite{Rychkov:Lorentzian}, as well as \cite{Streater:1989vi, Jost:1965yxu}). We can then analytically continue them to Euclidean signature, yielding the Euclidean correlators of interest. This procedure of analytic continuation consists of starting with a totally spacelike configuration (say, take them all at the time $x_i^0=0$), and giving imaginary parts to the time coordinates $x_i^0=-i\epsilon_i$, with $\epsilon_1>\cdots >\epsilon_n$ (see Section~2.5 of \cite{Simmons-Duffin:Lorentzian}).\label{footnote:analytic-continuation}} 
In the following, we present selected results for conformal scalars $O$, spin-1 currents $J$ and spin-2 currents $T$. We will omit irrelevant numerical factors, so all correlators should be seen as proportional to the explicit Penrose transform.

\begin{itemize}
	\item $\boldsymbol{\langle JOO\rangle}$: We first consider the three-point function of two conformal scalars and a spin-1 current.
	For both twistor integrals \eqref{equ:npt-ansatz} and \eqref{equ:npt-ansatz2}, we obtain the expected structure
	\be 
	\langle J_1O_2O_3\rangle=\langle \tilde J_1 \tilde O_2 \tilde O_3\rangle\propto\frac{P_{23}^{1/2}}{P_{12}^{3/2}P_{31}^{3/2}}\hs V_1\,,
	\ee 
	which is the unique structure allowed by conformal symmetry. 
	\item $\boldsymbol{\langle JJO\rangle}$: Similarly, for the three-point function of two spin-1 currents and a conformal scalar, we get
	\be 
	\langle J_1 J_2O_3\rangle=\langle \tilde J_1\tilde J_2\tilde O_3\rangle\propto\hs\frac{V_1 V_2-H_{1 2}}{P_{12}^{5/2} P_{23}^{1/2} P_{31}^{1/2} }\,,
	\ee 
	which is the unique structure allowed by conformal symmetry and conservation. 
	\item $\boldsymbol{\langle JJJ\rangle}$: Next, we consider the case of three identical spin-1 currents. The twistor integral~\eqref{equ:npt-ansatz} gives 
\be 
\langle  J_1 J_2 J_3\rangle\propto\hs\frac{(5 V_1 V_2 V_3+V_1 H_{2 3}+V_2 H_{3 1}+V_3 H_{1 2})}{\left(P_{1 2} P_{2 3} P_{3 1}\right)^{3/2}} \,, 
\ee 
which is proportional to the known result for the three-point function arising from a higher-derivative cubic term $F^3$~\cite{Giombi:2011rz, Zhiboedov:2012bm,osborn1994implications}.

	The twistor integral \eqref{equ:npt-ansatz2} instead gives 
\be \label{equ:3ptYM-tildeJ}
\langle \tilde J_1 \tilde J_2 \tilde J_3 \rangle\propto\hs\frac{ \left(V_1V_2V_3 +V_1 H_{2 3}+V_2 H_{3 1}+V_3H_{1 2}\right)}{ (P_{1 2} P_{2 3}
	P_{3 1})^{3/2}}\,, 
\ee 
which is proportional to the three-point function arising from the Yang--Mills vertex. 
\item $\boldsymbol{\langle TTT\rangle}$: Finally, we consider the case of three identical spin-2 currents. The twistor integral~\eqref{equ:npt-ansatz} gives
\begin{align}
\langle T_1 T_2 T_3\rangle&\propto  
\frac{ \left(2\hs V_1^2 H_{2,3}^2-16\hs V_2V_3 H_{31}H_{12}-52\hs H_{23}V_1^2V_2V_3+\text{cyclic}\right)-147\hs V_1^2V_2^2V_3^2}{(P_{12}P_{23}P_{31})^{5/2}} \,,
\end{align}
which is proportional to the known result~\eqref{equ:TTT-W-embedding} for the three-point function arising from a higher-derivative cubic term  $W^3$.

The twistor integral \eqref{equ:npt-ansatz2} instead gives
\begin{align}
\langle \tilde T_1 \tilde T_2 \tilde T_3 \rangle &\propto 
 \frac{\left(6\hs V_1^2 H_{2,3}^2+16\hs V_2V_3 H_{31}H_{12}+4\hs H_{23}V_1^2V_2V_3+\text{cyclic}\right)-9\hs V_1^2V_2^2V_3^2}{(P_{12}P_{23}P_{31})^{5/2}} \,,
\end{align}
which is proportional to the GR correlator \eqref{equ:TTT-E-embedding}.

\end{itemize}
We have therefore confirmed that these twistor integrals reproduce known three-point correlators involving conserved currents in embedding space. We found that the ``product-form" twistor integral \eqref{equ:npt-ansatz} gives the structure arising from higher-derivative interactions in the bulk ($F^3$ for spin~$1$ and $W^3$ for spin~$2$), while the ``derivative-form" twistor integral \eqref{equ:npt-ansatz2} give  the structure corresponding to the leading (minimal coupling) interactions in the bulk (Yang--Mills for spin $1$ and GR for spin~$2$). We stress that this formalism gives directly the conserved structures without first having to construct the larger space of conformally-invariant structures that may or may not be conserved.

\section{Back to Momentum Space}
\label{sec:FourierSpace}

In previous work~\cite{Maldacena:2011nz, Baumann:2020dch, Baumann:2021fxj}, some of us studied spinning correlators in Fourier space. 
The complexity of the results was somewhat unsatisfying, especially when compared to the dramatically simpler results for scattering amplitudes. In this section, we will relate our twistor-space correlators to these results in Fourier space via a half-Fourier transform \cite{Witten:2003nn}.

\subsection{Spinning Correlators in Fourier Space}

We begin with a quick review of the three-point functions of conserved currents in Fourier space~\cite{Maldacena:2011nz,Bzowski:2017poo,Bzowski:2018fql}.  As for massless scattering amplitudes, it is convenient to introduce spinor helicity variables via \footnote{There have been other proposals for defining spinor helicity variables in (A)dS, see \cite{Nagaraj:2018nxq,Nagaraj:2019zmk,Nagaraj:2020sji,David:2019mos,Basile:2024ydc}.}
\beq\label{equ:shv-sec4}
k_{\alpha\dot\alpha}=k_\mu (\sigma^\mu)_{\alpha\dot\alpha}=\lambda_\alpha \tilde\lambda_{\dot\alpha}\,,
\eeq 
where $k_\mu=(|\vec k|,\vec k)$ contains the three-momentum $\vec k$ and its energy $k\equiv |\vec k|$, which is contracted with the Pauli matrices with one lowered index $(\sigma^\mu)_{\alpha\dot\alpha}=(-\epsilon,-\sigma_z,i\mathbf{1}_2,\sigma_x)_{\alpha\dot\alpha}$. 
Unlike in four-dimensional Minkowski space, 
the symmetry group is just given by spatial rotations~SO$(3)$. This implies that we can contract dotted and undotted indices with a tensor $\tau^{\alpha\dot\beta}=\epsilon^{\alpha\dot\beta}$, such that $\tau^{\alpha\dot\beta}k_{\alpha\dot\beta}=2k$, which in turn allows us to trade any dotted index for an undotted index. In particular, we can define $\bar\lambda_\alpha=\tau_{\alpha}^{~\dot\alpha}\tilde\lambda_{\dot\alpha}$, and work only with spinors $\lambda_\alpha,\hs\bar\lambda_\alpha$ with undotted indices. Furthermore, we define the spinor brackets as $\langle ij\rangle \equiv \epsilon^{\beta \alpha}\lambda_\alpha^i \lambda^j_\beta$, $\langle \bar i \bar j\rangle \equiv  \epsilon^{\beta \alpha} \bar \lambda_\alpha^i \bar \lambda^j_\beta$ and $\langle i \bar j\rangle \equiv  \epsilon^{\beta \alpha}\lambda_\alpha^i \bar \lambda^j_\beta$. Notice that, unlike for amplitudes, we can have mixed brackets between holomorphic and anti-holomorphic spinors.

\vskip 4pt
To find the correlators of conserved currents in a CFT, we simultaneously solve the conformal Ward identity and the Ward--Takahashi identity of current conservation~\cite{Maldacena:2011nz, Baumann:2020dch}. For reference, we now present the results for the three-point functions of identical spin-$S$ currents.
\begin{itemize}

\item Let us first consider the contributions from {\it higher-derivative interactions}. For equal helicities, the three-point function of a spin-$S$ current is~\cite{Baumann:2020dch}
\beq
\frac{\langle J^+_1 J^+_2 J^+_3\rangle}{(k_1k_2k_3)^{S-1}}=\left(\frac{\langle\bar1 \bar2 \rangle \langle \bar2\bar3\rangle \langle \bar3\bar1\rangle}{ E^3}\right)^S\, ,
\label{equ:result1}
\eeq
where we defined $E \equiv k_1 +k_2+k_3$ as the sum of the external energies. 
For future convenience, the correlator was normalized in such a way that it has scaling dimension $2$.
Correlators with mixed helicities vanish, $\langle J^-_1 J^+_2 J^+_3 \rangle =0$.
For spins $S=1$ and $2$, the result reduces to the three-point functions arising from $F^3$ and $W^3$ interactions, respectively. 

\item For the {\it leading interactions}, the all-plus three-point function of a spin-$S$ current is
\be 
\frac{\langle J^+_1 J^+_2 J^+_3 \rangle}{(k_1k_2k_3)^{S-1}}=\left(\frac{\langle\bar1 \bar2 \rangle \langle \bar2\bar3\rangle \langle \bar3\bar1\rangle \hat{E}}{ k_1k_2k_3}\right)^S \frac{f_{[SSS]}(k_i)}{(k_1k_2k_3)^{S-1}}\,,
\label{equ:result2}
\ee
where $\hat{E}\equiv \frac{1}{2}(\langle \bar 1 1\rangle+\langle \bar 22\rangle+\langle \bar 33\rangle)$ and $f_{[SSS]}(k_i)$ are {\it form factors} that were derived in~\cite{Baumann:2020dch}. For spins $S=1$ and $2$, the result reduces to the three-point functions arising from Yang--Mills theory and Einstein gravity, respectively, with the form factors given by 
\begin{align}\label{equ:f111}
	f_{[111]}(k_i)&=\frac{1}{E}\,,\\
	f_{[222]}(k_i) &=\frac{Q(k_i)}{256\hs E^2}\,,\label{equ:f222}
\end{align}
where $Q(k_i)\equiv E^3-E(k_1k_2+k_2k_3+k_3k_1)-k_1k_2k_3$ is a fairly non-trivial numerator. Indeed, these form factors get more complicated for larger spin. 
Correlators with negative helicities can be obtained from the expression (\ref{equ:result2}) by interchanging barred and unbarred spinors, which changes some signs in $\hat{E}$ \cite{Baumann:2021fxj}.
\end{itemize}
These results at three points might not look so bad, but the complexity explodes at higher points~\cite{Baumann:2020dch, Baumann:2021fxj,Raju:2012zs,Albayrak:2018tam,Albayrak:2019yve,Albayrak:2023jzl,Albayrak:2023kfk}. Moreover, using these kinematic variables in momentum space, the different Feynman diagrams are still treated independently and there is no simple unified way to represent the physical correlators (without breaking them into unphysical (gauge-dependent) parts). This is morally wrong, since it hasn't absorbed the lesson of the amplitudes program that real simplicity arises only in physical on-shell amplitudes. Our hope (which we realized at three points) is that twistor space is the right arena to expose the hidden simplicity of spinning correlators in cosmology.

\subsection{Transforms of Twistor Correlators}
\label{ssec:tranforms}

In~\cite{Nair:1988bq,Witten:2003nn,Arkani-Hamed:2009hub,Mason:2009sa}, scattering amplitudes in four-dimensional momentum space were mapped to twistor space by performing a so-called {\it half-Fourier transform}. In the following, we will explain how to obtain the results (\ref{equ:result1}) and (\ref{equ:result2}) in Fourier space from an analogous (inverse) half-Fourier transform of our twistor correlators. 
\vskip 4pt
To define the half-Fourier transform that maps objects from twistor space to momentum space, we first write the twistor coordinates $Z_i^A$ in terms of two-component spinors
\be \label{equ:splitting-Z}
Z_i^A \equiv \begin{pmatrix} \lambda_{i}^{\alpha} \\
\mu_{i,\dot{\alpha}}
\end{pmatrix} .
\ee 
The half-Fourier transform with respect to $\mu_i$ is then defined as 
\be  \label{equ:fouriertransform-mu}
g(\lambda_i,\tilde{\lambda}_i) \equiv \int d^2\mu_i 
\hs \exp(i \tilde{\lambda}_i\cdot \mu_i)\,f(Z_i^A)\, ,
\ee 
where the spinors $\lambda_i$ and $\tilde{\lambda}_i$ are identified with the spinor helicity variables in momentum space, and we use the same convention as \eqref{equ:notation-contraction} for contractions of the spinor indices, $v\cdot u=v^\alpha u_\alpha$.

\vskip 4pt
Analogously, to define the half-Fourier transform that maps objects in dual twistor space to momentum space, we first write the dual twistor as 
\be 
W_{i,A} \equiv \begin{pmatrix}\tilde{\mu}_{i,\alpha}\\[4pt]
	\tilde{\lambda}_{i}^{\dot{\alpha}}
\end{pmatrix} ,
\ee 
and then define the transform with respect to $\tilde\mu_i$ as
\be  \label{equ:fouriertransform-mutilde}
\tilde g(\lambda_i,\tilde{\lambda}_i) \equiv \int d^2\tilde\mu_i 
 \hs \exp(-i \lambda_i\cdot \tilde\mu_i)\, \tilde f(W_{i,A})\,.
\ee 
This can be deduced by composing the map \eqref{equ:fouriertransform-mu} between momentum and twistor space with the map \eqref{equ:fouriertransform-F} between twistor and dual twistor space.

\vskip 4pt
Let us make some important observations. First, in order for these maps to make sense as Fourier transforms, we need the spinors $\lambda_i,\hs \tilde\lambda_i,\hs\mu_i,\hs \tilde\mu_i$ to be real and independent \cite{Witten:2003nn}. In the case of four-dimensional scattering amplitudes in flat space, this reality condition corresponds to the so-called split signature with a spacetime $\mathbb{R}^{2,2}$, while for our boundary correlators it corresponds to a physical position space $\mathbb{R}^{1,2}$ with Lorentzian signature. More explicitly, this amounts to a Wick rotation of the three-momenta in \eqref{equ:shv-sec4} as $\vec k=(k^1,i k^0,k^2)$, that must be spacelike,
so that the matrix in \eqref{equ:shv-sec4} is purely real. 
Second, the inverse map simplifies substantially the generator of special conformal transformations (SCT), as it maps a second-order differential operator in momentum space to a first-order operator in twistor space~\cite{Witten:2003nn}. For scattering amplitudes, this operator generates conformal transformations on the four-dimensional flat space that are not necessarily a symmetry of the amplitudes. On the other hand, for our boundary correlators, the operator that generates SCTs on the three-dimensional boundary,
\be 
K_{\alpha\beta}\equiv \frac{\partial^2}{\partial \lambda^{(\alpha} \partial \bar\lambda^{\beta)}}\,,
\ee  
is a symmetry of the theory. Finally, we notice that this operator acts more naturally on spin-$S$ conserved currents normalized as $J/k^{S-1}$ \cite{Maldacena:2011nz}, as well as on the half-Fourier transforms \eqref{equ:fouriertransform-mu} or~\eqref{equ:fouriertransform-mutilde}. Thus, we must normalize the correlators as in \eqref{equ:result1} or \eqref{equ:result2} before comparing to the half-Fourier transforms of the twistor correlators.

\subsubsection*{Higher-derivative interactions}

In Section~\ref{ssec:boot}, we bootstrapped the three-point functions of conserved spin-$S$ currents in twistor space. Using a product ansatz, we found
\begin{align}\label{equ:F-twistor-sec4}
F(Z_i \cdot Z_j)&=\int \frac{d^3c_{ij}}{(2\pi)^3}\hs \exp\left(-i c_{12} Z_1 \cdot Z_2+\text{cyclic}\right)\hs A(c_{ij})\,, 
\end{align} 
where $A(c_{ij}) = (c_{12} c_{23} c_{31})^S$. This twistor correlator has been mapped via \eqref{equ:npt-ansatz} to the three-point structures in embedding space arising from higher-derivative interactions in the bulk. Let us now map it to momentum space via the half-Fourier transform \eqref{equ:fouriertransform-mu}. 

\vskip 4pt
In Appendix~\ref{app:computations2}, we show that the half-Fourier transform applied to \eqref{equ:F-twistor-sec4} yields
\begin{align}\label{equ:half-FT-G}
G(\lambda_i, \tilde \lambda_i)  &=\left[\prod_{i=1}^3\int d^2\mu_i
 \hs \exp(i \tilde{\lambda}_i\cdot\mu_i)\right]F(Z_j\cdot Z_k)
= (2\pi)^3\delta^{(3)}\big(\vec{k}_1 + \vec{k}_2 + \vec{k}_3 \big)\, g(\lambda_i, \tilde \lambda_i)\,,
\end{align}
where the result after stripping off the momentum-conserving  delta function is 
\be \label{equ:half-FT-g}
g(\lambda_i, \tilde \lambda_i) =\frac{A(c_{ij})}{4} \Big|_{c_{ij}=\langle \bar{i}\bar{j}\rangle/E} =\frac{1}{4} \left(\frac{\langle \bar{1}\bar{2}\rangle\langle \bar{2}\bar{3}\rangle\langle \bar{3}\bar{1}\rangle}{E^3}\right)^S \,.
\ee 
We see that this is proportional to the all-plus three-point function of spin-$S$ currents in (\ref{equ:result1}) coming from higher-derivative interactions.

\subsubsection*{Leading interactions}

In Section~\ref{ssec:boot}, we also bootstrapped a second correlator using a derivative-ansatz in dual twistor space:
\be \label{equ:F-dualtwistor-sec4}
\tilde F(W_i \cdot W_j)=\int \frac{d^3c_{ij}}{(2\pi)^3}\hs \exp\left(-i c_{12} W_1 \cdot W_2+\text{cyclic}\right)\hs \tilde A(c_{ij})\,,
\ee
where $\tilde A(c_{ij}) = (c_{12} c_{23} c_{31})^{-S}$. This twistor correlator has been mapped via \eqref{equ:npt-ansatz2} to the three-point structures in embedding space that arise from the leading interactions (i.e.~YM and GR for spins $S=1$ and $S=2$, respectively). 

\vskip 4pt
Performing the same analysis as before,
the half-Fourier transform \eqref{equ:fouriertransform-mutilde} applied to \eqref{equ:F-dualtwistor-sec4} gives
\be\label{equ:half-FT-Gtilde}
	\tilde G(\lambda_i, \tilde \lambda_i)  =\left[\prod_{i=1}^3\int d^2\tilde\mu_i
	  \hs \exp(-i \lambda_i\cdot\tilde\mu_i)\right]\tilde F(W_j\cdot W_k) = (2\pi)^3\delta^{(3)}\big(\vec{k}_1 + \vec{k}_2 + \vec{k}_3 \big)\, \tilde g(\lambda_i, \tilde \lambda_i)\,,
\ee
where the result after stripping off the momentum-conserving delta function is 
\be \label{equ:half-FT-gtilde}
\tilde g(\lambda_i, \tilde \lambda_i) =\frac{\tilde A(c_{ij})}{4} \Big|_{c_{ij}=\langle ij\rangle/E} =\frac{1}{4} \left(\displaystyle\frac{\langle \bar 1\bar2\rangle\langle \bar2\bar3\rangle\langle \bar 3\bar 1\rangle}{(E-2k_1)(E-2k_2)(E-2k_3)}\right)^S.
\ee 
In the second equality, 
we used that $\langle ij\rangle/E = (E-2k_l)/\langle \bar i \bar j\rangle$, with $\{i,j,l\}$ a cyclic permutation of $\{1,2,3\}$, which is valid on the support of three-point kinematics \cite{Baumann:2020dch}.
Extracting the same factor from $\tilde g(\lambda_i, \tilde \lambda_i)$ as in~\eqref{equ:result2}, 
we can identify the associated form factor as
\be 
\label{equ:result3}
\tilde f_{[SSS]}(k_i)=\frac{1}{4}\frac{(k_1k_2k_3)^{2S-1}}{(E(E-2k_1)(E-2k_2)(E-2k_3))^S}\,.
\ee 
It turns out that this result is proportional to the  {\it discontinuity} of the form factor $f_{[SSS]}(k_i)$ defined in \eqref{equ:result2}: 
\be \label{equ:disc-fSSS}
\tilde f_{[SSS]}(k_i) 
\propto \text{Disc}_{k_1^2}\hs\text{Disc}_{k_2^2}\hs\text{Disc}_{k_3^2} f_{[SSS]}(k_i)\,,
\ee 
which we checked explicitly up to spin 6.

\vskip 4pt
We can derive the actual form factor $f_{[SSS]}(k_i)$ of the momentum-space correlator \eqref{equ:result2} 
by performing the following dispersive integral~\cite{Meltzer:2021bmb,Meltzer:2021zin}
\be \label{equ:fSSS-dispersive}
f_{[SSS]}(k_i)\propto\left[\prod_{i=1}^3 \frac{1}{2\pi i}\int_{-\infty}^{\infty} \frac{\omega_i \hs d\omega_i}{\omega_i^2-k_i^2}\right] \tilde f_{[SSS]}(\omega_j)\,.
\ee 
In Appendix~\ref{app:computations2}, we confirm explicitly that the correct form factors (\ref{equ:result2}) are obtained from this integral with the integrand (\ref{equ:result3}). It is rather remarkable how these complex form factors arise from the transformation of the extremely simple twistor correlators $A(c_{ij}) = (c_{12} c_{23} c_{31})^{-S}$.

\vskip 4pt
We believe that the discontinuity arises from the half-Fourier transform because we evaluate the correlators for Lorentzian kinematics restricted to some specific domain of spacelike momenta. This analytic continuation, though understood in position space at two points (see e.g.~\cite{Moschella:2024kvk}), is more challenging in momentum space \cite{Gillioz:2021sce,Bautista:2019qxj}. The results of this paper at three points suggest that the analytic continuation procedure in position space explained in Footnote~\ref{footnote:analytic-continuation} correspond to a dispersive integral in momentum space, which gives the correlator from its discontinuity. It would be interesting to develop this direction further. Finally, since discontinuities in momentum space obey simple cutting rules~\cite{Melville:2021lst,Goodhew:2021oqg,Goodhew:2020hob,Salcedo:2022aal}, they might even allow us to obtain higher-point twistor correlators from lower-point correlators via the half-Fourier transform.

\section{Conclusions and Outlook}
 \label{sec:Conclusions}
 
In 1986, Parke and Taylor discovered a remarkable formula describing the scattering of $n$ gluons in the maximally helicity violating configuration~\cite{Parke:1986gb}:
\beq
A(1^- 2^- 3^+ \cdots n^+) = \frac{\langle 12\rangle^4}{\langle 12\rangle \langle 23\rangle \cdots \langle (n-1) n \rangle \langle n1\rangle}\,,
\label{equ:ParkeTaylor}
\eeq
where the brackets $\langle ij \rangle$ represent the momenta of the external particles in terms of spinor helicity variables~\cite{Elvang:2015rqa}. At the time, this result was a stunning simplification of a very complex computation using Feynman diagrams, but subsequently it was shown that the Parke--Taylor formula can be bootstrapped from basic physical consistency requirements, avoiding the crippling complexity of Feynman diagrams. This started a revolution in our understanding of scattering amplitudes, with new on-shell methods allowing more direct computations of amplitudes, which then exposed hidden symmetries~\cite{Drummond:2008vq} and new geometrical objects~\cite{Arkani-Hamed:2013jha} underlying the physics of particle scattering.

\vskip 4pt
Cosmology has not yet had its Parke--Taylor moment, meaning that the analog of a formula like $\eqref{equ:ParkeTaylor}$ has not been discovered. Part of the problem is that, in the cosmological context, we don't have kinematic variables that describe the sum of Feynman diagrams as a unified physical object. Instead, cosmological correlators are still described in terms of their separation into Feynman diagrams, even for gauge theory and gravity where the individual diagrams themselves aren't even physical. This is against the spirit of the on-shell description of scattering amplitudes, so it isn't surprising that explicit results for gluon and graviton correlators in  (A)dS are very complicated (see e.g.~\cite{Baumann:2020dch, Baumann:2021fxj, Albayrak:2020fyp, Bonifacio:2022vwa,Sleight:2021iix}).

\vskip 4pt 
Although this paper has not yet realized the dream of a Parke--Taylor-like formula for spinning correlators in cosmology, it provides a new avenue towards finding such a structure. 
In particular, we believe that twistors are the right kinematic variables to expose the hidden simplicity of spinning correlators. Using twistors has allowed us to write integral expressions for these correlators that make both conformal symmetry and current conservation manifest. At the three-point level, the kinematic constraints completely fix the correlators and we have shown that our results in twistor space are consistent with known results in momentum space and embedding space, but arguably they are simpler.

\vskip 4pt
There are a few natural next steps to take in this adventure of using twistors in {\it cosmology}:
\begin{itemize}
\item First of all, it is important to ask how  the twistor representation of spinning correlators can be extended to higher points. While three-point functions are completely fixed by kinematics, determining higher-point functions requires additional dynamical input. For CFTs, these additional constraints come from consistency of the OPE, while for scattering amplitudes one imposes consistent factorization on all poles.  Already at four points, these constraints impose severe restrictions on the space of consistent theories~\cite{Benincasa:2007xk, McGady:2013sga}.  It will be interesting to explore how the space of consistent cosmological correlators is constrained by similar considerations.

\item For scattering amplitudes, the construction of higher-point amplitudes can be made systematic through the use of powerful recursion relations~\cite{Britto:2005fq}. It would be nice to find a similar recursive method for our correlators in twistor space. Indeed, in~\cite{Arkani-Hamed:2009hub, Mason:2009sa}, it was shown that BCFW recursion relations for amplitudes also have a natural representation in twistor space. 
It will be interesting to see how these insights translate to the cosmological context.

\item Finally, we would like to return to the ambitious goal of deriving (or guessing) a Parke--Taylor-like formula for all-multiplicity gluon correlators in de Sitter space. While until recently finding such a formula seemed like a distant dream, the simplicity of our twistor-based approach makes us optimistic that this is now within reach.
\end{itemize}
\newpage
\noindent
In addition, the work initiated in this paper may also have fruitful new applications to the study of {\it conformal field theories}:
\begin{itemize}
\item While we focused on 3d CFTs,  
presumably twistors will play a similar role for CFTs in higher dimensions. 
Of course, since spinors are different for each dimension \cite{Freedman:2012zz}, the twistor formalism will also be dimension-dependent. In particular, it should be possible to develop an analogous formalism for 4d CFTs, where we can use the spinor helicity variables of 6d flat space \cite{Cheung:2009dc} to define embedding-space spinors~\cite{Sinkovics:2004fm}. This could be applied to $\mathcal N=4$ SYM, and it would be interesting to explore how this twistor formalism can be translated to the gravity side of the (A)dS$_5/$CFT$_4$ conjecture \cite{Maldacena:1997re,Witten:1998qj,Gubser:1998bc,Aharony:1999ti}, and how it compares to twistors in (A)dS$_5$ \cite{Adamo:2016rtr}. 

\item In Footnote~\ref{footnoteLightray}, we speculated about an intriguing connection between twistor space and light-ray operators\footnote{Light-ray operators are powerful tools to investigate QFTs, as they are simpler than local operators, and yield information that local probes might not be able to access. They have been central to recent developments, see for example \cite{Hofman:2008ar,Hartman:2015lfa,Caron-Huot:2017vep,Kravchuk:2018htv}. They have also played a key role in various attempts to derive Einstein gravity from CFT through holographic duality \cite{Meltzer:2017rtf,Belin:2019mnx,Kologlu:2019bco,Belin:2020lsr}, and have far-reaching positivity properties~\cite{Faulkner:2016mzt,Hartman:2016lgu,Hartman:2023qdn,Hartman:2023ccw}. } in quantum field theory. This seems worth exploring further. In particular, a light-ray operator in Minkowski space, which is non-local, 
gets mapped to a point in twistor space, and vice versa. This suggests that light-ray operators could be studied as local operators in twistor space, and this seems like a fruitful mindset, where we could harness the power of CFT techniques for local operators. 

\item It would be very interesting to generalize the twistor formalism to CFT correlators involving both conserved currents and non-conserved tensors. This is analogous to the extension of spinor helicity variables to amplitudes with both massive and massless particles \cite{Conde:2016izb,Conde:2016vxs,Arkani-Hamed:2017jhn}. Since there is no holomorphicity requirement for  non-conserved tensors, we expect these mixed correlators to be described by  integrals where only the twistors 
of the conserved currents are integrated over.\footnote{For instance, an ansatz for the three-point function of a conserved spin-$S$ current and two scalars with arbitrary scaling dimensions $\Delta_2$ and $\Delta_3$ is
	\be 
	\langle J_1 O_2 O_3\rangle \propto \frac{1}{P_{23}^{\Delta-(S+1)}} \int DZ_1 \left(\Upsilon_1^*\cdot Z_1\right)^{2S} \delta^{[S]}\left(Z_1\cdot \slashed{P}_2\cdot \slashed{P}_3\cdot Z_1\right),
	\ee 
where we integrate only over the twistor $Z_1^A$ of the conserved current. Notice that this ansatz only makes sense if the scaling dimensions of the scalars coincide $\Delta_2=\Delta_3=\Delta$, otherwise it vanishes~\cite{Rychkov:2016iqz}. It is straightforward to compute this integral using \eqref{eq-Whittaker-contractions-Zlanguage} and show that it gives the known tensor structure $V_1^S$ for this correlator.}
\end{itemize}
Much of the revival of twistor methods has come from the study of {\it scattering amplitudes}.  It is therefore also natural to use our twistor-formulation of cosmological correlators to make closer connections with some of the remarkable structures that have been found for amplitudes:

\begin{itemize}
\item One interesting structure lurking inside scattering amplitudes are {\it double-copy relations} between gauge-theory and gravity amplitudes~\cite{Bern:2019prr}. For cosmological correlators, however, the double copy is not as straightforward when formulated in momentum or position space~\cite{Lipstein:2023pih,Armstrong:2023phb,Diwakar:2021juk,Herderschee:2022ntr,Cheung:2022pdk}, even at low number of points~\cite{Lee:2022fgr}. It would be interesting to investigate whether twistor space makes the double copy more manifest. As we have shown, at three points, the twistor double copy requires just squaring the Schwinger parametrization of the correlator.  We would like to explore if this simple structure still holds at higher points.

%\item Another fascinating structure found for amplitudes are {\it positive geometries}, which are objects defined in kinematic space whose volumes are scattering amplitudes. In searching for such positive geometries, an important first step is to elucidate the ``best kinematic description" of the observables, and then search for a geometry in this kinematic space. So far, most of the structures found for cosmological correlators either refer to a single diagram or to the combinatorics of the sum of diagrams  \cite{Arkani-Hamed:2017fdk,Benincasa:2019vqr,Arkani-Hamed:2023bsv,Arkani-Hamed:2023kig,De:2023xue}. Moreover, being restricted to individual diagrams, these positive geometries are only applicable to scalar theories. Twistors open up the possibility of studying spinning theories in cosmology, combining all diagrams in rigid formulas, and thus making the search for positive geometries  feasible. \fr{What if we replace this item by the next one in color, used in the Bad-Honnef notes?}

\item Finally, twistor space can potentially allow us to interpret cosmological correlators as volumes of {\it positive geometries} in kinematic space, as it happens for scattering amplitudes. In fact, writing amplitudes in twistor space was instrumental to relate them with Grassmanians and positive geometries 
	\cite{Arkani-Hamed:2009ljj}. In searching for such geometries, an important first step is to elucidate the ``best kinematic description" of the observables, and then search for a geometry in this kinematic space. So far, most of the structures found for cosmological correlators either refer to a single diagram or to the combinatorics of the sum of diagrams in scalar theories  \cite{Arkani-Hamed:2017fdk,Benincasa:2019vqr,Arkani-Hamed:2023bsv,Arkani-Hamed:2023kig,De:2023xue}. 
	Twistors open up the possibility of studying spinning theories in cosmology, combining all diagrams in rigid formulas, and thus making the search for positive geometries  feasible.
\end{itemize}
We will pursue these ideas in future work.

 \vspace{0.2cm}
 \paragraph{Acknowledgments} We have benefited greatly from the feedback and advice of many colleagues. Thanks for  insightful discussions to Tim Adamo, Soner Albayrak, Nima Arkani-Hamed, Jan de Boer, Laura Engelbrecht, Harry Goodhew, Aidan Herderschee, Yu-tin Huang,  Austin Joyce, Hayden Lee,  Lionel Mason, Ugo Moschella,  Hugh Osborn, Justinas Rumbutis, Slava Rychkov, Kamran Salehi Vaziri, Chia-Hsien Shen, David Skinner, Charlotte Sleight, Zimo Sun, Massimo Taronna and Alessandro Vichi.  A special thanks to Mariana Carrillo Gonz\'alez who was the first to suggest the connections to mini-twistor space to us. 
 GLP presented an earlier version of this work at the ``50+$\varepsilon$ Years of Conformal Bootstrap" conference, and is grateful for the questions and feedback received from its participants.
 
 \vskip 4pt
The research of DB is funded by the European Union (ERC,  \raisebox{-2pt}{\includegraphics[height=0.9\baselineskip]{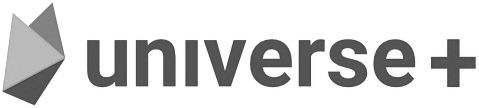}}, 101118787). Views and opinions expressed are, however, those of the author(s) only and do not necessarily reflect those of the European Union or the European Research Council Executive Agency. Neither the European Union nor the granting authority can be held responsible for them. DB is further supported by a Yushan Professorship at National Taiwan University funded by the Ministry of Education (MOE) NTU-112V2004-1. DB thanks the Max-Planck-Institute for Physics (MPP) in Garching for its hospitality while this work was being completed. He is grateful to the Alexander von Humboldt-Stiftung and the Carl Friedrich von Siemens-Stiftung for supporting his visits to the~MPP.

\vskip 4pt 
 GM is supported by the Simons Foundation through grant 488649 (Simons Collaboration on the Nonperturbative Bootstrap), by the Swiss National Science Foundation through the project 200020 197160 and by
the National Centre of Competence in Research SwissMAP.

\vskip 4pt
GLP is supported by Scuola Normale, by a Rita-Levi Montalcini Fellowship from the Italian Ministry of Universities and Research (MUR), and by INFN (IS GSS-Pi).

\newpage
\appendix
\section{Discrete Symmetries}
\label{sec-discrete-symm}

In Section~\ref{ssec:boot}, we proposed a power-law ansatz for twistor correlators at three points:
\be
F(Z_i\cdot Z_j) \stackrel{?}{=}  i^{S_T}\hs
\frac{1}{(Z_1\cdot  Z_2)^{n_3+1}}\frac{1}{(Z_2\cdot Z_3)^{n_1+1}}\frac{1}{(Z_3\cdot Z_1)^{n_2+1}}\,,
\label{equ:Ansatz1}
\ee
where $S_T \equiv S_1 + S_2 +S_3$ and $n_k\equiv S_i+S_j-S_k$ for $\{i,j,k\}$ a cyclic permutation of $\{1,2,3\}$. In this appendix, we will show that the correlators  obtained from this ansatz are odd under PT (i.e.~parity composed with time reversal), which is inconsistent with the CPT theorem. 
We will also demonstrate that the ansatz (\ref{equ:integrandF}),
\be 
F(Z_i \cdot  Z_j)= i^{S_T}\hs
\delta^{[n_3]}(Z_1\cdot  Z_2)\hs\delta^{[n_1]}(Z_2\cdot Z_3)\hs\delta^{[n_2]}(Z_3\cdot Z_1)\,,
\label{equ:Ansatz2}
\ee 
gives the correct PT-even correlators.

\subsection{Time Reversal and Parity} 

We will first study the action of parity and time reversal on the embedding-space spinors $\Lambda_a^A$.
We begin by noting that the PT operator acts on the 3d Lorentzian positions as 
\be \label{equ:PT-xmu}
(x^0,x^1,x^2)\ \mapsto\ \tilde{x}^\mu\equiv (-x^0,x^1,-x^2)\,,
\ee 
flipping the sign of one spatial and one temporal component. Hence, this operation may be written in terms of the spinors on the Poincaré slice as
\be \label{equ:PT-Lambda}
\Lambda_{a,A}(x^\mu)\ \mapsto\ \Lambda_{a,A}(\tilde{x}^\mu)= (\sigma_1)_a^{~b} \hs \left(\Gamma_{[20]}\right)_A^{~B}\hs\Lambda_{b,B}(x^\mu)\,,\quad(\sigma_1)_a^{~b} \equiv \begin{pmatrix}
	0&1\\1&0
\end{pmatrix},
\ee 	
where we defined $\Gamma_{[20]}\equiv \Gamma_2\cdot \Gamma_0$ to avoid clutter, which is symmetric (once we lower its indices) and squares to the identity. In \eqref{equ:PT-Lambda}, we are acting on both the little group and the spinor indices. Moreover, performing a little group transformation, $\Lambda_{a,A}=M_a^{~b}\Lambda_{b,A}$, where $M_a^{~b}\in \hs \hs$GL$(2,\mathbb R)$, we find that the PT operator acts on a generic slice as
\be
\Lambda_{a,A}\, \mapsto\, q_a^{~b}\hs \left(\Gamma_{[20]}\right)_A^{~B}\hs\Lambda_{b,B}\,,
\ee
where the matrix $q_a^{~b}\equiv (M \sigma_1 M^{-1})_a^{~b} $ satisfies $\det(q)=-1$, $\text{Tr}(q)=0$,  $q_a^{~b}q_b^{~c}=\delta_a^{c}$ and $q_{ba}=q_{ab}$. 
This induces the following  transformation for the embedding-space position
\be \label{equ:PT-Pslash}
\slashed{P}_A^{~B}=\Lambda_{a,A}\Lambda^{a,B}\ \mapsto\ (\Gamma_{[20]} \cdot\slashed{P}\cdot\Gamma_{[20]}	)_A^{~B}\, ,
\ee 
Writing $\slashed{P}=P^M \Gamma_M$, this becomes
\begin{align}\label{eq-transform-PM-PT}
	P^M\, \mapsto\, (-P^0,P^1,-P^2,P^3,P^4)\,,
\end{align}
which is precisely as expected, because it reduces to (\ref{equ:PT-xmu}) after projecting to the Poincaré slice.

\vskip 4pt
Furthermore, the PT operator acts on the embedding-space polarization vector $W^M$ in the same way as on $P^M$, i.e.~just by flipping the signs of $W^0$ and $W^2$.\footnote{After projecting to the Poincaré slice, $W^M$ reads~\cite{Costa:2011mg}
	\be 
	(W^\mu,W^+,W^-)=(\epsilon^\mu,0,2x_\mu \epsilon^\mu)\,,
	\ee 
	where $\epsilon^\mu=(\epsilon^0,\epsilon^1,\epsilon^2)$ is the polarization vector in position space, and thus the PT transformation simply flips the signs of $\epsilon^0$ and $\epsilon^2$ as in \eqref{equ:PT-xmu}.}  Like in  \eqref{equ:PT-Pslash}, this transformation can also be written as  
\be \label{equ:PT-Wslash}
\slashed{W}\, \mapsto\, \Gamma_{[20]} \cdot\slashed{W}\cdot \Gamma_{[20]}\,.
\ee 
We also need to understand how this operator acts on the spinors in  \eqref{equ:polvector-W}:
\be 
\slashed{W}_A^{~B}=\Upsilon_A \Upsilon^{*,B}-\Upsilon_A^* \Upsilon^{B}\,,
\label{equ:polvector-W-2}
\ee 
where $\Upsilon_A=\zeta^a\Lambda_{a,A}$. Recalling that PT acts on $\Lambda_{a,A}$ as \eqref{equ:PT-Lambda}, we can take $\zeta^a\mapsto \zeta^b\hs q_b^{~a}$, so that the transformation of $\Upsilon_A$ does not depend on the choice of the slice encoded in $ q_b^{~a}$:
\be \label{equ:PT-Upsilon}
\Upsilon_A\, \mapsto\, \left(\Gamma_{[20]}\right)_A^{~B} \hs\Upsilon_B\,.
\ee 
To preserve the relation $\Upsilon_A=\slashed{P}_A^{~B}\Upsilon^*_B$, PT must act on the dual spinor $\Upsilon^*_B$ as
\be \label{equ:PT-Upsilonstar}
\Upsilon_A^*\, \mapsto\, \left(\Gamma_{[20]}\right)_A^{~B} \hs\Upsilon_B^*\,,
\ee 
ignoring possible gauge transformations $\Upsilon^*_B\sim \Upsilon^*_B+c^b\Lambda_{b,B}$. Recalling the definition (\ref{equ:polvector-W-2}), 
we notice that the transformation law implied by \eqref{equ:PT-Upsilon} and \eqref{equ:PT-Upsilonstar} differs from the expected law~\eqref{equ:PT-Wslash} by a minus sign. To fix this, we must add one minus sign per polarization vector $W^M$, which gives a factor $(-1)^{S}$ after assuming that $S$ is integer.
In this way, the embedding-space polarization vector $W^M$ transforms as it should under PT.

\vskip 4pt
We conclude that the PT operation corresponds to  the transformations \eqref{equ:PT-Lambda}, \eqref{equ:PT-Upsilon} and \eqref{equ:PT-Upsilonstar} for the spinors $\Lambda_{a,A}$, $\Upsilon_{A}$ and $\Upsilon^*_A$, respectively, together with multiplying the spin-$S$ field by a factor $(-1)^{S}$. If we apply this to a correlator with multiple fields, the sign factor will become $(-1)^{S_1+S_2+\cdots +S_n}$, where $S_i$ are the spins of each field.

\subsection{Penrose Transforms} 

Next, we consider the effect of PT on a Penrose transform.
Given the considerations of the previous subsection, 
the composition of parity and time reversal acts on a twistor $Z^A$ as
\be\label{equ:PT-ZA} 
Z^A=\pi^a \Lambda_a^A\ \mapsto\  \pi^a\hs q_a^{~b} \hs\Lambda_b^B \hs\left(-\Gamma_{[20]}\right)_B^{~A}=\tilde Z^B\hs \left(\Gamma_{[20]}\right)_B^{~A}\,,
\ee 
where we defined $\tilde Z^B\equiv \tilde\pi^b \Lambda_b^B$, with $\tilde \pi^b\equiv -\pi^a\hs q_a^{~b}$. We can use this to derive the transformation for the Penrose transform \eqref{equ:twistorintegral-product}:
\be 
J(\Lambda,\Upsilon^*)=i^{-S}\int DZ \hs (\Upsilon^*\cdot Z)^{2 S}\hs F(Z^A)\,.
 \label{equ:PT-penrose-prev}
\ee 
First, note that the product $\Upsilon^*\cdot Z \equiv -Z^A\Upsilon^*_A$ transforms as
\be 
\Upsilon^*\cdot Z \, \mapsto\, -(\tilde Z\cdot \Gamma_{[20]})\cdot (\Gamma_{[20]}\cdot \Upsilon^*)=-\tilde Z\cdot \Upsilon^*=\Upsilon^*\cdot\tilde Z\,.
\ee 
We can easily change variables from $DZ=D\pi$ to $D\tilde\pi=D \tilde Z$ in the Penrose transform, where the Jacobian is trivial since $|\det (-q)|=1$. Hence, PT acts on~\eqref{equ:PT-penrose-prev} as
\be \label{equ:PT-penrose-product}
J(\Lambda,\Upsilon^*) \  \mapsto\   i^{-S}\int DZ  \left(\Upsilon^*\cdot Z\right)^{2S}~(-1)^{S} \hs F\left( (Z\cdot \Gamma_{[20]})^{A}\right),
\ee
where we included the $(-1)^S$ factor of the transformation law, assumed that $S$ is integer, and renamed $\tilde Z$ as $Z$.

\vskip4pt
Similarly, we can determine the effect of PT 
on the Penrose transform \eqref{equ:alt} in dual twistor space:
\be 
\tilde J(\Lambda,\Upsilon^*)=i^S\hs\int DW \left(\Upsilon^*\cdot\frac{\partial }{\partial W}\right)^{2S}\hs \tilde F(W_A)\,.
\label{equ:alt-2}
\ee 
We first recall that the dual twistor transforms as
\be 
W_A=\pi^a\Lambda_{a,A} \ \mapsto\  \left(\Gamma_{[20]}\right)_A^{~B} \tilde W_B\,,
\ee 
where we again defined $\tilde W_B\equiv \tilde \pi^b \Lambda_{b,B}$, with $\tilde \pi^b\equiv -\pi^a \hs q_a^{~b}$. The differential operator $\Upsilon^*\cdot\partial_W$ then transforms as
\begin{align}
	\Upsilon^*\cdot \frac{\partial}{\partial W}=
	\Upsilon^*_A \frac{\partial}{\partial W_A}\ \mapsto\ &\Upsilon^*_B \left(\Gamma_{[20]}\right)_A^{~B}\frac{\partial}{\partial (\Gamma_{[20]}\cdot\tilde W)_A}=\Upsilon^*_B \frac{\partial(\Gamma_{[20]}\cdot\tilde W)_A}{\partial \tilde W_B}\frac{\partial}{\partial (\Gamma_{[20]}\cdot\tilde W)_A}\nonumber\\
	&=\Upsilon^*_B \frac{\partial}{\partial \tilde W_B}=\Upsilon^*\cdot \frac{\partial}{\partial \tilde W}\,.
\end{align}
Hence, PT acts on the dual Penrose transform \eqref{equ:alt-2} as
\be\label{equ:PT-penrose-der}
\tilde J(\Lambda,\Upsilon^*)\ \mapsto\ i^{S}\int DW  \left(\Upsilon^*\cdot \frac{\partial}{\partial W}\right)^{2S} ~(-1)^{S} \hs F\left( ( \Gamma_{[20]}\cdot W)_{A}\right),
\ee
where we performed the same steps as in \eqref{equ:PT-penrose-product}.

\subsection{Twistor Correlators} 

Finally, let us impose that the $n$-point correlator coming from a Penrose transform like \eqref{equ:npt-ansatz} is even under PT. Together with \eqref{equ:PT-penrose-product}, this implies that the corresponding twistor correlator must satisfy
\be \label{equ:PT-F-generic}
(-1)^{S_1+\cdots +S_n}\hs F\left( (Z_i\cdot\Gamma_{[20]})^{A}\right)=F(Z_i^A)\,.
\ee 
Notice first that the inner products of the twistors transform as
\be 
Z_i\cdot Z_j=Z_i^A\Omega_{AB}Z_j^B\ \mapsto\ 
-Z_i\cdot Z_j\,,
\ee 
after using that $\left(\Gamma_{[20]}\right)_C^{~A}\Omega_{AB} \left(\Gamma_{[20]}\right)_D^{~B}=-\Omega_{CD}$. 
Assuming that the twistor correlators depend only on inner products, the requirement \eqref{equ:PT-F-generic} then becomes
\be \label{equ:PT-F}
(-1)^{S_1+\cdots +S_n}\hs F( -Z_i\cdot Z_j)=F(Z_i\cdot Z_j)\,.
\ee 
Focusing on the case of three-point functions, it is easy to see that the power-law ansatz \eqref{equ:Ansatz1} does not satisfy this constraint, while the ansatz with delta functions in \eqref{equ:Ansatz2} does because $\delta^{[n]}(-x)=(-1)^{n}\delta^{[n]}(x)$. We thus conclude that the power-law and delta function ans\"atze are odd and even under PT, respectively. 
The same is true for a three-point ansatz $\tilde F(W_i\cdot W_j)$ in dual twistor space: It must be a product of delta functions rather than power laws in order to be invariant under the composition of parity and time reversal.

\newpage
\section{Computing the Twistor Integrals}
\label{app:computations}

In this appendix, we explicitly perform the twistor integrals introduced in Section~\ref{ssec:boot}, showing that this leads to the known results in embedding space.

\subsection{Preliminaries}
\label{ssec:prelim}

We will start by quoting---and proving---some preliminary results that will be very helpful in our computations of the twistor integrals of interest. 
 
\paragraph{Integral of delta function} The integral of the Dirac delta distribution $\delta(Z\cdot U)$ against a test function $f(Z^A)$ gives
\begin{align}\label{eq-delta-Zlanguage}
	\int DZ \, \delta(Z\cdot U)\hs f(Z^A)=f(Z^A)\Big|_{Z=\slashed{P} \cdot U}\,,
\end{align}
where the right-hand side is evaluated at  $Z^A=\slashed{P}^{AB} U_B$. 

\begin{framed}
{\small \noindent {\bf Proof}\ \ We start by parameterizing the twistor coordinate as $Z^A \equiv \pi^a \Lambda_a^A$. The left-hand side of \eqref{eq-delta-Zlanguage} then becomes
\begin{align}\label{equ:delta-middlestep}
	\int DZ \, \delta(Z\cdot U)\hs f(Z^A) = \int D\pi \ \delta(\pi^a\xi_a)\hs \tilde{f}(\pi^a)\,,
\end{align}
where we defined  $\xi_a\equiv \Lambda_a^AU_A$, such that $Z\cdot U=\pi^a\xi_a$, and $\tilde{f}(\pi^a)\equiv f(\pi^a\Lambda_a^A)$. It is further convenient to write $\pi^a$ as
\be \label{equ:pi-omega}
\pi^a \equiv r(u^a+\omega\hs v^a)\,,
\ee 
where the basis is normalized as $v^au_a \equiv 1$ and
the integral measure is $DZ=r^2\hs d\omega$. Since the projective integral will not depend on the overall scale, we may take $r=1$ and
 choose the basis $\{u^a,v^a\}$ such that $\xi_v\equiv v^a\xi_a\neq0$,. The integrand \eqref{equ:delta-middlestep} is then not singular in the ray $\pi^a=r\hs v^a$ missing in this parameterization.  
Defining $\xi_u\equiv u^a\xi_u$, we have $\xi^a=u^a \xi_v-\xi_u v^a$ and the integral becomes
\begin{align}
	\int d\omega \hs \delta(\xi_u+\omega \xi_v )\hs \tilde{f}(u^a+\omega v^a) &=\int d\tilde{\omega}\hs\frac{1}{|\xi_v|} \hs \delta(\xi_u+\tilde{\omega} ) \hs \tilde{f}(u^a+\tilde{\omega} v^a/\xi_v)\nonumber\\
	&=\tilde{f}(u^a \xi_v-\xi_u v^a) \equiv \tilde{f}(\xi^a)\,,
\end{align}
where, in the first line, we changed variables to $\tilde{\omega}=\omega \xi_v$, and, in the second line, we used that $\tilde{f}(\pi^a/r )=|r|\hs\tilde{f}(\pi^a)$. The result is equal to the right-hand side of \eqref{eq-delta-Zlanguage}:
\be 
\tilde{f}(\xi^a)=f(\xi^a \Lambda_a^A)=f( \Lambda_a^A \hs(\Lambda^a\cdot U))= f(\slashed{P}\cdot U)\,,
\ee 
which completes the proof.}
\end{framed}

\paragraph{Integral of derivatives of delta function}
The integral of the $n$-th derivative of the Dirac delta distribution $\delta^{[n]}(Z\cdot U)$, with $n \geq 0$, against a test function $f(Z^A)$ gives
\be\label{equ:distribution-der-delta}
	\int DZ \hs \delta^{[n]}(Z\cdot U)\hs f(Z^C) 
	=\frac{(-1)^n}{(V\cdot U)^n}\left(V\cdot \frac{\partial}{\partial Z}\right)^nf(Z^C)\bigg|_{Z=\slashed{P}\cdot U}\,,
\ee 
where $V^A$ is a reference spinor satisfying $V^A \slashed{P}_A^{~B}=0$ and $V\cdot U\neq0$, and the result is evaluated at $Z^C=\slashed{P}^{CD} U_D$ after applying the derivatives with respect to $Z^A$.

\begin{framed}
{\small \noindent {\bf Proof}\ \ 
To compute twistor integrals involving derivatives of delta functions like $\delta^{[n]}(Z\cdot U)$, we first need to understand how to integrate by parts. 
Consider an integral of the form
\be  
I_A \equiv
\int
 DZ \hs  \frac{\partial}{\partial Z^{A}}\hs h(Z^C)\,,
\label{equ:INT}
\ee
where $Z^A\slashed{P}_A^{~B}=0$. 
Contracting the free index with $\slashed{P}^{BA}$, we get 
\begin{align}
\slashed{P}^{BA} I_A =	\int  DZ \hs \slashed{P}^{BA}\hs \frac{\partial}{\partial Z^{A}}\hs h(Z^C)&=\int  (Z^{B}d Z^{A}-Z^{A}d Z^{B})\hs \frac{\partial}{\partial Z^{A}} h(Z^C)\nonumber\\
	&=\int  \left(Z^B d(h(Z))-dZ^B \left(Z^A\frac{\partial}{\partial Z^A}\right)h(Z^C)  \right) ,
	\end{align}
where we used the expression for the measure in Footnote~\ref{footnote-defmeasure}. Recalling that $h(r Z^C)=r^{-1}\hs h(Z^C)$ for the integral \eqref{equ:INT} to be projectively invariant, we can show that this vanishes
\begin{align}\label{equ:int-parts-totalder}
	\slashed{P}^{BA} I_A &=\int  \left(Z^B d(h(Z^C))+dZ^B h(Z^C)  \right)=\int  d(Z^B h(Z^C))=0\,,
\end{align}
because the integration region, isomorphic to $\mathbb{RP}^1$, has no boundary.\footnotemark~This has a simple corollary: Let $V^A$ be a spinor satisfying $V^A \slashed{P}_A^{~B}=0$. Writing $V^{A}=\slashed{P}^{AB}V^*_B$, for some dual spinor $V^*_B$, we thus get
\begin{align}
V^A I_A =	\int DZ \hs V^A\frac{\partial}{\partial Z^A}h(Z^C)=V^{*}_B\slashed{P}^{AB}\int DZ \hs \frac{\partial}{\partial Z^A}h(Z^C) = V^{*}_B\slashed{P}^{AB} I_A =0\,,
\end{align}
where we used \eqref{equ:int-parts-totalder} in the last step. This means that we can perform integrations by parts as
\begin{align}\label{equ:int-parts-Vderivative}
	\int DZ \hs f(Z^C)\hs V\cdot\frac{\partial}{\partial Z}h(Z^C)=-\int DZ\hs h(Z^C)\hs V\cdot\frac{\partial}{\partial Z}f(Z^C)\,.
\end{align}
Notice also that we can write a derivative of a function $h'(Z\cdot U)$ 
as
\be  
	h'(Z\cdot U)=\frac{1}{(V\cdot U)}V\cdot \frac{\partial}{\partial Z}h(Z\cdot U)\,.
\ee
where $V^A$ is a reference spinor. 
If we further assume that $V^A$ satisfies $V^A\slashed{P}_A^{~B}=0$, we can integrate $h'(Z\cdot U)$ by parts  as
\be
\int DZ \hs h'(Z\cdot U) f(Z^C) 
=-\int DZ \hs h(Z\cdot U)\frac{1}{(V\cdot U)}V\cdot \frac{\partial}{\partial Z}f(Z^C)\,.
\ee
This implies that we can easily perform integrals with derivatives of delta functions as
\be
\int DZ \hs \delta^{[n]}(Z\cdot U)\hs f(Z^C) 	 
=\frac{(-1)^n}{(V\cdot U)^n}\int DZ \hs\delta(Z\cdot U) \left(V\cdot \frac{\partial}{\partial Z}\right)^nf(Z^C)\, .
\ee 
Using \eqref{eq-delta-Zlanguage} to perform the last integral, we then obtain the right-hand side of \eqref{equ:distribution-der-delta}.
}
\end{framed}
\footnotetext{It can be proven explicitly that the boundary term vanishes by parameterizing $Z^A=\pi^a(\omega) \Lambda_a^A$, where we further expand $\pi^a(\omega)=u^a+\omega \hs v^a$ in a basis chosen such that $\tilde{h}(\pi^a)\equiv h(Z^A(\pi^a))$ isn't singular for $\pi^a=v^a$. This allows \eqref{equ:int-parts-totalder} to be written as a linear combination of two integrals $\int_{-\infty}^\infty d\omega\hs \partial_\omega \tilde{h}$ and $\int_{-\infty}^\infty d\omega\hs \partial_\omega (\omega\hs\tilde{h})$, with $\tilde{h}=\tilde{h}(u^a+\omega\hs v^a)$. It is then straightforward to show that both of them vanish due to the scaling of $\tilde{h}(r \pi^a)=r^{-1}\tilde{h}(\pi^a)$, implying that the boundary term \eqref{equ:int-parts-totalder} vanishes.}

\paragraph{An example} The following integral will later be useful
\be \label{eq-intDZ1-generic-n}
\int DZ_1\hs \delta^{[-n]}(Z_1\cdot Z_2)\delta^{[n]}(Z_3\cdot Z_1)=\left(\frac{V_1\cdot Z_2 }{V_1\cdot Z_3}\right)^n \delta(Z_3\cdot \slashed{P}_1\cdot Z_2)\,,
\ee 
where $V_1$ is a reference spinor satisfying $V_1^A\slashed{P}_{1,A}^{~~~B}=0$, and $n$ is an arbitrary integer. 

\begin{framed}
{\small \noindent {\bf Proof}\ \ 
To prove \eqref{eq-intDZ1-generic-n}, we consider separately the cases where $n$ is positive or negative: If $n\geq0$, we can perform the integral over $Z_1$ using the distribution $\delta^{[n]}(Z_3\cdot Z_1)$ as in~\eqref{equ:distribution-der-delta}, while, if $n<0$, we do the integral by using $\delta^{[-n]}(Z_1\cdot Z_2)$. In both cases, we end up with the right-hand side of  \eqref{eq-intDZ1-generic-n}, which does not actually depend on the choice of the reference spinor $V_{1,A}\in \rm{ker}(\slashed{P}_1)$. This can be proven by noticing that we are in the support of the delta function on the right-hand side, where
\be \label{equ:support-Z3P1Z2}
Z_3\cdot \slashed{P}_1\cdot Z_2=(Z_3\cdot \Lambda_{1,a})(\Lambda_1^a\cdot Z_2)=0\,,
\ee 
which immediately implies that the spinors in parenthesis are linearly dependent:
\be \label{eq-x123-def}
\Lambda_{1,a}\cdot Z_2=x_{1,23}\hs \Lambda_{1,a}\cdot Z_3\,,
\ee 
for some factor $x_{1,23}$. Contracting this equation with another spinor $-\Lambda_{1,A}^a$, we may write it as 
\be 
\slashed{P}_1\cdot Z_2=x_{1,23}\hs \slashed{P}_1\cdot Z_3 \,,
\ee
and hence the factor depends on $x_{1,23}(\slashed{P}_1,Z_2,Z_3)$. Furthermore, contracting \eqref{eq-x123-def} with a certain $v_1^a$, the equation only depends on $V_1\equiv v_1^a\Lambda_{1,a} $, and assuming $V_1\cdot Z_3\neq0$, we can isolate $x_{1,23}$ as
\be 
x_{1,23}=\frac{V_1\cdot Z_2}{V_1\cdot Z_3}\,,
\ee 
at the expense of introducing an arbitrary vector $V_{1,A}\in \rm{ker}(\slashed{P}_1)$. However, we see explicitly from the definition \eqref{eq-x123-def} that the factor $x_{1,23}$ doesn't depend on the choice of $V_1$ on the support of the delta function that sets $Z_3\cdot \slashed{P}_1\cdot Z_2=0$.}
\end{framed}

\paragraph{Whittaker integrals}

The {\it scalar Whittaker integral} is
\be \label{eq-whittaker-result}
\int DZ \hs  \delta(Z\cdot Y \cdot Z)
=\frac{\sqrt{2}}{\sqrt{\text{Tr}(\slashed{P} Y \slashed{P} Y)}}\,,
\ee 
where $Y_{AB}=Y_{BA}$ is a symmetric and real matrix, the argument of the delta function is $Z\cdot Y \cdot Z\equiv -Z^A Y_{AB} Z^B$, and we assume that $\text{Tr}(\slashed{P} Y \slashed{P} Y)>0$. 

\begin{framed}
{\small \noindent {\bf Proof}\ \ The integral  \eqref{eq-whittaker-result} can be performed by parameterizing the twistor as $Z^A(\pi)=\pi^a\Lambda_a^A$ and expanding $\pi^a=r(u^a+\omega \hs v^a)$ in a basis as in \eqref{equ:pi-omega}. The measure is $DZ=r^2d\omega$, and we can take $r=1$ because there is no dependence on the overall scale.
Let us define $y_{ab}\equiv \Lambda_a\cdot Y\cdot \Lambda_b$, such that $Z\cdot Y \cdot Z=\pi^ay_{ab}\pi^b$, which is also symmetric and real. We consider a basis $\{u^a,v^a\}$ that satisfies $y_{vv}\equiv v^ay_{ab}v^b\neq0$, so that the integrand is not singular in the ray $\pi^a=r\hs v^a$ that is missing in this parameterization $\pi^a(\omega)$.
Further defining $y_{uu}\equiv u^a y_{ab}u^b$ and $y_{uv}\equiv u^a y_{ab}v^b=v^a y_{ab}u^b$, the integral on the left-hand side of \eqref{eq-whittaker-result} reads 
\begin{align} \label{equ:whittaker-middlestep}
	\int d\omega \hs \delta\Big(\overbrace{y_{uu}+2y_{uv}\omega +y_{vv} \omega^2}^{\displaystyle\equiv q_2(\omega)}\Big) &=\int d\omega \hs\left(\frac{1}{|q_2'(\omega_+)|}\delta(\omega-\omega_+)+\frac{1}{|q_2'(\omega_-)|}\delta(\omega-\omega_-)\right)\nonumber\\
	&
	=\frac{2}{\sqrt{\Delta_q}}\,,
\end{align}
where we defined the quadratic polynomial in the argument of the delta function as $q_2(\omega)$, and its roots as $\omega_\pm$. Moreover, we used, in the second line, that $|q_2'(\omega_\pm)|=\sqrt{\Delta_q}$, where $\Delta_q$ is the discriminant of the quadratic polynomial in the argument of the delta function
\be \label{equ:discriminant-whittaker}
\Delta_q=4y_{uv}^2-4y_{uu}y_{vv}=-2y_{ab}y^{ab}=2\hs(\Lambda^a\cdot Y \cdot \Lambda_b) (\Lambda^b\cdot Y \cdot \Lambda_a)=2\hs\text{Tr}(\slashed{P} Y \slashed{P} Y)\,.
\ee  
Since we assumed that $\text{Tr}(\slashed{P} Y \slashed{P} Y)>0$, the discriminant $\Delta_q$ is positive  and thus the roots $\omega_\pm$ are real---otherwise the integral \eqref{equ:whittaker-middlestep} would vanish. Hence, we can substitute \eqref{equ:discriminant-whittaker} into \eqref{equ:whittaker-middlestep}, and therefore arrive to the right-hand side of \eqref{eq-whittaker-result}.
}
\end{framed}

The {\it spinning Whittaker integral} is
\begin{align}\label{eq-Whittaker-contractions-Zlanguage}
	&\int DZ \hs(\Upsilon^{*}_1\cdot Z)\cdots (\Upsilon^{*}_{2S}\cdot Z)\hs  \delta^{[S]}(Z\cdot Y \cdot Z)\nonumber\\
	&=\frac{\sqrt{2}}{(\text{Tr}(\slashed{P} Y \slashed{P} Y))^{(2S+1)/2}}\hs \left(\wick{\c1 \Upsilon_1\c1 \Upsilon_2}\cdots \wick{\c1 \Upsilon_{2S-1}\c1 \Upsilon_{2S}} +\rm{contractions}\right),
\end{align}
where $\Upsilon^{*,A}_i$ are elements of the dual vector space ker$(\slashed{P})^*$ and we defined the contraction
\be \label{equ:contraction-generic}
\wick{\c1 \Upsilon_i \c1 \Upsilon_j}\equiv 
\Upsilon_i^*\cdot \slashed{P}\cdot Y\cdot \slashed{P}\cdot \Upsilon_j^*\,.
\ee 
On the right-hand side of \eqref{eq-Whittaker-contractions-Zlanguage}, we sum over all $(2S-1)!!$ inequivalent contractions. As in~\eqref{eq-whittaker-result}, we assume that the real and symmetric matrix $Y_{AB}$ satisfies $\text{Tr}(\slashed{P} Y \slashed{P} Y)>0$.

\begin{framed}
	{\small \noindent {\bf Proof}\ \ Let $\Upsilon^*$ be an element of the dual vector space ker$(\slashed{P})^*$. Applying the operator
\be 
 -\Upsilon^{*}_A \Upsilon^{*,B} \displaystyle\frac{\partial}{\partial Y_A^{~B}}=-\Upsilon^{*}_A \Upsilon^{*}_B \displaystyle\frac{\partial}{\partial Y_{AB}}\,,
\ee 
to the integral \eqref{eq-whittaker-result}, we then get  
\be 
\int DZ \hs(\Upsilon^{*}\cdot Z)^{2}\hs  \delta'(Z\cdot  Y\cdot Z)=\sqrt{2}\hs\frac{( \Upsilon^{*}\cdot \slashed{P}\cdot Y\cdot \slashed{P}\cdot \Upsilon^*)}{(\text{Tr}(\slashed{P} Y \slashed{P} Y))^{3/2}}\,.
\ee 
Furthermore, if we apply this operator $S$ times, it is easy to show by induction that
\be \label{eq-basicintegral-spins}
\int DZ \hs(\Upsilon^{*}\cdot Z)^{2S}\hs  \delta^{[S]}(Z\cdot  Y\cdot Z)=\sqrt{2}\hs(2S-1)!! \frac{ (\Upsilon^{*}\cdot \slashed{P}\cdot Y\cdot \slashed{P}\cdot \Upsilon^*)^S}{(\text{Tr}(\slashed{P} Y \slashed{P} Y))^{(2S+1)/2}}\,,
\ee 
where $\delta^{[S]}(X)$ is the $S$-th derivative of the delta function and $(2S-1)!!$ is the double factorial (i.e.~the product of all odd natural numbers less or equal than $2S-1$). 
Defining $\tilde Y^{A_1A_2}$ as
\be 
\Upsilon^{*}\cdot \slashed{P}\cdot Y\cdot \slashed{P}\cdot \Upsilon^*=\Upsilon^{*}_{A_1}\Upsilon^{*}_{A_2} (-\slashed{P}\cdot Y\cdot \slashed{P})^{A_1A_2}\equiv \Upsilon_{*,A_1}\Upsilon_{*,A_2}\tilde{Y}^{A_1A_2}\,,
\ee  
we easily deduce the integral with explicit indices to be
\begin{align}\label{equ:spinningwhittaker-middlestep}
	&\int DZ \hs Z^{A_1}\cdots Z^{A_{2S}}\hs  \delta^{[S]}(Z\cdot  Y\cdot Z) =\frac{\sqrt{2}}{(\text{Tr}(\slashed{P} Y \slashed{P} Y))^{(2S+1)/2}}\hs(2S-1)!! \hs  \tilde{Y}^{(A_1A_2}\cdots \tilde{Y}^{A_{2S-1}A_{2S})}\nonumber\\
	&\hspace{4.94cm}=\frac{\sqrt{2}}{(\text{Tr}(\slashed{P} Y \slashed{P} Y))^{(2S+1)/2}}\hs \bigg(\tilde{Y}^{A_1A_2}\cdots \tilde{Y}^{A_{2S-1}A_{2S}}+\rm{perms}\bigg)\,.
\end{align}
In the last step, we noticed that, 
since $\tilde Y^{AB}=\tilde  Y^{BA}$, there are precisely $(2S-1)!!$ inequivalent permutations of the indices of the tensor $\tilde{Y}^{A_1A_2}\cdots \tilde{Y}^{A_{2S-1}A_{2S}}$, and we are summing over these inequivalent permutations. For example, for $S=2$, we have 
\begin{equation*}
	3!!\hs \tilde{Y}^{(A_1A_2}\tilde{Y}^{A_3A_4)}=\tilde{Y}^{A_1A_2}\tilde{Y}^{A_3A_4}+\tilde{Y}^{A_1A_3}\tilde{Y}^{A_2A_4}+\tilde{Y}^{A_1A_4}\tilde{Y}^{A_2A_3}\,.
\end{equation*}
This is the same combinatorial problem as in Wick's theorem, when we ask how many inequivalent contractions we have to consider to compute the $2S$-point function $\braket{\phi_1 \cdots \phi_{2S}}$ of a free theory. 
Thus, after contracting the free indices of \eqref{equ:spinningwhittaker-middlestep} with different spinors $\Upsilon^{*}_{i,A_i}$, we obtain precisely \eqref{eq-Whittaker-contractions-Zlanguage}, where the contractions
\be 
\wick{\c1 \Upsilon_i \c1 \Upsilon_j}\equiv \Upsilon^{*}_{i,A_i}\Upsilon^{*}_{j,A_j} \tilde{Y}^{A_iA_j}= \Upsilon_i^*\cdot \slashed{P}\cdot Y\cdot \slashed{P}\cdot \Upsilon_j^*\,,
\ee 
appear due to the inequivalent permutations of the indices on the right-hand side of \eqref{equ:spinningwhittaker-middlestep}. }
\end{framed}

\paragraph{Relation between Penrose transforms} 
The ``product-based" Penrose transform is related to the ``derivative-based" Penrose transform as
\be \label{equ:penroseoffourier}
i^S\int DW \left(\Upsilon^*\cdot \frac{\partial}{\partial W}\right)^{2S}\hs \tilde F(W_A)=i^{-S}\int DZ \left(\Upsilon^*\cdot Z\right)^{2S}\hs F(Z^A)\,,
\ee 
where $\tilde F(W_A)$ is the Fourier transform of $F(Z^A)$.

\begin{framed}
	{\small \noindent {\bf Proof}\ \ The Fourier transform of the function $F(Z^A)$ is
\be 
\tilde{F}(W_A)=\int \frac{d^4 Z}{(2\pi)^2} \hs  e^{i Z\cdot W}\hs F(Z^A)\,.
\ee 
	The left-hand side of \eqref{equ:penroseoffourier} can then be written as
		\begin{align}
			i^{S}\int DW \left(\Upsilon^*\cdot \frac{\partial}{\partial W}\right)^{2S}\hs \tilde F(W_A)&=i^S\int DW \int \frac{d^4 Z'}{(2\pi)^2} \hs\left(-i\Upsilon^*\cdot Z'\right)^{2S} \hs  e^{i Z'\cdot W}\hs F(Z^{\prime A})\nonumber\\
			&=i^{-S}\int \frac{D^3 Z}{(2\pi)^2} \hs\int dt\hs |t|^3  \left(t\Upsilon^*\cdot Z\right)^{2S}\hs F(t\hs Z^A)\int DW  e^{it\hs Z\cdot W}\,,
		\end{align}
		where, in the first line, we applied the $W$ derivatives to $e^{i Z'\cdot W}$, while, in the second line, we changed variables to $Z^{\prime A}=t\hs Z^A$, with $Z^A$ being a projective spinor. We further used that the measure in these new variables is $d^4Z'=D^3Z\wedge dt\hs |t|^3$, where $D^3Z$ is the natural projective measure in $\mathbb{RP}^3$ \cite{Arkani-Hamed:2009hub}. Recalling also that $F(t\hs Z^A)=t^{-2S-2} \hs F(Z^A)$, we obtain
		\be\label{equ:penrose-fourier-middlestep}
		i^{S}\int DW \left(\Upsilon^*\cdot \frac{\partial}{\partial W}\right)^{2S}\hs \tilde F(W_A)=i^{-S}\int D^3 Z \hs  \left(\Upsilon^*\cdot Z\right)^{2S}\hs F(Z^A)\left(\int\frac{DW}{(2\pi)^2}  \int dt\hs |t|\hs e^{i t\hs Z\cdot W}\right) .
		\ee
		To evaluate the integral in the parenthesis, we write
	$W_A=\pi^a \Lambda_{a,A}$ and  $DW=D\pi$. Redefining $\pi^{\prime a}=t\hs \pi^a$, so that $D\pi\wedge dt\hs |t|=d^2\pi'$, we then get 
		\be \label{equ:delta2-D3Z}
		\int \frac{DW}{(2\pi)^2} \int dt\hs |t|\hs e^{i t\hs Z\cdot W}=\int \frac{d^2\pi'}{(2\pi)^2}\hs e^{i \pi^{\prime a}\hs Z\cdot \Lambda_a}=\delta^{(2)}(Z\cdot \Lambda_a)\,.
		\ee 
		Consider now a basis $\{u^a,v^a\}$ normalized as $v^a u_a \equiv 1$.	Let us parameterize
		\be \label{equ:param-Zprime}
		Z^{\prime A}(r,\omega,\chi_u,\chi_v) \equiv r\hs Z^A(\omega,\chi_u,\chi_v)=r\left(\Lambda_u^A +\omega \Lambda_v^A+\chi_v \Lambda_u^{*,A}-\chi_u \Lambda_v^{*,A}\right),
		\ee 
		where $\Lambda_b^{*,A}$ is a spinor satisfying $\Lambda_b^{*,A}\slashed{P}_A^{~B}=\Lambda_b^B$. We also defined $\Lambda_u^A\equiv u^a \Lambda_a^A$, and similarly for $\Lambda_v^A$, $\Lambda_u^{*,A}$ and $\Lambda_v^{*,A}$. These four spinors form a basis in which we expand our twistor in \eqref{equ:param-Zprime}. Thus, the measure can be written as
		\be \label{equ:measure-D3Z}
		d^4 Z'=D^3Z\wedge dr\hs |r|^3= J\, d\omega\wedge d\chi_u\wedge d\chi_v \wedge dr\,,
		\ee 
		where the Jacobian is
		\be 
			J=\left|\det\left(\frac{\partial Z^{\prime A}}{\partial (r,\omega,\chi_u,\chi_v)}\right)\right|
			%=\left|\det\begin{pmatrix}\ldots  \Lambda_u^A\ldots\\
			%	\ldots  r\Lambda_v^A\ldots\\
			%	\ldots  (-r)\Lambda_v^{*,A}\ldots\\
			%	\ldots  r\Lambda_u^{*,A}\ldots\\\end{pmatrix}\right|
				=|r|^3\hs\left|\epsilon_{ABCD}\Lambda_u^A\Lambda_v^B\Lambda_v^{*,C}\Lambda_u^{*,D}\right|\,.
		\ee 		
		%Here, we used that determinant does not change when we add multiples of a row to another row. 
		%We also defined the 4d Levi-Civita symbol as the anti-symmetric tensor with $\epsilon_{1234}=1$.
		It is straightforward to check that the contraction in the previous equation can be written as 
		\be 
		\epsilon_{ABCD}\Lambda_u^A\Lambda_v^B\Lambda_v^{*,C}\Lambda_u^{*,D}=-(\Lambda_u\cdot\Lambda_v)( \Lambda_v^*\cdot \Lambda_u^{*})+(\Lambda_u\cdot\Lambda_v^{*})( \Lambda_v\cdot \Lambda_u^{*})-(\Lambda_u\cdot\Lambda_u^{*})( \Lambda_v\cdot \Lambda_v^{*})\,.
		\ee 
		Furthermore, notice that $\Lambda_b^{*,A}\slashed{P}_A^{~B}=\Lambda_b^B$ implies $\Lambda_b^*\cdot \Lambda_a=\epsilon_{ba}$ and thus $\Lambda_v^*\cdot \Lambda_u=-\Lambda_u^*\cdot \Lambda_v=1$ and $\Lambda_v^*\cdot \Lambda_v=\Lambda_u^*\cdot \Lambda_u=0$. Recalling also the constraint $\Lambda_a\cdot \Lambda_b=0$, we can write the Jacobian as
		\be 
		J=|r|^3\hs\left|\epsilon_{ABCD}\Lambda_u^A\Lambda_v^B\Lambda_v^{*,C}\Lambda_u^{*,D}\right|=|r|^3\hs\left|(\Lambda_u\cdot\Lambda_v^{*})( \Lambda_v\cdot \Lambda_u^{*})\right|=|r|^3\,.
		\ee  
		Plugging this into \eqref{equ:measure-D3Z} yields the projective measure written in this parameterization as
		\be \label{equ:measure-D3Z-simple}
		D^3Z=d\omega\wedge d\chi_u\wedge d\chi_v\,,
		\ee 	
		while the  delta function in \eqref{equ:delta2-D3Z} becomes
		\be 
		\delta^{(2)}(Z\cdot \Lambda_a)=\delta^{(2)}(\chi_v \Lambda_u^*\cdot \Lambda_a-\chi_u \Lambda_v^*\cdot \Lambda_a)=\delta^{(2)}(-\chi_v u_a+\chi_u v_a)=\delta(\chi_v)\delta(\chi_u)\,.
		\ee 
		The latter simply trivializes the integrals over $d\chi_u d\chi_v$ in \eqref{equ:measure-D3Z-simple}, leaving only the $d\omega$ integral as
		\be 
		D^3Z\hs \delta^{(2)}(Z\cdot \Lambda_a)=d\omega=D\pi=DZ\,,
		\ee 
		where we further recalled that this is precisely the measure $DZ=D\pi$ used in \eqref{equ:measure-omega} after parameterizing $Z^A=\pi^a\hs\Lambda_a^A$, with $\pi^a=r(u^a+\omega v^a)$, and modding out the scale $r$. Coming back to \eqref{equ:penrose-fourier-middlestep}, we can put everything together to obtain
		\begin{align}
			i^{S}\int DW \left(\Upsilon^*\cdot \frac{\partial}{\partial W}\right)^{2S}\hs \tilde F(W_A)=i^{-S}\int DZ\left(\Upsilon^*\cdot Z\right)^{2S}\hs F(Z^A)\,,
		\end{align}
		which concludes the proof.}
\end{framed}

\subsection{Twistor Integrals}

We will now compute the twistor integrals \eqref{equ:npt-ansatz2} and \eqref{equ:npt-ansatz} at three points for any integer or half-integer spins $S_i$, whose sum $S_T\equiv S_1+S_2+S_3$ is an integer. Choosing the integrands to be given by \eqref{equ:integrandFt} and \eqref{equ:integrandF}, will give the correlators for the leading and higher-derivative interactions in the bulk, respectively.

\paragraph{Leading interactions} To simplify the computation of the integral \eqref{equ:npt-ansatz2}, for $n=3$, 
we will consider the twistor integral of the Fourier transform $\tilde F_{23}(W_1,Z_2,Z_3)$ of $\tilde F(W_i)$ in \eqref{equ:integrandFt} with respect to $W_2$ and $W_3$. As proved in the box below \eqref{equ:penroseoffourier}, 
this coincides with the integral \eqref{equ:npt-ansatz2} of interest:
\begin{align} \label{equ:WZZ-representation}
\langle \tilde J_1\tilde J_2\tilde J_3\rangle&=i^{S_T}\left[ \prod_{j=1}^3 \int DW_i \left(\Upsilon_j^*\cdot \frac{\partial}{\partial W_j}\right)^{2S_j} \right]
\tilde F(W_{i,A})\nonumber\\
&=i^{-n_1}\int DW_1 DZ_2DZ_3\left(\Upsilon_1^*\cdot \frac{\partial}{\partial W_1}\right)^{2S_1}\left(\Upsilon_2^*\cdot Z_2\right)^{2S_2}\left(\Upsilon_3^*\cdot Z_3\right)^{2S_3}\tilde F_{23}\, . 
\end{align}
The twistor integral in the second line is simpler than that in the first line because is has less derivatives. 
It is straightforward to show that the function $\tilde{F}_{23}(W_1,Z_2,Z_3)$ is given by
\begin{align}
\tilde F_{23}(W_1,Z_2,Z_3)&=\int \frac{d^4 W_2}{(2\pi)^2}\hs  e^{-i Z_2\cdot W_2} \int \frac{d^4 W_3}{(2\pi)^2}\hs  e^{-i Z_3\cdot W_3}\hs \tilde F(W_{i})\nonumber\\
&= \int d^4 W_2 \hs e^{-i Z_2\cdot W_2}\hs \int \frac{d^3c_{ij}}{(2\pi)^3}\hs e^{-i c_{12}W_1\cdot W_2}\hs \delta^{(4)}(Z_3^A+c_{23}W_{2}^A-c_{31}W_{1}^A)\hs \tilde A(c_{ij})\nonumber\\
&= \int \frac{d^3c_{ij}'}{(2\pi)^3} \exp\left(-ic_{12}' Z_2\cdot W_1-i c_{23}' Z_3\cdot Z_2-i c_{31}' Z_3\cdot W_1\right)A'(c_{ij}')\,,
\end{align}
where in the last line we changed variables to $c_{23}' \equiv 1/c_{23}$, $c_{12}' \equiv c_{31}/c_{23}$ and $c_{31}' \equiv c_{12}/c_{23}$, and defined
\be 
A'(c_{ij}')\equiv \tilde A(c_{ij}(c_{ij}'))= \frac{c_{23}^{\prime\, S_T}}{c_{12}^{\prime\, n_2}c_{31}^{\prime\, n_3}}\,.
\ee 
Plugging this into the second line of \eqref{equ:WZZ-representation}, we get
\begin{align}\label{equ:WZZ-deltas}
\langle \tilde J_1\tilde J_2\tilde J_3\rangle&= \int DW_1 DZ_2DZ_3\left(\Upsilon_1^*\cdot \frac{\partial}{\partial W_1}\right)^{2S_1}\left(\Upsilon_2^*\cdot Z_2\right)^{2S_2}\left(\Upsilon_3^*\cdot Z_3\right)^{2S_3}\nonumber\\
&\quad\quad\times \delta^{[-n_2]}(Z_2\cdot W_1) \hs\delta^{[S_T]}(Z_3\cdot Z_2)\hs\delta^{[-n_3]}(Z_3\cdot W_1)\,, 
\end{align} 
and using the generalized Leibniz rule 
to apply the $2S_1$ derivatives to the integrand, we find
\begin{align}\label{eq-tripleintegral-++step2}
	\langle \tilde J_1\tilde J_2\tilde J_3\rangle &= \int DZ_3\hs  \left(\Upsilon_3^*\cdot Z_3\right)^{2S_3}  \int DZ_2\hs \left(\Upsilon_2^*\cdot Z_2\right)^{2S_2}\hs \sum_{k=0}^{2S_1}\binom{2S_1}{k}(-\Upsilon^*_1\cdot Z_2)^{2S_1-k}(-\Upsilon^*_1\cdot Z_3)^k\nonumber\\
	&\quad\quad\times\delta^{[S_T]}(Z_3\cdot Z_2)\int DW_1\hs
	\delta^{[n_3-k]}(Z_2\cdot W_1) \hs \delta^{[-n_3+k]}(Z_3\cdot W_1)\,.
\end{align}
To compute the integral over $DW_1$ in the last line, we use the identity \eqref{eq-intDZ1-generic-n}. This gives
\begin{align}
	\langle \tilde J_1\tilde J_2\tilde J_3\rangle &= \int DZ_3\hs  \left(\Upsilon_3^*\cdot Z_3\right)^{2S_3}  \int DZ_2\hs \left(\Upsilon_2^*\cdot Z_2\right)^{2S_2}\hs \delta^{[S_T]}(Z_3\cdot Z_2)\nonumber\\
	&\quad\times\sum_{k=0}^{2S_1}\binom{2S_1}{k}(-\Upsilon^*_1\cdot Z_2)^{2S_1-k}(-\Upsilon^*_1\cdot Z_3)^k\left(\frac{-V_1\cdot Z_2 }{V_1\cdot Z_3}\right)^{-n_3+k} \delta(Z_3\cdot \slashed{P}_1\cdot Z_2)\,,
\end{align}
where we introduced a reference spinor $V_1$, such that $\slashed{P}_1\cdot V_1=0$. 
With the help of the identity~\eqref{eq-delta-Zlanguage}, we then perform the integral over $Z_2$:
\begin{align}
	\langle \tilde J_1\tilde J_2\tilde J_3\rangle &=  (2\hs P_{12})^{-n_3}\int DZ_3\hs  \left(\Upsilon_3^*\cdot Z_3\right)^{2S_3}   \left(\Upsilon_2^*\cdot\slashed{P}_2\cdot\slashed{P}_1\cdot Z_3\right)^{2S_2}\hs \delta^{[S_T]}( Z_3\cdot\slashed{P}_2\cdot\slashed{P}_1\cdot Z_3)\nonumber\\
	&\quad\quad\times\sum_{k=0}^{2S_1}\binom{2S_1}{k}(-\Upsilon^*_1\cdot\slashed{P}_2\cdot\slashed{P}_1\cdot Z_3)^{2S_1-k}(-2\hs P_{12}\hs\Upsilon^*_1\cdot Z_3)^k\,,
	\label{equ:LastLine}
\end{align}
where the factors of $2\hs P_{12}$ come from 
\be 
-\frac{V_1\cdot Z_2}{V_1\cdot Z_3}\bigg|_{Z_2=\slashed{P}_2\cdot\slashed{P}_1\cdot Z_3}=-\frac{V_1\cdot \slashed{P}_2\cdot \slashed{P}_1\cdot Z_3}{V_1\cdot Z_3}=-2P_1\cdot P_2=2\hs P_{12}\,.
\ee 
As expected, the result does not depend on the choice of $V_1$. From the anti-commutation relations between the $\Gamma^M$ matrices, we easily deduce the anti-commutator  $\{\slashed{P}_1,\slashed{P}_2\}=-2P_{12}$. Hence, writing the last line in (\ref{equ:LastLine}) as 
\begin{align}
	\sum_{k=0}^{2S_1}\binom{2S_1}{k}(-\Upsilon^*_1\cdot\slashed{P}_2\cdot\slashed{P}_1\cdot Z_3)^{2S_1-k}(-2\hs P_{12}\hs\Upsilon^*_1\cdot Z_3)^k
	&=\left(\Upsilon^*_1\cdot(-\slashed{P}_2\cdot\slashed{P}_1+\{\slashed{P}_1,\slashed{P}_2\})\cdot Z_3\right)^{2S_1}\nonumber\\
	&=\left(\Upsilon^*_1\cdot\slashed{P}_1\cdot\slashed{P}_2\cdot Z_3\right)^{2S_1}\,,
\end{align}
and recalling from \eqref{equ:dual} that $\Upsilon_i=\slashed{P}_i\cdot \Upsilon_i^*$, we get
\begin{align}
	\langle \tilde J_1\tilde J_2\tilde J_3\rangle &=  (2\hs P_{12})^{-n_3}(-1)^{S_T}\int DZ_3\hs  \left(\Upsilon_3^*\cdot Z_3\right)^{2S_3}   \left(\Upsilon_2\cdot\slashed{P}_1\cdot Z_3\right)^{2S_2}\hs\left(\Upsilon_1\cdot\slashed{P}_2\cdot Z_3\right)^{2S_1}\nonumber\\
	&\quad\quad\times\delta^{[S_T]}( Z_3\cdot\slashed{P}_1\cdot\slashed{P}_2\cdot Z_3)\,,
\end{align}
where the sign factor $(-1)^{S_T}$ appeared because we changed the sign of the argument of the delta function, using  \eqref{eq-scaling-diracdelta-n}. Notice that we have kept the dual spinors $\Upsilon_i^*$ generic (without fixing its gauge redundancy) and now we can see explicitly that $\langle \tilde J_1\tilde J_2\tilde J_3\rangle$ is invariant under a gauge transformation $\Upsilon^*_i\mapsto \Upsilon^*_i+U_i$, for any $U_i\in\rm{ker}(\slashed{P}_i)$, as both $\Upsilon_i=\slashed{P}_i\cdot \Upsilon_i^*$ and $\Upsilon_i^*\cdot Z_i$, with $Z_i\in\rm{ker}(\slashed{P}_i)$, are invariant under this transformation. 

\vskip4pt
 Finally, we use \eqref{eq-Whittaker-contractions-Zlanguage} to
 perform the last integral over $Z_3$, with $Y_{AB}=(\slashed{P}_1\slashed{P}_2)_{(AB)}$ being the symmetric part of the matrix $(\slashed{P}_1\slashed{P}_2)_{AB}$, or equivalently $Y_{AB}=[\slashed{P}_1,\slashed{P}_2]_{AB}/2$. This yields
\begin{align}\label{equ:sumovercontractions}
	\langle \tilde J_1\tilde J_2\tilde J_3\rangle &=\frac{ \sqrt{2}\hs (2\hs P_{12})^{-n_3} (-1)^{S_T}}{(2 A)^{(2S_T+1)/2}}  \left(\sum \text{contractions}\right) ,
\end{align}
where we defined
\begin{align}\label{eq-def-2A}
	2A\equiv\text{Tr}(\slashed{P}_3 Y\slashed{P}_3 Y) &=\text{Tr}(\slashed{P}_3 \slashed{P}_1\slashed{P}_2\slashed{P}_3 \slashed{P}_1\slashed{P}_2)
	=16\hs P_{12}\hs P_{23}\hs P_{31}\,.
\end{align}
Since we are restricting to  totally spacelike point configurations with $P_{ij}>0$ (recall Footnote~\ref{footnote:analytic-continuation}), the quantity $A$ is positive, and thus we are able to compute the integral over $Z_3$ following \eqref{eq-Whittaker-contractions-Zlanguage}. In the formula \eqref{equ:sumovercontractions}, we need to sum over all the $(2S_T-1)!!$ contractions between the $2S_i$ factors of $\Upsilon_i$, for each $i=1,2,3$. The most general contraction is
\be \label{eq-contractions-mi}
{\cal C}_{m_i} \equiv \left(\wick{\c1 \Upsilon_1\c1 \Upsilon_1}\right)^{m_1}\left(\wick{\c1 \Upsilon_2\c1\Upsilon_2}\right)^{m_2} \left(\wick{\c1 \Upsilon_3\c1 \Upsilon_3}\right)^{m_3} \left(\wick{\c1 \Upsilon_1\c1 \Upsilon_2}\right)^{r_{12}}\left(\wick{\c1 \Upsilon_2\c1\Upsilon_3}\right)^{r_{23}} \left(\wick{\c1 \Upsilon_3\c1 \Upsilon_1}\right)^{r_{31}}, 
\ee 
where $0\leq m_i\leq \floor{S_i}$ and the exponents $r_{ij}$ are given by
\be 
r_{ij}\equiv S_i-m_i+S_j-m_j-(S_k-m_k)\,,
\ee 
for $\{i,j,k\}$ a cyclic permutation of $\{1,2,3\}$, so that there are always $2S_i$ factors of $\Upsilon_i$ in \eqref{eq-contractions-mi}. We further require that $r_{ij}\geq 0$, imposing stricter bounds on the possible values of $m_i$. It is then a straightforward combinatorial exercise to determine the number $c_{m_i}$ of contractions of the form of  \eqref{eq-contractions-mi}: 
\begin{align}
	c_{m_i} 
	&=\frac{(2S_1)!\hs (2S_2)!\hs (2S_3)! }{2^{m_1+m_2+m_3}\hs m_1!\hs m_2!\hs m_3!\hs r_{12}!\hs r_{23}!\hs r_{31}!}\,.
\end{align}
Using this, the precise formula for the twistor integral \eqref{equ:WZZ-representation} is
\begin{align}\label{equ:formula-JJJtilde}
	\langle \tilde J_1\tilde J_2\tilde J_3\rangle &=\frac{ \sqrt{2}\hs (2\hs P_{12})^{-n_3}(-1)^{S_T} }{(2A)^{S_T+1/2}}  \sum_{m_1=0}^{\floor{S_1}}\sum_{m_2=0}^{\floor{S_2}}\sum_{m_3=0}^{\floor{S_3}}c_{m_i}\hs {\cal C}_{m_i}\,,
\end{align}
where configurations with $r_{ij}<0$ don't contribute, because $c_{m_i}\propto 1/r_{ij}!=0$.
	Using $\slashed{P}_3\cdot \Upsilon^{*}_3=\Upsilon_3$ and the definition \eqref{equ:contraction-generic}, we can compute all contractions appearing in \eqref{eq-contractions-mi}:
\be\label{eq-contractions-++}
\begin{split}
	\wick{\c1 \Upsilon_1\c1 \Upsilon_1} &=(\Upsilon_1\cdot \slashed{P}_2)\cdot \slashed{P}_3\cdot \slashed{P}_1\cdot\slashed{P}_2\cdot\slashed{P}_3\cdot (\slashed{P}_2\cdot \Upsilon_1)=16 P_{23}^2\hs P_{12} \hs V_1\,,\\
	\wick{\c1 \Upsilon_2\c1 \Upsilon_2} &=\left(\Upsilon_2\cdot \slashed{P}_1\right)\cdot \slashed{P}_3\cdot \slashed{P}_1\cdot\slashed{P}_2\cdot\slashed{P}_3\cdot \left(\slashed{P}_1\cdot \Upsilon_2\right)=16\hs P_{31}^2\hs P_{12} \hs V_2
	\,,\\
	\wick{\c1 \Upsilon_3\c1 \Upsilon_3} &=\Upsilon_3\cdot \slashed{P}_1\cdot\slashed{P}_2\cdot\Upsilon_3=4\hs P_{12}\hs V_3
	\,,\\
	\wick{\c1 \Upsilon_1\c1 \Upsilon_2} &=\wick{\c1 \Upsilon_2\c1 \Upsilon_1}=\left(\Upsilon_2\cdot \slashed{P}_1\right)\cdot \slashed{P}_3\cdot \slashed{P}_1\cdot\slashed{P}_2\cdot\slashed{P}_3\cdot \left(\slashed{P}_2\cdot \Upsilon_1\right)=A\hs R_{12}\,,\\
	\wick{\c1 \Upsilon_2\c1 \Upsilon_3} &=(\Upsilon_2\cdot \slashed{P}_1)\cdot \slashed{P}_3\cdot \slashed{P}_1\cdot\slashed{P}_2\cdot\Upsilon_3= 4\hs P_{12}\hs  P_{31}\hs R_{23}\,,\\
	\wick{\c1 \Upsilon_3\c1 \Upsilon_1} &=\Upsilon_3\cdot\slashed{P}_1\cdot\slashed{P}_2\cdot\slashed{P}_3\cdot( \slashed{P}_2\cdot \Upsilon_1)=4\hs P_{23}\hs  P_{12} \hs R_{31}\,,
\end{split}
\ee
where we defined $R_{ij}\equiv\Upsilon_i\cdot \Upsilon_j$
 and recalled \eqref{equ:structure-V} to write the result in terms of $R_{ij}$ and $V_i$. Notice that for $3$d CFTs these structures are related by~\cite{Giombi:2011rz} 
\be \label{eq-structures-relation-RRR}
R_{12}R_{23}R_{31}=V_{1} R_{23}^2+V_{2} R_{31}^2+V_{3} R_{12}^2-4 V_1 V_2V_3\,.
\ee 
Hence, we can compute the final result of this twistor integral by writing each individual contraction in \eqref{equ:formula-JJJtilde} in terms of the structures $R_{ij}$ and $V_i$ in \eqref{eq-contractions-++}, and then using \eqref{eq-structures-relation-RRR} to replace $R_{12}R_{23}R_{31}$ whenever it appears. This procedure yields the final expression for our twistor integral without any redundancies, which can easily be evaluated for specific spins as we do in Section~\ref{ssec:boot}.

\paragraph{Higher-derivative interactions}

Next, we compute  the twistor integral \eqref{equ:npt-ansatz}, for $n=3$:
\be\label{eq-tripleintegral+++}
\langle J_1 J_2 J_3\rangle= \left[\prod_{j=1}^3\int  DZ_j \hs \left(\Upsilon_j^*\cdot Z_j\right)^{2S_j}\right] \delta^{[n_3]}(Z_1\cdot Z_2) \hs \delta^{[n_1]}(Z_2\cdot Z_3)\hs \delta^{[n_2]}(Z_3\cdot Z_1)\,,
\ee
with the integrand given by \eqref{equ:integrandF}. Let us assume, without loss of generality, that $S_1\leq S_2\leq S_3$, which implies $n_{2}\geq0$. Hence, we can integrate over $DZ_1$ by using the distribution $\delta^{[n_2]}(Z_3\cdot Z_1)$ as in  \eqref{equ:distribution-der-delta}: 
\begin{align}
	\langle J_1 J_2 J_3\rangle &=  \int DZ_2\hs  \left(\Upsilon_2^*\cdot Z_2\right)^{2S_2}\int DZ_3\hs  \left(\Upsilon_3^*\cdot Z_3\right)^{2S_3}\delta^{[n_1]}(Z_2\cdot Z_3)\hs \frac{1}{(V_1\cdot Z_3)^{n_2}} \nonumber\\
	&   \quad \quad\times\left(V_1\cdot\frac{\partial}{\partial Z_1}\right)^{n_2}  \left(\left(\Upsilon_1^*\cdot Z_1\right)^{2S_1}\delta^{[n_3]}(Z_1\cdot Z_2)\right)\Big|_{Z_1=\slashed{P}_1\cdot Z_3}\,,
\end{align}
where $V_{1,A}$ is an arbitrary spinor of ker$(\slashed{P}_1)$. 
For convenience, we choose $V_1=\Upsilon_1$, so that the operator $V_1\cdot \partial_{Z_1}=\Upsilon_1\cdot \partial_{Z_1}$ does not hit the contraction $\Upsilon_1^*\cdot Z_1$, simply because $\Upsilon_1^*\cdot \Upsilon_1=0$. This derivative therefore acts only on $\delta^{[n_3]}(Z_1\cdot Z_2)$, yielding
\begin{align}
	\langle J_1 J_2 J_3\rangle &=  \int DZ_2\hs  \left(\Upsilon_2^*\cdot Z_2\right)^{2S_2}\int DZ_3\hs  \left(\Upsilon_3^*\cdot Z_3\right)^{2S_3}\delta^{[n_1]}(Z_2\cdot Z_3)\hs (\Upsilon_1\cdot Z_3)^{n_3} \nonumber\\
	&  \quad \quad\times\left(\Upsilon_1\cdot Z_2\right)^{n_2}  \delta^{[2S_1]}(Z_3\cdot \slashed{P}_1\cdot Z_2)\,,
\end{align}
where we used that $\Upsilon_1=\slashed{P}_1\cdot \Upsilon_1^*$. We can subsequently integrate over $DZ_2$ by using $\delta^{[2S_1]}(Z_3\cdot \slashed{P}_1\cdot Z_2)$ as in \eqref{equ:distribution-der-delta}. Choosing $V_2=\Upsilon_2$ as the arbitrary spinor, so that the derivative operator $\Upsilon_2\cdot \partial_{Z_2}$ does not act on the contraction $\Upsilon_2^*\cdot Z_2$, we get
\begin{align}
	\langle J_1 J_2 J_3\rangle 
	&=  \int  DZ_3\hs  \left(\Upsilon_3^*\cdot Z_3\right)^{2S_3}\hs (\Upsilon_1\cdot Z_3)^{n_3} \left(\Upsilon_2\cdot\slashed{P}_1\cdot Z_3\right)^{2(S_2-S_1)}\nonumber\\
	&\quad \quad\times\left(\Upsilon_2\cdot \frac{\partial}{\partial Z_2}\right)^{2S_1}\left(  \left(\Upsilon_1\cdot Z_2\right)^{n_2}  \delta^{[n_1]}(Z_2\cdot Z_3)\right)\bigg|_{Z_2=\slashed{P}_2\cdot \slashed{P}_1\cdot Z_3} \, ,
\end{align}
where we recalled that $\Upsilon_2=\slashed{P}_2\cdot \Upsilon_2^*$ to simplify the first line. 
The derivative in the second line can then be computed with the generalized Leibniz rule as
\begin{align}
	\langle J_1 J_2 J_3\rangle &= \sum_{k=0}^{\min\{ 2S_1,n_2\}}\binom{2S_1}{k}\frac{(n_2)!}{(n_2-k)!}(\Upsilon_1\cdot \Upsilon_2)^k  \hs (-2 \hs P_{12})^{n_2-k}\nonumber\\
	& \quad \quad\times\int DZ_3\hs\left(\Upsilon_1\cdot Z_3\right)^{2S_1-k}  (\Upsilon_2\cdot Z_3)^{2S_1-k}  \left(\Upsilon_3^*\cdot Z_3\right)^{2S_3}\hs  \left(\Upsilon_2\cdot \slashed{P}_1\cdot Z_3\right)^{2(S_2-S_1)}\nonumber\\
	&\quad \quad\times\delta^{[S_T-k]}(Z_3\cdot \slashed{P}_1\cdot \slashed{P}_2\cdot Z_3)\,,
\end{align}
where we used that $\Upsilon_1\cdot Z_2=-2 \hs P_{12}\hs \Upsilon_1\cdot Z_3$ after evaluating in $Z_2=\slashed{P}_2\cdot \slashed{P}_1\cdot Z_3$, and reordered the terms for convenience. Finally, we perform the last integral over $Z_3$ using  \eqref{eq-Whittaker-contractions-Zlanguage}, which yields
\begin{align}
	\langle J_1 J_2 J_3\rangle &= \sum_{k=0}^{\min\{ 2S_1,n_2\}}\binom{2S_1}{k}\frac{(n_2)!}{(n_2-k)!}(\Upsilon_1\cdot \Upsilon_2)^k  \hs \frac{\sqrt{2} \hs (-2 \hs P_{12})^{n_2-k} }{(2A)^{S_T-k+1/2}}  \left(\sum \text{contractions} \right) .
\end{align}
Let us recall that $Y=[\slashed{P}_1,\slashed{P}_2]/2$ is the symmetric part of $\slashed{P}_1\slashed{P}_2$, and that $A$, defined in \eqref{eq-def-2A}, is positive because we only consider totally spacelike point configurations with $P_{ij}>0$.

\vskip 4pt
In the previous formula, we need to sum over all contractions between the factors of $\Upsilon_{1}$, $\Upsilon_2$, $\Upsilon_3$ and $\Upsilon_2'\equiv \slashed{P}_1\cdot\Upsilon_2$. Since
\be 
\wick{\c2\Upsilon_2'\c2\Upsilon_2} =\Upsilon_2\cdot \slashed{P}_1\cdot  \slashed{P}_3\cdot \slashed{P}_1\cdot\slashed{P}_2\cdot\slashed{P}_3\cdot\Upsilon_2=4\hs P_{12}\hs P_{31}\hs  (\Upsilon_2\cdot\slashed{P}_3 \cdot\Upsilon_2)=0\,,
\ee
the most general (non-vanishing) contraction is
\begin{align} \label{eq-contractions-mi-+++}
 {\cal C}_{m_i,m_2',r_{12}} \,\equiv\,	&\left(\wick{\c1 \Upsilon_1\c1 \Upsilon_1}\right)^{m_1}\left(\wick{\c1 \Upsilon_2\c1\Upsilon_2}\right)^{m_2} \left(\wick{\c1 \Upsilon_3\c1 \Upsilon_3}\right)^{m_3}\left(\wick{\c1 \Upsilon_2'\c1\Upsilon_2'}\right)^{m_2'}  \left(\wick{\c1 \Upsilon_1\c1 \Upsilon_2}\right)^{r_{12}}\nonumber\\
	&  \times\left(\wick{\c1 \Upsilon_2\c1\Upsilon_3}\right)^{r_{23}} \left(\wick{\c1 \Upsilon_3\c1 \Upsilon_1}\right)^{r_{31}} \left(\wick{\c1 \Upsilon_2'\c1 \Upsilon_1}\right)^{r_{12}'}\left(\wick{\c1 \Upsilon_2'\c1\Upsilon_3}\right)^{r_{23}'}\,.  
\end{align} 
Notice that the exponents of the first line satisfy $0\leq m_1,\hs m_2\leq \floor{S_1-k/2}$, $0\leq m_3\leq \floor{S_3}$, $0\leq m_2'\leq \floor{S_2-S_1}$, and $0\leq r_{12}\leq r_{12}^{\rm max} \equiv \min\{2S_1-k-2m_1,2S_1-k-2m_2\}$, while the exponents of the second line are given by
\begin{align} 
	r_{23}&\equiv 2S_1-k-2m_2-r_{12}\,,\\
	r_{31}&\equiv n_2-m_3+m_2-m_1+m_2' \,,\\
	r_{23}'&\equiv n_1-m_2'-m_3-(2S_1-k-m_1-m_2)+r_{12}\,,\\ 
	r_{12}'&\equiv n_3-m_2'+m_3-k-m_1-m_2-r_{12}\,,
\end{align}
so that there are always $2S_1-k$ factors of $\Upsilon_1$ and $\Upsilon_2$, $2S_3$ factors of $\Upsilon_3$, and $2(S_2-S_1)$ factors of $\Upsilon_2'$ in \eqref{eq-contractions-mi-+++}. We further require that the exponents in the second line of \eqref{eq-contractions-mi-+++} are non-negative, imposing stricter bounds on the allowed values of $m_1,m_2,m_3,m_2',r_{12}$. 
 It is a straightforward combinatorial exercise to determine the number of contractions of the form of \eqref{eq-contractions-mi-+++}:
\be
	c_{m_i,m_2',r_{12}}=\frac{((2S_1-k)!)^2\hs (2S_3)!\hs (2(S_2-S_1))! }{2^{m_1+m_2+m_3+m_2'}\hs m_1!\hs m_2!\hs m_3!\hs m_2'!\hs  r_{12}!\hs r_{23}!\hs r_{31}!\hs r_{12}'!\hs r_{23}'!}\,.
\ee
Hence, we can write down a precise formula for the result of this twistor integral:
\begin{align}\label{equ:formula-JJJ}
	\langle J_1 J_2 J_3\rangle &= \sum_{k=0}^{\min\{ 2S_1,n_2\}}\binom{2S_1}{k}\frac{(n_2)!}{(n_2-k)!}(\Upsilon_1\cdot \Upsilon_2)^k  \hs \frac{(- P_{12})^{n_2-k}}{2^{2S_2}\hs A^{S_T-k+1/2}}\nonumber\\
	&\quad \quad \times \sum_{m_1=0}^{\floor{S_1-k/2}}\sum_{m_2=0}^{\floor{S_1-k/2}}\sum_{m_3=0}^{\floor{S_3}}\sum_{m_2'=0}^{\floor{S_2-S_1}}\sum_{r_{12}=0}^{ r_{12}^{\rm max}}c_{m_i,m_2',r_{12}} \, {\cal C}_{m_i,m_2',r_{12}}\,,
\end{align}
where the configurations for which one exponent in the second line of \eqref{eq-contractions-mi-+++} is a negative integer don't contribute, because $c_{m_i,m_2',r_{12}}=0$.

\vskip 4pt
Using $\slashed{P}_3\cdot \Upsilon^{*}_3=\Upsilon_3$, we can compute each individual contraction in \eqref{eq-contractions-mi-+++} as
\beq \label{eq-contractions+++}
\begin{split}
	\wick{\c1 \Upsilon_3\c1 \Upsilon_3} &=\Upsilon_3\cdot \slashed{P}_1\cdot\slashed{P}_2\cdot\Upsilon_3=4\hs P_{12}\hs V_3\,,
	\\
	\wick{\c1 \Upsilon_3\c1 \Upsilon_1} &=\wick{\c1 \Upsilon_1\c1 \Upsilon_3}=\Upsilon_1\cdot\slashed{P}_3\cdot\slashed{P}_1\cdot\slashed{P}_2\cdot\Upsilon_3=-2\hs P_{31}\hs (\Upsilon_1\cdot\slashed{P}_2\cdot\Upsilon_3)\,, \\
	\wick{\c1 \Upsilon_2\c1 \Upsilon_3} &=\wick{\c1 \Upsilon_3\c1 \Upsilon_2}=\Upsilon_3\cdot \slashed{P}_1\cdot\slashed{P}_2\cdot\slashed{P}_3\cdot\Upsilon_2=-2\hs P_{23}\hs (\Upsilon_3\cdot\slashed{P}_1\cdot  \Upsilon_2)\,,\\
	\wick{\c1 \Upsilon_1\c1 \Upsilon_1} &=\Upsilon_1\cdot \slashed{P}_3\cdot \slashed{P}_1\cdot\slashed{P}_2\cdot\slashed{P}_3\cdot  \Upsilon_1=-8\hs P_{23}\hs P_{31}\hs V_1
	\,,\\
	\wick{\c1 \Upsilon_2\c1 \Upsilon_2} &=\Upsilon_2\cdot  \slashed{P}_3\cdot \slashed{P}_1\cdot\slashed{P}_2\cdot\slashed{P}_3\cdot  \Upsilon_2=-8\hs P_{23}\hs P_{31}\hs V_2
	\,,\\
	\wick{\c1 \Upsilon_1\c1 \Upsilon_2} &=\Upsilon_1\cdot \slashed{P}_3\cdot \slashed{P}_1\cdot\slashed{P}_2\cdot\slashed{P}_3\cdot  \Upsilon_2=4\hs P_{23}\hs P_{31}\hs R_{12}\,,\\
	\wick{\c2\Upsilon_2'\c2\Upsilon_2'} &=\Upsilon_2\cdot \slashed{P}_1\cdot  \slashed{P}_3\cdot \slashed{P}_1\cdot\slashed{P}_2\cdot\slashed{P}_3\cdot  \slashed{P}_1\cdot\Upsilon_2=16\hs P_{31}^2 \hs P_{12}\hs  V_2\,,\\
	\wick{\c2\Upsilon_2'\c2\Upsilon_3} &=\Upsilon_2\cdot \slashed{P}_1\cdot  \slashed{P}_3\cdot \slashed{P}_1\cdot\slashed{P}_2\cdot\Upsilon_3=4\hs P_{12}\hs P_{31}\hs  R_{23}\,,\\
	\wick{\c2\Upsilon_2'\c2\Upsilon_1} &=\Upsilon_2\cdot \slashed{P}_1\cdot  \slashed{P}_3\cdot \slashed{P}_1\cdot\slashed{P}_2\cdot\slashed{P}_3\cdot\Upsilon_1=4\hs P_{12}\hs P_{31}\hs   (\Upsilon_2\cdot\slashed{P}_3 \cdot\Upsilon_1)\,,
\end{split}
\eeq
where we used $R_{ij}=\Upsilon_i\cdot \Upsilon_j$ and \eqref{equ:structure-V} to express the results in terms of the structures $R_{ij}$ and~$V_i$. Note that the contractions in \eqref{eq-contractions+++} are different from those defined  in \eqref{eq-contractions-++} even though we denote them in the same way.  Besides the structures $R_{ij}$ and $V_i$, there is also the parity-odd structure $\Upsilon_k\cdot \slashed{P}_i\cdot \Upsilon_j$ in these contractions, where $\{i,j,k\}$ is a cyclic permutation of $\{1,2,3\}$. It can be written in terms of the usual parity-odd structure in $3$d CFTs \cite{Costa:2011mg} as
	\be 
	\epsilon(W_j,W_k,P_1,P_2,P_3)
	=\frac{1}{4}R_{jk}\hs (\Upsilon_k\cdot \slashed{P}_i\cdot \Upsilon_j)\,.
	\ee 
	Note that the product of two parity-odd structures can always be written in terms of parity even structures as \cite{Giombi:2011rz}
\begin{align}\label{eq-structures-relation-SS}
	(\Upsilon_k\cdot \slashed{P}_i\cdot \Upsilon_j)^2 &=\frac{2 \hs P_{ij}\hs P_{ki}}{P_{jk}}\left(R_{jk}^2-4V_{j} V_{k}\right) ,\\
	(\Upsilon_k\cdot \slashed{P}_i\cdot \Upsilon_j)(\Upsilon_i\cdot \slashed{P}_j\cdot \Upsilon_k) &=2 \hs P_{ij}\left(2V_{k}R_{ij}-R_{jk}R_{ki}\right) .\label{eq-structures-relation-SS-2}
\end{align}
Hence, we can compute the final result of this twistor integral by writing each individual contraction in \eqref{equ:formula-JJJ} in terms of the corresponding structures $R_{ij}$, $V_i$ or $(\Upsilon_k\cdot \slashed{P}_i\cdot \Upsilon_j)$. Subsequently, we can use  \eqref{eq-structures-relation-SS} and \eqref{eq-structures-relation-SS-2} to replace the product of two parity-odd structures every time they appear, and use  \eqref{eq-structures-relation-RRR} to replace the structure $R_{12}R_{23}R_{31}$ every time it appears. As before, this procedure gets rid of redundancies, and we can easily evaluate the final result for specific spins as we do in Section~\ref{ssec:boot}.

\newpage
\section{Fourier Transforms and Dispersive Integrals}
\label{app:computations2}

In this appendix, we provide details of the half-Fourier transforms and dispersive integrals discussed in Section~\ref{ssec:tranforms}.

\subsection{Half-Fourier Transform}
\label{ssec:FT}

In order to compute the half-Fourier transform 
of the twistor correlator $F(Z_i\cdot Z_j)$ in~\eqref{equ:F-twistor-sec4}, we first need to express the products $Z_i\cdot Z_j$ in terms of the spinors $\lambda_i$, $\mu_i$ used in \eqref{equ:splitting-Z}. Notice that the twistors are contracted with the tensor
\be \label{equ:Omega-tau}
\Omega_{AB}=\left(\begin{array}{cc}
	0 &  \tau_\alpha^{~\dot{\beta}} \\
	-\tau_\beta^{~\dot{\alpha}} &  0 \\
\end{array}\right) ,
\ee 
where $\tau_\alpha^{~\dot{\beta}}=\delta_\alpha^{~\dot{\beta}}$. 
The inner products between twistors can therefore be written as
	\be \label{equ:product-Z}
	Z_i\cdot Z_j=Z_i^A\Omega_{AB}Z_j^B
	=\lambda_i\cdot \bar{\mu}_j-\lambda_j\cdot \bar{\mu}_i\,,
	\ee 
	where $\bar{\mu}_{\beta}\equiv \tau^{~\dot{\alpha}}_{\beta} \mu_{\dot{\alpha}}$ and $v\cdot u=v^\alpha u_\alpha$.

\vskip 4pt
We then substitute \eqref{equ:product-Z}  into~\eqref{equ:F-twistor-sec4} and compute its half-Fourier transform as 
\begin{align}\label{eq-ZZZ-firststep}
	&G(\lambda_i,\tilde\lambda_i) =\left[\prod_{i=1}^3\int d^2\mu_i 
	\hs \exp(i \tilde{\lambda}_i\cdot\mu_i)\right]F(Z_j\cdot Z_k) \\[4pt]
	&\hspace{-0.1cm}=(2\pi)^3 \int \hspace{-2pt}d^3c_{ij}\hs \delta^{(2)}(\bar{\lambda}_1^\alpha+c_{12}\lambda_2^\alpha-c_{31}\lambda_3^\alpha) \hs \delta^{(2)}(\bar{\lambda}_2^\alpha+c_{23}\lambda_3^\alpha-c_{12}\lambda_1^\alpha) \hs \delta^{(2)}(\bar{\lambda}_3^\alpha+c_{31}\lambda_1^\alpha-c_{23}\lambda_2^\alpha)\hs A(c_{ij})\,, \nonumber
\end{align}
where we changed variables from $\mu_{i,\alpha}$ to $\bar\mu_{i,\alpha}$ by using $\tilde{\lambda}_i\cdot\mu_i=\bar{\lambda}_i\cdot\bar\mu_i$, and noticed that the Fourier transforms  of the exponential in \eqref{equ:F-twistor-sec4} with respect to $\bar\mu_i$ yield the product of delta functions. Defining $r_{ij}\equiv \langle k \bar{i}\rangle/\langle  jk\rangle$, this product can be written as
  \begin{align}
&\delta^{(2)}(\bar{\lambda}_1^\alpha+c_{12}\lambda_2^\alpha-c_{31}\lambda_3^\alpha) \hs \delta^{(2)}(\bar{\lambda}_2^\alpha+c_{23}\lambda_3^\alpha-c_{12}\lambda_1^\alpha) \hs \delta^{(2)}(\bar{\lambda}_3^\alpha+c_{31}\lambda_1^\alpha-c_{23}\lambda_2^\alpha)\hs\label{equ:momentumdiracdelta-expanded}\\
&\quad =\frac{1}{|\langle 12\rangle\langle 23\rangle\langle 31\rangle|}\delta(r_{21}+r_{12}) \hs \delta(r_{13}+r_{31}) \hs \delta(r_{32}+r_{23}) \times \delta(c_{12}-r_{12}) \hs \delta(c_{23}-r_{23}) \hs \delta(c_{31}-r_{31})\, ,\nonumber
  \end{align}
which can be derived by expanding $\bar\lambda_i$ as a linear combination of $\lambda_j$ and $\lambda_k$, for $\{i,j,k\}$ a cyclic permutation of $\{1,2,3\}$, and performing some changes of variables.\footnote{
	We further used that, after expanding the argument of  $\delta^{(2)}(u^\alpha)$ in a certain basis,  $u^\alpha \equiv a_1 v_1^\alpha + a_2 v_2^\alpha$, the delta function becomes
	\be 
	\delta^{(2)}(u^\alpha) = \delta^{(2)}(a_1v_1^\alpha + a_2v_2^\alpha) = \delta^{(2)}\left((v_1^\alpha, v_2^\alpha)\begin{pmatrix}a_1\\a_2\end{pmatrix} \right) = \frac{1}{|v_1\cdot v_2|}\delta(a_1)\delta(a_2)\, ,
	\ee
	where $v_1\cdot v_2=v_1^\alpha v_2^\beta\hs \epsilon_{\alpha\beta}$ is the determinant of the $2\times 2$ matrix $(v_1^\alpha, v_2^\alpha)$.
} Employing similar manipulations, 
 we find that the factors on the right-hand side of \eqref{equ:momentumdiracdelta-expanded} that do not involve $c_{ij}$ yield
precisely the momentum-conserving delta function
 \begin{align} 
 \frac{\delta(r_{21}+r_{12}) \hs \delta(r_{13}+r_{31}) \hs \delta(r_{32}+r_{23})}{|\langle 12\rangle\langle 23\rangle\langle 31\rangle|} &=\int dE \,\delta^{(4)}\bigg(-E\epsilon^{\alpha\beta}+\sum_{i=1}^3\lambda_i^\alpha\bar{\lambda}_i^\beta\bigg)\nonumber\\
 &= \frac{1}{4} \hs \delta^{(3)}\big(\vec{k}_1 + \vec{k}_2 + \vec{k}_3 \big) \, .
 \end{align}
The remaining delta functions in \eqref{equ:momentumdiracdelta-expanded} trivialize the integral over $d^3c_{ij}$, as we only need to evaluate its integrand at
\be \label{equ:cij}
c_{ij}=r_{ij}=\frac{\langle k\bar{i}\rangle}{\langle  jk\rangle}=\frac{\langle \bar i\bar j\rangle}{E}\,,
\ee 
where we used that we are in the support of three-point kinematics (see equation (C.35) in \cite{Baumann:2020dch}).  
Putting everything together, we identify the result of \eqref{eq-ZZZ-firststep} as
\begin{align}\label{eq-halfFT-F}
	G(\lambda_i,\tilde\lambda_i) &= (2\pi)^3\delta^{(3)}\big(\vec{k}_1 + \vec{k}_2 + \vec{k}_3 \big)  \hs g(\lambda_i,\tilde\lambda_i)\,,\quad\text{where}\quad g(\lambda_i,\tilde\lambda_i)=\frac{A(c_{ij})}{4} \Big|_{c_{ij}=\langle \bar{i}\bar{j}\rangle/E}\,,
\end{align}
with $A(c_{ij})$ given by \eqref{equ:A-cij} for any spins $S_1$, $S_2$, $S_3$.  An analogous computation gives \eqref{equ:half-FT-Gtilde} for the half-Fourier transform of $\tilde F(W_i\cdot W_j)$. 

\subsection{Dispersive Integrals}

 In Section~\ref{ssec:tranforms}, we showed that the half-Fourier transform of the correlator \eqref{equ:F-dualtwistor-sec4} leads to the {\it discontinuity} of the momentum-space form  factor \eqref{equ:disc-fSSS}. In the following, we derive the actual form factor \eqref{equ:result2} by performing a dispersive integral of this discontinuity.

\begin{figure}[t!]
	\centering
	\includegraphics[width=1\textwidth]{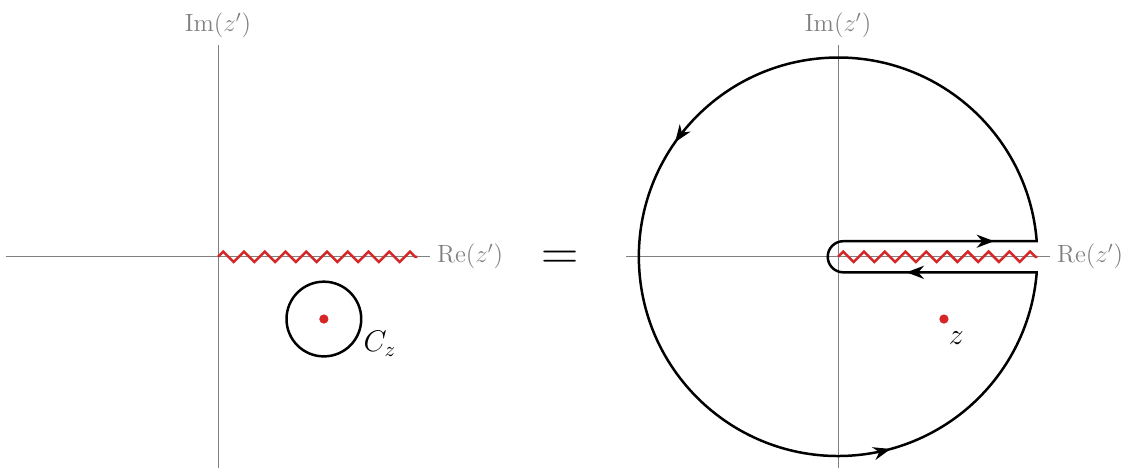}
	\caption{Deformation of the integration contour used in \eqref{equ:ContourIntegral}.}
	\label{fig:dispersion-contours}
\end{figure}

\paragraph{Complex analysis} We start with some general preliminaries.
First, notice that in order for the propagators to obey adequate boundary conditions in the early-time limit $\eta\to-\infty$, we need the complexified energy $k=|\vec{k}|$ to satisfy  Im$(k)<0$. 
Suppose that we have a function $\tilde{f}(k)$ which is analytic in the lower half-plane, i.e.~for $k=re^{i\delta}$, with $\delta\in(-\pi,0)$. Viewed as a function  of $z\equiv k^2$, i.e.~$f(k^2)\equiv\tilde{f}(k)$, 
it has a branch cut on the positive real axis. 
The discontinuity along this branch cut is defined as 
\be\label{equ:def-disc}
\text{Disc}_{k^2}f(k^2)\equiv  f(k^2+i\epsilon)-f(k^2-i\epsilon)\,.
\ee
Since the function $f(z)$ is analytic everywhere except on the branch cut, we can write
\be
f(z)=\frac{1}{2\pi i}\oint_{C_z} \frac{dz'}{z'-z}f(z') =\frac{1}{2\pi i}\int_0^\infty \frac{dz'}{z'-z}\displaystyle\text{Disc}_{z'} f(z')\, ,
\label{equ:ContourIntegral}
\ee
where the contour $C_z$ encircles $z$ counter-clockwise.
To obtain the result in the second equality, we have deformed
the contour of integration as in Figure \ref{fig:dispersion-contours}, and assumed that the integral over the large arc vanishes.   
Changing variables to $\omega^2=z'$ and $k^2=z$, we then get
\be
f(k^2)=\frac{1}{\pi i}\int_{0}^{\infty} \frac{\omega \hs d\omega}{\omega^2-k^2}\text{Disc}_{\omega^2}f(\omega^2)\,.
\ee
If $\text{Disc}_{\omega^2}f(\omega^2)$ is an odd function of $\omega$---which is the case for the discontinuity of the form factor in \eqref{equ:result3}---the integrand is an even function of $\omega$ and we can extend the integration to the entire real axis
\be
f(k^2)=\frac{1}{2\pi i}\int_{-\infty}^{\infty} \frac{\omega \hs d\omega}{\omega^2-k^2}\text{Disc}_{\omega^2}f(\omega^2)\,.
\label{equ:dispersive}
\ee
To compute this integral, we close the contour in the lower half-plane, and write it as a sum of residues of the poles $\omega_P$, with Im$(\omega_P)<0$. 

\paragraph{Einstein form factor}
We will now use this reasoning to recover the form factor \eqref{equ:f222} of the Einstein correlator from its discontinuity \eqref{equ:disc-fSSS}. Using \eqref{equ:result3} for $S=2$, the dispersive integral~(\ref{equ:dispersive}) implies
\be\label{equ:dispersive-f222}
f_{[222]}(k_i)\propto\left[\prod_{i=1}^3 \frac{1}{2\pi i}\int_{-\infty}^{\infty} \frac{\omega_i \hs d\omega_i}{\omega_i^2-k_i^2}\right] \underbrace{\displaystyle\frac{1}{4}\frac{(\omega_1\omega_2\omega_3)^3}{(E_\omega (E_\omega-2 \omega_1)(E_\omega-2 \omega_2)(E_\omega-2 \omega_3))^2}}_{\displaystyle=\tilde f_{[222]}(\omega_j)}\,,
\ee
where $E_\omega\equiv \omega_1+\omega_2+\omega_3$. The factors $E_\omega$ and $(E_\omega-2\omega_i)$ in the denominator are defined through the principal value prescription
\be 
\frac{1}{E_\omega^2}=\frac{1}{2}\left(\frac{1}{(E_\omega+i\epsilon)^2}+\frac{1}{(E_\omega-i\epsilon)^2}\right),
\ee 
which is inherited from the same prescription for the Schwinger parameters in the twistor correlator (see Footnote~\ref{footnote-pv-prescription}). 
Considering that each $k_i$ has a small negative imaginary part, it is straightforward to compute the integral over $\omega_3$ by summing over the residues of the five poles $\omega_{3,P}\in\{k_3,\pm\omega_1\pm\omega_2-i \epsilon\}$ in the lower half-plane: 
\begin{align} 
r_3(\omega_1,\omega_2,k_3)&\equiv\frac{1}{2\pi i}\int_{-\infty}^{\infty} \frac{\omega_3 \hs d\omega_3}{\omega_3^2-k_3^2} \tilde f_{[222]}(\omega_j)=-\sum_{\omega_{3,P}}\text{Res}_{\omega_3=\omega_{3,P}}\left[\frac{ \omega_3}{\omega_3^2-k_3^2} \tilde f_{[222]}(\omega_j)\right]\nonumber\\
&=-\frac{(\omega_1 \hs\omega_2\hs k_3)^3}{
	8 (k_3  + \omega_ 1 - \omega_ 2)^2 (k_3  - \omega_ 1 + \omega_ 2)^2 (-k_3  
	+ \omega_ 1 + \omega_ 2)^2 (k_3  + \omega_ 1 + \omega_ 2)^2}\,.
\end{align} 
Analogously, we can compute the integral over $\omega_2$ by adding up the residues of the three poles $\omega_{2,P}\in\{k_2,k_3+\omega_1,k_3-\omega_1\}$ with negative imaginary part:
\begin{align} 
r_{23}(\omega_1,k_2,k_3)&\equiv\frac{1}{2\pi i}\int_{-\infty}^{\infty} \frac{\omega_2 \hs d\omega_2}{\omega_2^2-k_2^2}r_3(\omega_1,\omega_2,k_3)=-\sum_{\omega_{2,P}}\text{Res}_{\omega_2=\omega_{2,P}}\left[\frac{ \omega_2}{\omega_2^2-k_2^2} r_3(\omega_1,\omega_2,k_3)\right]\nonumber\\
&=\frac{\omega_ 1^3 (k_2 ^2 + 4 k_2  k_3  + k_3 ^2 - \omega_ 1^2)}{256	(k_2  + 
	k_3  - \omega_ 1)^2 (k_2  + k_3  + \omega_ 1)^2}\,.
\end{align} 
Finally, the last integral over $\omega_1$ 
is evaluated by summing the residues at $\omega_{1,P} \in \{k_1,k_2+k_3\}$:
\begin{align}
f_{[222]}(k_i)&\propto 
\frac{1}{2\pi i}\int_{-\infty}^{\infty} \frac{\omega_1 \hs d\omega_1}{\omega_1^2-k_1^2}r_{23}(\omega_1,k_2,k_3)=-\sum_{\omega_{1,P}}\text{Res}_{\omega_1=\omega_{1,P}}\left[\frac{ \omega_1}{\omega_1^2-k_1^2} r_{23}(\omega_1,k_2,k_3)\right]\nonumber\\
&=\frac{Q(k_i)}{512\hs E^2}\,,
\end{align}
where the function $Q(k_i)$ was defined below \eqref{equ:f222}. This is precisely 
 the form factor \eqref{equ:f222} for the three-point function in Einstein gravity.

\subsection{Mixed-Helicity Correlators}

So far, we have only derived equal-helicity correlators. 
It is thus natural to ask how to recover the results for mixed-helicity correlators.

\vskip 4pt
Recall that, in the ``product-based" integrals like \eqref{equ:twistorintegral-product}, we take the spinor to be a twistor~$Z^A$, while, in the ``derivative-based" integrals like \eqref{equ:alt}, we use the dual twistor $W_A$. 
However, this is an arbitrary choice, because we can change variables from $W_A$ to $Z^A=\Omega^{AB}W_B$ in the derivative-based integral, without changing the result of \eqref{equ:alt}.  
Similarly, the product-based integral \eqref{equ:twistorintegral-product} does not change if we change variables from $Z^A$ to $W_A=\Omega_{AB}Z^B$. 
However, the result of the half-Fourier transform {\it does} change, because the map depends on whether the spinor is a twistor~$Z^A$ or a dual twistor $W_A$: It is given by  \eqref{equ:fouriertransform-mu} or \eqref{equ:fouriertransform-mutilde}, respectively. Indeed, we will show that the choice determines the helicities of the resulting half-Fourier transforms.

\paragraph{Relations between transforms} We will first derive the relation between the half-Fourier transform of $\tilde F(W_A)$ in dual twistor space and the transform of $\tilde F(\Omega_{AB}Z^B)$ in twistor space. Writing $Z^B \equiv (\lambda^\beta,\mu_{\dot\beta})$, and using  \eqref{equ:Omega-tau}, we have 
\be 
\Omega_{AB}Z^B=
- \begin{pmatrix}\tilde \mu_{\alpha}\\[4pt]
	\tau_\beta^{~\dot\alpha}\lambda^{\beta}
\end{pmatrix}  ,
\ee 
where $\tilde\mu_\alpha \equiv - \tau_\alpha^{~\dot\beta}\mu_{\dot\beta}$.
The half-Fourier transform of $\tilde F(\Omega_{AB}Z^B)$ then reads  
\be 
\label{equ:G}
G(\lambda,\tilde\lambda) = \int d^2\mu\, 
\hs e^{i\tilde\lambda\cdot \mu}\hs \tilde F(\Omega_{AB}Z^B)=\int d^2\tilde\mu\,
\hs e^{-i\bar\lambda\cdot \tilde\mu}\hs \tilde F(\Omega_{AB}Z^B)\,,
\ee 
where we changed of variables from $\mu_{\dot\beta}$ to $\tilde\mu_\alpha$ and recalled the definition  $\bar\lambda_\alpha \equiv \tau_\alpha^{~\dot\beta}\tilde\lambda_{\dot\beta}$. We further assumed that the spin $S$ is an integer, so that $\tilde F(-W_A)=\tilde F(W_A)$, otherwise there would be an extra factor of $(-1)^{2S}$  coming from the scaling  in \eqref{equ:scaling-Ftilde}. The right-hand side of (\ref{equ:G}) coincides precisely with the half-Fourier transform of $\tilde F(W_A)$, given by \eqref{equ:fouriertransform-mutilde}, after exchanging $\lambda_\alpha\leftrightarrow \bar\lambda_\alpha$. To be explicit,  the relation between these two transforms is 
\be
\int  d^2\mu\, 
\hs e^{i\tilde\lambda\cdot \mu}\hs \tilde F(\Omega_{AB}Z^B)=\hat{T} \left[\int d^2\tilde\mu\,
\hs e^{-i\bar\lambda\cdot \tilde\mu}\hs \tilde F(W_A) \right] ,
\ee 
where the operator $\hat{T}$ exchanges $\lambda_\alpha\leftrightarrow \bar\lambda_\alpha$. 
Similarly, the half-Fourier transform of $F(\Omega^{AB}W_B)$ can be obtained by applying this operator to the half-Fourier transform of $F(Z^A)$. 

\vskip 4pt
While the half-Fourier transforms of $F(Z^A)$ and $\tilde F(W_A)$ have both positive helicity, the transforms of $F(\Omega^{AB}W_B)$ and $\tilde F(\Omega_{AB}Z^B)$ have negative helicity. This is simply because the operator $\hat{T}$ flips the helicity in momentum space. 

\paragraph{Mixed-helicities correlators}

Using the operators $\hat T_i$, for each field $i$, we can now derive the results for mixed-helicity correlators from our previous results for the all-plus correlators.

\vskip 4pt
As an example, 
let us consider the correlators of identical spin-$S$ currents. We can obtain the half-Fourier transform of $\tilde F(\Omega_{AB}Z_1^B,W_{2,A},W_{3,A})$  by applying the operator $\hat{T}_1$ to the result for the half-Fourier transform of $\tilde F(W_{1,A},W_{2,A},W_{3,A})$.  
Since $\hat{T}_1$ maps $k_{1,\mu}=(k_1,\vec k_1)\mapsto (-k_1,\vec k_1)$, it flips only the sign of the energy $k_1$, while keeping the three-momentum $\vec k_1$ fixed. It does not change the delta function of momentum conservation, and we can therefore apply this operator directly to the result \eqref{equ:half-FT-gtilde} after stripping this delta function:
\be \label{equ:half-FT-gtilde-++}
\tilde g_{-++}(\lambda_i,\tilde\lambda_i)\equiv\hat{T}_1\hs \tilde g(\lambda_i, \tilde \lambda_i) =\frac{1}{4} \left(\displaystyle\frac{\langle  1\bar2\rangle\langle \bar2\bar3\rangle\langle \bar 3 1\rangle}{E(E-2k_2)(E-2k_3)}\right)^S \,.
\ee  
The definition of the form factor 
associated to a given correlator depends on the helicity configuration, as we need to exchange barred and unbarred spinors in the corresponding prefactor. In particular, the form factor $\tilde f_{[SSS]}^{-++}(k_i)$ associated to \eqref{equ:half-FT-gtilde-++} is
\be 
\tilde g_{-++}(\lambda_i,\tilde\lambda_i) \equiv \left(\frac{\langle1 \bar2 \rangle \langle \bar2\bar3\rangle \langle \bar31\rangle (E-2k_1)}{ k_1k_2k_3}\right)^S \frac{\tilde f_{[SSS]}^{-++}(k_i)}{(k_1k_2k_3)^{S-1}}\,.
\ee 
Using this definition, 
we find that the form factor coincides precisely with that associated to the all-plus result in~\eqref{equ:result3}, i.e.~$\tilde f_{[SSS]}^{-++}(k_i)=\tilde f_{[SSS]}(k_i)$.

\vskip 4pt
Interestingly, the result of the half-Fourier transform $\tilde g(\lambda_i,\tilde\lambda_i)$ in \eqref{equ:half-FT-g} can be obtained by combining the correlators \eqref{equ:result2} with different helicities as
\begin{align}\label{equ:half-FT-That-combination}
\tilde g(\lambda_i,\tilde\lambda_i) \propto\frac{\langle J^+_1 J^+_2 J^+_3\rangle}{(k_1k_2k_3)^{S-1}} &+\left(\hat{T}_1 \frac{\langle J^-_1 J^+_2 J^+_3\rangle}{(k_1k_2k_3)^{S-1}}+\text{cyclic}\right)\nonumber\\
&+\left(\hat{T}_1\hat{T}_2 \frac{\langle J^-_1 J^-_2 J^+_3\rangle}{(k_1k_2k_3)^{S-1}}+\text{cyclic}\right)+\hat{T}_1\hat{T}_2\hat{T}_3 \frac{\langle J^-_1 J^-_2 J^-_3\rangle}{(k_1k_2k_3)^{S-1}}\,,
\end{align}
where we have applied appropriate factors of $\hat T_i$ to change negative helicities to positive helicities. We interpret the combination on the right-hand side as a discontinuity of the momentum-space correlators.
A similar formula exists also for $\tilde g_{-++}(\lambda_i,\tilde\lambda_i)$. 

\vskip 4pt
Finally, we can find the half-Fourier transform of $ F(\Omega^{AB}W_{1,B},Z_2^A,Z_3^A)$ by applying $\hat{T}_1$ to the result for  the half-Fourier transform of $F(Z_1^A,Z_2^A, Z_3^A)$. After stripping off the delta function of momentum conservation, and using the result \eqref{equ:half-FT-g}, this gives
\be 
g_{-++}(\lambda_i,\tilde\lambda_i)\equiv \hat{T}_1\hs g(\lambda_i,\tilde\lambda_i)=\frac{1}{4}\left(\frac{\langle 1\bar{2}\rangle\langle \bar{2}\bar{3}\rangle\langle \bar{3}1\rangle}{(E-2k_1)^3}\right)^S \,.
\ee  
Although the momentum-space correlator with mixed helicities vanishes, $\langle J_1^- J_2^+ J_3^+ \rangle = 0$, the result of this half-Fourier transform is non-zero, $g_{-++}(\lambda_i,\tilde\lambda_i) \ne 0$. We can interpret this by thinking of $g_{-++}$ as a discontinuity of the momentum correlators as in \eqref{equ:half-FT-That-combination}, which combines all possible helicity configurations. Since the correlator $\langle J_1 J_2 J_3 \rangle$ 
does not vanish for all-plus and all-minus configurations, there is no reason for $g_{-++}$ to vanish. 

\newpage
\section{Bootstrapping Two-Point Functions}\label{sec:TwoPoints}

In this appendix, we will study the two-point functions of conserved currents in twistor space, as well as their maps to both embedding and momentum space.

\subsection{Computing the Twistor Integral} 

Let us consider the ansatz \eqref{equ:npt-ansatz}, for $n=2$:
\beq
\langle J_1J_2 \rangle = \left[ \prod_{j=1}^2 i^{-S_j}\int DZ_j \hs (\Upsilon_j^*\cdot Z_j)^{2S_j} \right]
F(Z_1 \cdot Z_2) \, . \label{equ:2pt-ansatz}
\eeq
 The integrand $F(Z_1\cdot Z_2)$ must scale in the same way with respect to both twistors $Z_1$ and $Z_2$, which implies that the spins must be equal $S_1=S_2=S$.  
Moreover, in order for the integral \eqref{equ:2pt-ansatz} to be invariant under rescalings, the integrand must be  $F\propto (Z_1\cdot Z_2)^{-2S-2}$, which is even under PT as it satisfies \eqref{equ:PT-F}. For later convenience, we normalize $F$ as
\be \label{equ:2pt-F}
F(Z_1\cdot Z_2)=\int\frac{dc_{12}}{2\pi} 
\hs c_{12}^{2S+1}\hs\text{sign}(c_{12})\hs e^{-ic_{12} Z_1\cdot Z_2}=\frac{i^{2S+2}\hs (2S+1)!}{\pi} 
\frac{1}{(Z_1\cdot Z_2)^{2S+2}}\,,
\ee 
where the integral over the Schwinger parameter $c_{12}$ was computed following equation (7.6) of~\cite{Mason:2009sa}. Ignoring an overall constant, the twistor integral  \eqref{equ:2pt-ansatz} then gives 
\be 
\langle J_1J_2\rangle \propto \int DZ_1 \left(\Upsilon_1^*\cdot Z_1\right)^{2S}\int DZ_2 \left(\Upsilon_2^*\cdot Z_2\right)^{2S}\hs \frac{1}{(Z_1\cdot Z_2)^{2S+2}}
\propto \frac{H_{12}^S}{P_{12}^{2S+1}}\,,
\label{equ:2pt-twistorintegral}
\ee 
which agrees with the known two-point function \eqref{equ:2pt} in embedding space.

\begin{framed}
{\small \noindent {\bf Proof}\ \ First, we write the integrand of the $Z_2$ integral as
\be 
i^{-2S}\left(\Upsilon_2^*\cdot Z_2\right)^{2S}\hs F(Z_1\cdot Z_2)=\left(\Upsilon_2^*\cdot \frac{\partial}{\partial Z_1}\right)^{2S}\hs \int \frac{dc_{12}}{2\pi} 
\hs |c_{12}|\hs e^{-ic_{12} Z_1\cdot Z_2}\,.
\ee 
It is convenient to
 choose the reference spinor $\Upsilon_2^*$ to be
\be \label{equ:upsilon2-2pt}
\Upsilon_2^* \equiv \frac{\slashed{P}_1\cdot \Upsilon_2}{2 P_1\cdot P_2}\,,
\ee
so that $\Upsilon_2^{*,A}\slashed{P}_{1,A}^{~~B}=0$. Following~\eqref{equ:int-parts-Vderivative}, we then integrate the $Z_1 $ derivatives by parts:
\begin{align}
\langle J_1J_2\rangle&=\int DZ_1 \left(\Upsilon_1^*\cdot Z_1\right)^{2S}\int DZ_2 \left(\Upsilon_2^*\cdot \frac{\partial}{\partial Z_1}\right)^{2S}\hs \int \frac{dc_{12}}{2\pi}
\hs |c_{12}|\hs e^{-ic_{12} Z_1\cdot Z_2}\nonumber\\
&=\int DZ_1 \int DZ_2  \int \frac{dc_{12}}{2\pi}
\hs |c_{12}|\hs e^{-ic_{12} Z_1\cdot Z_2}\left(-\Upsilon_2^*\cdot \frac{\partial}{\partial Z_1}\right)^{2S}\hs\left(\Upsilon_1^*\cdot Z_1\right)^{2S}\nonumber \\[4pt]
&=\left(\int DZ_1 \int DZ_2  \int \frac{dc_{12}}{2\pi}
 \hs |c_{12}|\hs e^{-ic_{12} Z_1\cdot Z_2}\right) \hs(2S)!\hs \left(\frac{R_{12}}{2 \hs P_{12}}\right)^{2S} \,,\label{equ:2pt-middlestep}
\end{align} 
where we recalled our choice \eqref{equ:upsilon2-2pt} for $\Upsilon_2^*$, and used that $\Upsilon_1=\slashed{P}_1\cdot \Upsilon_1^*$ and $R_{ij}\equiv \Upsilon_i\cdot \Upsilon_j$. 

\vskip 4pt
It then only remains to compute the integral in the parenthesis in \eqref{equ:2pt-middlestep}, which we define as $2\pi \hs I_0$. 
We combine the projective measure $DZ_2$ with the integral over $c_{12}$ to get the non-projective measure $DZ_2\wedge dc_{12} \hs |c_{12}|=d^2Z_2$. Recalling the parameterization $Z_2^A \equiv \pi_2^b\hs\Lambda_{2,b}^A$, the integral over $d^2Z_2=d^2\pi_2$ then becomes
\begin{align}
I_0&\equiv\int DZ_1 \int DZ_2  \int \frac{dc_{12}}{(2\pi)^2} \hs |c_{12}|\hs e^{-ic_{12} Z_1\cdot Z_2}=\int DZ_1 \int \frac{d^2\pi_2}{(2\pi)^2}\hs e^{-i Z_1\cdot \Lambda_{2,b}\pi_2^b}\nonumber\\
&=\int DZ_1\hs \delta^{(2)}(Z_1\cdot \Lambda_{2,b}) \,.
\end{align}
Further using $Z_1^A \equiv \pi_1^a\hs\Lambda_{1,a}^A$, the remaining integral over $DZ_1=D\pi_1$ can be written as
\begin{align}\label{equ:2pt-I0}
I_0=\int D\pi_1\hs \delta^{(2)}(\pi_1^a\hs\Lambda_{1,a}\cdot \Lambda_{2,b})=\frac{1}{|2\hs P_{12}|}\hs  \int D\pi_1\hs\delta^{(2)}(\pi_1^a) \,,
\end{align}
where we performed a change of variables in the delta function, using
\be 
\det(\Lambda_{1,a}\cdot \Lambda_{2,b})=\frac{1}{2}\epsilon^{ab}\epsilon^{cd}(\Lambda_{1,a}\cdot \Lambda_{2,c})(\Lambda_{1,b}\cdot \Lambda_{2,d})=\frac{1}{2}\text{Tr}(\slashed{P}_1\cdot \slashed{P}_2)=-2P_{12}\,.
\ee 
Since the factor $\int D\pi_1\hs\delta^{(2)}(\pi_1^a)$ is just an overall kinematic-independent constant,\footnotemark~we can plug this back into \eqref{equ:2pt-middlestep} to obtain 
\be 
\langle J_1 J_2\rangle\propto \frac{R_{12}^{2S}}{P_{12}^{2S+1}}\propto \frac{H_{12}^S}{P_{12}^{2S+1}}\,,
\ee 
where we assumed $P_{12}>0$ (see  Footnote~\ref{footnote:analytic-continuation}).
}
\end{framed}
\footnotetext{Notice that
	\be 
	1=\int d^2\pi\hs\delta^{(2)}(\pi^a)=\left(\int D\pi\hs\delta^{(2)}(\pi^a)\right) \left(\int \frac{dt}{|t|}\right) ,
	\ee 
	where we wrote the non-projective measure as $d^2\pi=D\pi\wedge dt\hs |t|$. Since the integral $ \int dt/|t|$ along the real line diverges, the projective integral $\int D\pi_1\hs\delta^{(2)}(\pi_1^a)$ should technically vanish. However, it has been argued in~\cite{Neiman:2017mel} that the projective and non-projective integrals differ by an overall factor like $\int dt/|t|$ that can be treated as finite. Hence, the projective integral should be a non-vanishing constant given this prescription, which can then be safely ignored. In any case, this factor is an overall constant that can be absorbed in the normalization of the two-point function, as it does not depend on the kinematic variables.}

Finally, let us mention that there is a way to represent \eqref{equ:2pt-twistorintegral} as an integral of the form \eqref{equ:npt-ansatz2} by Fourier transforming the integrand with respect to $Z_1$ and $Z_2$: 
\be 
\tilde F(W_1\cdot W_2)=\int \frac{d^4Z_1}{(2\pi)^2}\hs e^{iZ_1\cdot W_1}\int \frac{d^4Z_2}{(2\pi)^2}\hs e^{iZ_2\cdot W_2}\hs F(Z_1\cdot Z_2)\propto \int dc_{12} \frac{\text{sign}(c_{12})}{c_{12}^{2S-1}}e^{-ic_{12}\hs W_1\cdot W_2}\,,
\ee 
where the integral over $c_{12}$ has been computed in equation (7.7) of \cite{Mason:2009sa}. The twistor integral~\eqref{equ:npt-ansatz2}, for $n=2$, with this integrand gives the same result as~\eqref{equ:2pt-twistorintegral}. 

\newpage
\subsection{Transform to Momentum Space}

In the following, we will map the two-point function $F(Z_1\cdot Z_2)$ given by \eqref{equ:2pt-F} to momentum space via the half-Fourier transform \eqref{equ:fouriertransform-mu}. We will show that
\be\label{equ:2pt-halfFT}
\left[\prod_{j=1}^2\int d^2\mu_j
\hs e^{i\hs\tilde\lambda_j\cdot \mu_j}\right]F(Z_1\cdot Z_2)=(2\pi)^3\delta^{(3)}(\vec k_1+\vec k_2)~\frac{1}{4}\frac{\langle \bar 1 \bar 2\rangle^{2S}}{E^{2S-1}}\,,
\ee
where $E=k_1+k_2$ is the total energy, which is taken to be positive. This result is proportional to the known two-point function in momentum space (see e.g.~equation (3.5) of \cite{Maldacena:2011nz}).

\begin{framed}
	{\small \noindent {\bf Proof}\ \ Plugging the result~\eqref{equ:2pt-F} into the integrand of the half-Fourier transform \eqref{equ:2pt-halfFT}, we obtain
	\be \label{equ:2pt-halfFT-middlestep}
	\left[\prod_{j=1}^2\int d^2\mu_j
	\hs  e^{i\hs\tilde\lambda_j\cdot \mu_j}\right]F(Z_1\cdot Z_2)=(2\pi)^3\int dc_{12}\hs \delta^{(2)}(\bar\lambda_1^\alpha+c_{12}\lambda_2^\alpha)\hs \delta^{(2)}(\bar\lambda_2^\alpha-c_{12}\lambda_1^\alpha)\hs c_{12}^{2S+1}\hs\text{sign}(c_{12})\,,
	\ee 
where we used \eqref{equ:product-Z} and changed variables from $\mu_{i,\alpha}$ to $\bar\mu_{i,\alpha}$ by using $\tilde{\lambda}_i\cdot\mu_i=\bar{\lambda}_i\cdot\bar\mu_i$. The integrals over $\bar\mu_j^\alpha$ simply give the product of the two delta functions. Expanding $\bar\lambda_1^\alpha \equiv a_1\lambda_1^\alpha+b_1\lambda_2^\alpha$ and $\bar\lambda_2^\alpha \equiv b_2\lambda_1^\alpha+a_2\lambda_2^\alpha$, with the coefficients given by $a_1=\langle \bar 1 2 \rangle/\langle 12\rangle$, $b_1=-2k_1/\langle 12\rangle$, $b_2=2k_2/\langle 12\rangle$, and $a_2=\langle 1\bar 2 \rangle/\langle 12\rangle$, we can write these delta functions as
\be \label{equ:2pt-product-Dirac}
\delta^{(2)}(\bar\lambda_1^\alpha+c_{12}\lambda_2^\alpha)\hs \delta^{(2)}(\bar\lambda_2^\alpha-c_{12}\lambda_1^\alpha)=\frac{1}{|\langle 12\rangle|^2}\delta(a_1)\delta(a_2)\delta(b_1+b_2)\delta(c_{12}-b_2)\,.
\ee 
Performing the same expansion, it is straightforward to show that the three-dimensional delta function of momentum conservation can be written as
\be
\frac{1}{4}\delta^{(3)}\big(\vec{k}_1 + \vec{k}_2  \big) =\int dE \,\delta^{(4)}\bigg(-E\epsilon^{\alpha\beta}+\sum_{i=1}^2\lambda_i^\alpha\bar{\lambda}_i^\beta\bigg)=\frac{1}{|\langle 12\rangle|^3}\delta(a_1)\delta(a_2)\delta(b_1+b_2)\, .
\label{equ:XXXX}
\ee
Comparing \eqref{equ:2pt-product-Dirac} to \eqref{equ:XXXX}, we can express the half-Fourier transform \eqref{equ:2pt-halfFT-middlestep} as

\be 
\left[\prod_{j=1}^2\int d^2\mu_j 
\hs e^{i\hs\tilde\lambda_j\cdot \mu_j}\right]F(Z_1\cdot Z_2)=
(2\pi)^3\delta^{(3)}\big(\vec{k}_1 + \vec{k}_2  \big)\hs\frac{1}{4} |\langle 12\rangle|\hs \int dc_{12}\hs\delta(c_{12}-b_2) \hs c_{12}^{2S+1}\hs \text{sign}(c_{12})\,.
\ee 
Hence, the integral over $c_{12}$ can be easily computed by evaluating the integrand for
\be 
c_{12}=b_2=\frac{2k_2}{\langle 12\rangle}=\frac{E}{\langle 12\rangle}=\frac{\langle\bar 1\bar2\rangle}{E}\,,
\ee 
where we defined $E=k_1+k_2=2k_2$ and used that momentum conservation implies the last equality. We therefore obtain
\be 
\left[\prod_{j=1}^2\int d^2\mu_j 
\hs e^{i\hs\tilde\lambda_j\cdot \mu_j}\right]F(Z_1\cdot Z_2)=(2\pi)^3\delta^{(3)}\big(\vec{k}_1 + \vec{k}_2  \big)~\frac{1}{4}\frac{\langle \bar 1 \bar 2\rangle^{2S}}{E^{2S-1}}  \hs\text{sign}(E)\,,
\ee 
where the sign of $E$ is simply $+1$ provided we take $E$ to be positive. This concludes the proof of~\eqref{equ:2pt-halfFT}.
}
\end{framed} 
\newpage
%%%%%%%%%%%%%%%%%%%%%%%%%
\phantomsection
%\enlargethispage{\baselineskip}
%\addtocontents{toc}{\protect\enlargethispage{\baselineskip}}
\addcontentsline{toc}{section}{References}
\bibliographystyle{utphys}
{\linespread{1.075}
	\bibliography{TwistorSpace-Refs}
}

\end{document}